\documentclass{aa}
\usepackage{amssymb,amsmath,amsfonts}
\usepackage{graphicx}
\usepackage{color}
\usepackage{ulem}             % \sout
\usepackage{multicol}
\usepackage{caption}
\usepackage{pgf,tikz}
\usetikzlibrary{arrows}
\usepackage{colortbl}
\usepackage{xcolor}

\usepackage{amssymb}
\usepackage{latexsym}
\usepackage{pstricks}
\usepackage{pst-plot}
\usepackage{tabularx}
\usepackage{mathtools}
\usepackage{bm}
\usepackage{soul}
\usepackage{cancel}

	\usetikzlibrary[patterns]

\newcommand{\me}{\, {\rm M}_{\oplus}}
\newcommand{\au}{\, {\rm au}}

\newcommand\eps{{\varepsilon} }

\newcommand\ab{{\overline a}_\Gamma }

\newcommand\ol{\overline  }
\newcommand\xt{{\tilde x}}
\newcommand\yt{{\tilde y}}

\newcommand\cG{{\cal G} }

\newcommand\gO{{\cal O} }

\newcommand\gH{{\cal H} }

\newcommand\cL{{\cal L} }
\newcommand\Dv{{\Delta \varpi} }

\newcommand{\be}{\begin{equation}}
\newcommand{\ee}{\end{equation}}

 {\everymath{\displaystyle\everymath{}}\array}%
 {\endarray}

\begin{document}

\title{Stability of the co-orbital resonance under dissipation}
\subtitle{Application to its evolution in protoplanetary discs}
\titlerunning{Evolution of the co-orbital resonance in protoplanetary discs}

\author{
Adrien Leleu$^1$\thanks{\email{adrien.leleu@space.unibe.ch}, CHEOPS fellow}
\and Gavin A.\ L.\ Coleman$^1$
\and Sareh Ataiee$^2$
}
%%%% AENDERN %%%%%%
\authorrunning{A. Leleu, G. Coleman and S. Ataiee}

\institute{
$^1$ Physikalisches Institut, Universit\"at Bern, Gesellschaftsstr.\ 6, 3012 Bern, Switzerland.\\
$^2$ Universität Tübingen, Institut für Astronomie und Astrophysik,  
Computational Physics, Auf der Morgenstelle 10, D-72076 Tübingen,  
Germany.
}
%
%Despite the existence of co-orbital bodies in the solar system, and the prediction of the formation of coorbital planets by planetary system formation models, no co-orbital exoplanets (also called trojans) have been detected so far. We investigate how a pair of trojan exoplanets would fare during their migration in a protoplanetary disc. To this end, we start with an analytical study of the evolution of two planets near the Lagrangian equilibria L4 and L5, identifying for which values of the parameters these equilibria are either attractive or repulsive. We then compare these results to hydrodynamical simulations. Finally, we study the evolution of co-orbital configurations using a planetary system formation model that simulates the orbital evolution of the planet over the disc lifetime. Depending on the parameters of the disc, and the orbital parameters and masses of the planets, the system can either evolve toward the Lagrangian equilibrium, or tend to increase its amplitude of libration, possibly all the way to horseshoe orbits or even exiting the resonance. The stability in the direction of the eccentricities and the inclinations is also studied.

\abstract
{Despite the existence of co-orbital bodies in the solar system, and the prediction of the formation of co-orbital planets by planetary system formation models, no co-orbital exoplanets (also called trojans) have been detected thus far. In this paper we investigate how a pair of co-orbital exoplanets would fare during their migration in a protoplanetary disc. 

To this end, we computed a stability criterion of the Lagrangian equilibria $L_4$ and $L_5$ under generic dissipation and slow mass evolution. Depending on the strength and shape of these perturbations, the system can either evolve towards the Lagrangian equilibrium, or tend to increase its amplitude of libration, possibly all the way to horseshoe orbits or even exiting the resonance. 
%We then compare these results to a planetary system formation model that simulates the orbital evolution of the planets over the disc lifetime, and to hydrodynamical simulations on much shorter time scale. 
We estimated the various terms of our criterion using a set of hydrodynamical simulations, and show that the dynamical coupling between the disc perturbations and both planets have a significant impact on the stability: the structures induced by each planet in the disc perturb the dissipative forces applied on the other planets over each libration cycle.

Amongst our results on the stability of co-orbitals, several are of interest to constrain the observability of such configurations: long-distance inward migration and smaller leading planets tend to increase the libration amplitude around the Lagrangian equilibria, while leading massive planets and belonging to a resonant chain tend to stabilise it. We also show that, depending on the strength of the dissipative forces, both the inclination and the eccentricity of the smaller of the two co-orbitals can be significantly increased during the inward migration of the co-orbital pair, which can have a significant impact on the detectability by transit of such configurations. 
%We then compare these results to the evolution of a pair of co-orbital planets in an evolving 1-D disc model, and to hydrodynamical simulations. Although our criterion correctly predict the stability of co-orbitals in the 1-D disc model, we show through the hydrodynamical simulations that 2-dimensional effects require to be properly modelled 

%we illustrate that in type II migration the $L_4$ and $L_5$ equilibria can be either attractive or repulsive depending on the disc parameters. The stability in the direction of the eccentricities and the inclinations is also studied.
}
%\textit{Context}
%{ Absence of detected co-orbital exoplanets so far despite the theoretical prediction of their existence.}
%\textit{Aims}
%{ We present a detailed exploration of possible formation pathways of \hipb\ within the core-accretion model. }
%    %
%    %\textit{Methods}
%    { Formation-model-dependent and -independent constraints on the planet's post-formation entropy are derived from its luminosity.
%    We use the Coleman et al.\ formation model which includes $N$-body dynamics with several embryos per disc to study the migration history,
%    taking the upper limits on further companions into account. We also look at the high rotation velocity of the star.}
%    %
%    %\textit{Results}
%    { Formation close-in/far-out is likely/not likely. It is easy/not easy to form it with several companions in the dics.
%    Rotation of star yields/does not yield information on formation.}
%    %
%    %\textit{Conclusions}
%    {
%    Interesting system. Great paper.}


\maketitle

\section{Introduction}

Among the known multiplanetary systems, a significant number contain bodies in (or close to) first and second order mean-motion resonances (MMR) \citep{Fabrycky2014}. However, no planets were found in a zeroth order MMR, also called trojan or co-orbital configuration, despite several dedicated studies \citep{MaWi2009,Ja2013,LiBo2017,LiBo2018}. The formation of the first and second order resonances is generally explained by the convergent migration of two planets under the dissipative forces applied by the protoplanetary disc \citep[see for example][]{LePe2002}. In the co-orbital case, the resonance is surrounded by a chaotic area due to the overlap of first-order MMRs \citep{Wi1980,DePaHo2013}. The crossing of this area generally results in the excitation of the bodies' eccentricities, leading to a collision or scattering instead of the capture into the co-orbital resonance.% Co-orbital bodies are nonetheless common in our solar system \citep[the Earth, Mars, Jupiter, Uranus and Neptune have known co-orbitals, along with several Saturnian moons][]{MPC2018}, albeit there is generally a huge mass difference between the co-orbital bodies, except for the Saturnian moons Janus and Epimetheus ($m_{Janus}/m_{Epimetheus} \approx 3.6$). %Jupiter's trojan were proposed to be captured during mean-motion crossing with Saturn \citep{MoLeTsiGo2005}.  

Two processes that can form co-orbital exoplanets were proposed by \cite{LauCha2002}: either planet-planet gravitational scattering, or accretion in situ at the $L_4/L_5$ points of a primary. The assumptions made on the gas-disc density profile in the scattering scenario can either lead to systems with a high diversity of mass ratios \citep{CreNe2008,CreNe2009}, or to equal mass co-orbitals when a density jump is present \citep{GiuBe2012}. In their model, \citet{CreNe2008} form co-orbitals in over $30\%$ of the generated planetary systems, with very low inclinations and eccentricities ($e<0.02$). In several hydrodynamical simulations of the formation of the outer part of the solar system by \cite{Crida2009}, Uranus and Neptune ended up in co-orbital configuration, while both were trapped in the same MMR with Saturn. In the in situ scenario, different studies yielded different upper limits to the mass that can form at the $L_4/L_5$ equilibrium point of a giant planet: \citet{BeSa2007} obtained a maximum mass of $\sim 0.6 M_\oplus$, while \citet{LyJo2009} obtained $5 -15 M_\oplus$ planets in the tadpole area of a Jupiter-mass primary. 

The growth and evolution of co-orbitals have also been studied. For existing co-orbitals, \citet{CreNe2009} found that gas accretion increases the mass difference between the planets, with the more massive of the two reaching Jovian mass while the starving one stays below $70 M_\oplus$. They also found that inward migration tends to slightly increase the amplitude of libration of the co-orbital, while remaining within the tadpole domain. The divergence from the resonance accelerates during late migrating stages with low gas friction, which may lead to instabilities. Another study from \citet{PiRa2014} shows that equal mass co-orbitals (from super-Earths to Saturns) are heavily disturbed during the gap-opening phase of their evolution. \citet{RoGiMi2013} have also shown that in some cases long-lasting tidal evolution may perturb equal mass close-in systems.

In this work we aim to better understand what causes the stability or instability of the co-orbital configuration during the protoplanetary disc phase.
To do this we study the effects that planet migration, through interactions with a protoplanetary gas disc, has on the evolution of the co-orbital resonance angle.
We also examine the evolution of the eccentricities and inclinations of the co-orbitals as the planets migrate throughout the disc.
Both the type I (when a planet is fully embedded in the protoplanetary disc), and the type II (when the planet is massive enough to significantly perturb the disc, i.e. open a gap in the disc) migration regimes are studied.

This paper is laid out as follows.
In Sect. \ref{sec:coorbcons}, we describe the co-orbital dynamics in the absence of dissipation. In Sect. \ref{sec:H3bp}, we develop an integrable analytical model of the co-orbital resonance under a generic dissipation and mass change, in the coplanar-circular case. The application to type I migration is performed in Sect. \ref{sec:type1}, first using a simple analytical model for the disc torque, then comparing to the long-term evolution in an evolving protoplanetary disc. The result of population synthesis and the effect of resonant chains on the co-orbital configuration will be discussed in that Sect. as well. We then estimate the forces that are actually applied on the coorbital by performing hydrodynamical simulation in Sect. \ref{sec:type2}. Finally, in Sect. \ref{sec:stabei}, we discuss the stability of the co-orbital configuration in the direction of the eccentricities and inclinations.
We then draw our conclusions in Sect. \ref{sec:conclusions}.

%While Jupiter's trojan were proposed to be captured during mean-motion crossing with Saturn \citep{MoLeTsiGo2005}, formation processes   Co-orbital planets can also be disturbed in the presence of additional planetary companions, in particular by a significantly massive planet in another MMR with the co-orbitals \citep{MoLeTsiGo2005, RoBo2009}. 

\section{Dynamics of the co-orbital resonance in the non-dissipative case}
\label{sec:coorbcons}
%We expose in this section  of two planets orbiting a star in the same plane and without any dissipative force. This problem is usually referred to as the planar planetary three body problem. We note with a subscript 1 the internal planet and 2 the external one. The star is referred to as body 0. Masses are noted mi. For both planets we define: μi = G(m0+mi), and βi = m0mi/(m0+mi), where G is the gravitational constant.We note ai the semi-major axis of planet i, and ei its eccentricity.  

In this section we describe the co-orbital motion of two planets of masses $m_1$ and $m_2$ around a central star of mass $m_0$ without any dissipative forces. For both planets we define: $\mu_j= \cG (m_0+m_j)$, and $\beta_j= m_0 m_j/(m_0+  m_j)$, where $\cG$ is the gravitational constant. We note $a_j$ the semi-major axis of planet $j$, $e_j$ its eccentricity, and $I_j$ its inclination. We use Poincar\'e astrocentric coordinates for both planets:
\begin{equation}
\begin{aligned}
\Lambda_j &=\beta_j\sqrt{\mu_ja_j}\, , \vspace{2cm} &\lambda_j &=\lambda_j\, , \\ 
x_j & =\sqrt{\Lambda_j}\sqrt{1-\sqrt{1-e_j^2}}\operatorname{e}^{i\varpi_j}\, , \vspace{2cm} &\tilde{x}_j &=-i\bar{x}_j\, , \\
y_j & =\sqrt{\Lambda_j}\sqrt{\sqrt{1-e_j^2}(1-\cos I)}\operatorname{e}^{i\Omega_j}\, , \vspace{2cm} &\tilde{y}_j &=-i\bar{y}_j\, , \\
\end{aligned}
\label{eq:poincvar}
\end{equation}
where $\lambda_j$, $\varpi_j$ and $\Omega_j$ are its mean longitude, longitude of the pericenter, and ascending node of each planet, and $\bar x_j$ and $\bar y_j$ are the complex conjugates of  $x_j$ and  $y_j$, respectively. %
The Hamiltonian of the system reads, in these coordinates:
\begin{equation}
\begin{aligned}
H = & H_K(\Lambda_1,\Lambda_2) \\
& + H_P(\lambda_1,\lambda_2,\Lambda_1,\Lambda_2,x_1,x_2,\xt_1,\xt_2,y_1,y_2,\yt_1,\yt_2)\, ,
\end{aligned}
\label{eq:Hpoinc}
\end{equation}
where $ H_K$ is the Keplerian component and $ H_P$ is the perturbative component due to planet-planet interactions taking into account both direct
and indirect effects. The Keplerian component depends only on $\Lambda_j$:
\begin{equation}
H_K =- \sum^2_{j=1}  \left( \frac{\mu_j^2 \beta_j^3}{2 \Lambda_j^2} \right)\, , 
\label{eq:keppoinc}
\end{equation}
whereas the perturbative component depends on all twelve Poincar\'e coordinates. We do not need to express the explicit form of $H_P$ at
this point but it could be seen as an expansion of the $x_j$ and $y_j$ variables around 0.  $H_P$ can be obtained for example using the algorithm developed in \cite{LaRo1995}.\\

As we study here the $1:1$ mean motion resonance, we place ourselves in the neighbourhood of the exact Keplerian resonance defined by $\dot \lambda_1=\dot \lambda_2$. The equations canonically associated with the Hamiltonian (\ref{eq:Hpoinc}) hence read:
\begin{equation}
\frac{\partial H_K}{\partial \Lambda_1}(\Lambda_1,\Lambda_2) = \frac{\partial H_K}{\partial \Lambda_2}(\Lambda_1,\Lambda_2) \, ,
\label{eq:defeta1}
\end{equation}

We note $\Lambda_1^0$ and $\Lambda_2^0$ are the solutions of Eqs. (\ref{eq:defeta1}) and (\ref{eq:keppoinc}). $\Lambda_1^0$ and $\Lambda_2^0$ are uniquely determined if we choose the exact Keplerian resonance which has the same total angular momentum than the studied orbit. At first order in $e$ and $I$, the total angular momentum reads:
\begin{equation}
L = \Lambda_1+\Lambda_2= \Lambda^0_1+\Lambda^0_2 \, .
\label{eq:defGamma}
\end{equation}
We can thus express $\Lambda_1^0$ and $\Lambda_2^0$ as a function of $L$:
\begin{equation}
\begin{aligned}
\Lambda^0_{jL}=m_j L/(m_1+m_2) + \gO(\eps) \, , \\
\end{aligned}
\label{eq:LA0j}
\end{equation}
where $\eps=\gO((m_1+m_2)/m_0)$. We can also define the average mean-motion $\eta$ associated to the exact Keplerian resonance by, at order $0$ in $\eps$:
\begin{equation}
\eta_L=  \frac{\mu^2_j \beta_j^3}{(\Lambda^0_j)^3}=\mu_0^2\left( \frac{m_1+m_2}{L} \right)^3\, ,
\label{eq:defeta}
\end{equation}
where $\mu_0=\cG m_0$. We also define the associated semi-major axis:
 \begin{equation}
\ab= \frac{L^2}{\mu_0(m_1+m_2)^2} \, .
\label{eq:defab}
\end{equation}

\subsection{Averaging the Hamiltonian near the co-orbital resonance}

Since the mean motions $n_j$ of the two bodies are close at any given time, the quantity $\zeta=\lambda_1-\lambda_2$ evolves slowly with respect to the longitudes. The Hamiltonian (\ref{eq:Hpoinc}) hence possesses $3$ time-scales: a fast one, associated with the mean motion $\eta$ and the mean longitudes, a semi-fast one, associated with the resonant frequency $\nu=\gO(\sqrt{\eps})$ and the libration of the resonant angle $\zeta$, and a slow time-scale (called secular), which is associated with the orbital precession and the variables $x_j$, $\xt_j$, $y_j$ and $\yt_j$.
%We note $\nu \propto \sqrt{ \eps} n_1$ the fundamental frequency associated with the resonant angle $\zeta$. 
To emphasise the separation of these time-scales, we process to the following canonical change of variables ($x_j$ and $y_j$ remain unchanged):
\begin{equation}
\begin{aligned}
 \begin{pmatrix}
  \zeta \\
  \zeta_2
   \end{pmatrix} &=
   \begin{pmatrix}
  1 & -1 \\
  0 & \phantom{-}1
   \end{pmatrix} 
      \begin{pmatrix}
  \lambda_1 \\
  \lambda_2
   \end{pmatrix}\, , 
       \begin{pmatrix}
  Z \\
  Z_2
   \end{pmatrix} &=
   \begin{pmatrix}
  1 & \phantom{-} 0 \\
  1 & \phantom{-}1
   \end{pmatrix} 
      \begin{pmatrix}
  \Lambda_1 \\
  \Lambda_2
   \end{pmatrix}\, .
   \end{aligned}
   \label{eq:tl1}
   \end{equation}
%
%to obtain the following Hamiltonian:

The Hamiltonian now reads:
\begin{equation}
\begin{aligned}
\gH =& \gH_K(Z,Z_2)\\
    & +  \gH_P(\zeta,\zeta_2,Z,Z_2,x_1,x_2,\xt_1,\xt_2,y_1,y_2,\yt_1,\yt_2)\, .
\end{aligned}
\label{eq:Hts}
\end{equation}

%The dissipative terms introduced in equation (\ref{eq:syshamd}) are slow as well with respect to the semi-fast motion, as  $1/\tau = \gO(\eps)$. 
%

The separation between the time-scales allows for the averaging over the rapid angle $\zeta_2$. 
%$\zeta_2$ does not appear in the dissipative part of the equation, while the conservative part is transformed in the same way than in previous works 
Following \citet{RoPo2013,RoNi2015}, we obtain the Hamiltonian:
\begin{equation}
\ol \gH = \gH_K(Z,Z_2) + \ol \gH_P(\zeta,Z,Z_2,x_j,\xt_j,y_j,\yt_j)\, . 
\label{eq:Hbp}
\end{equation}

\subsection{Circular coplanar case}

 \begin{figure}
\begin{center}
\includegraphics[width=0.99\linewidth]{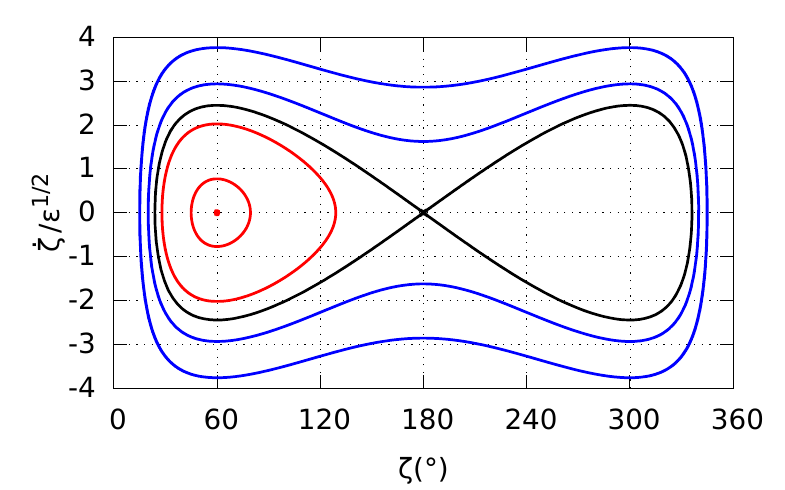}
\caption{\label{fig:zeta} Phase space of Eq. (\ref{eq:eqerdi}). The black line represents the separatrix between tadpole orbits (3 examples in red) and horseshoe orbits (in blue). The phase space is symmetric with respect to $\zeta=180^\circ$.}
\end{center}
\end{figure}

In the coplanar circular case, 1D models of the $1:1$ mean-motion (co-orbital) can be obtained taking $x_j=y_j=0$ and developing the Hamiltonian (\ref{eq:Hbp}) at second order in $Z-\Lambda^0_1$ and  $Z_2-(\Lambda^0_1+\Lambda^0_2)$ \citep{RoNi2015}. The equation canonically associated with that Hamiltonian can be rewritten as a 2nd order differential equation \citep{Ed1977,RoNi2015} :
\begin{equation}
\ddot{\zeta}=-3 \eta^2 \frac{m_1+m_2}{m_0} \left( 1- (2- 2\cos \zeta)^{-3/2} \right) \sin \zeta\, ,
\label{eq:eqerdi}
\end{equation}
%
%where $\eta$ is the average mean-motion of the two bodies in resonance, and $m_0$ is the mass of the central body. 
The phase space of Eq. (\ref{eq:eqerdi}) is shown in Fig. \ref{fig:zeta}. 

Out of the four fixed points of Eq. (\ref{eq:eqerdi}), the collision ($\zeta=0^\circ$) is not in the validity domain of the equation and will be ignored. $\zeta=180^\circ$ is the hyperbolic (unstable) $L_3$ Lagrangian equilibrium, while $\zeta=\pm60^\circ$ are elliptic (stable) configurations, the $L_4$ and $L_5$ Lagrangian equilibria. Orbits that librate around these elliptic equilibria are called tadpole, or trojan (in reference to Jupiter's trojan swarms). Examples of trojan orbits are shown in red in Fig. \ref{fig:zeta}. The separatrix emanating from $L_3$ (black curve) delimits trojan orbits from horseshoe orbits (examples are shown in blue), for which the system undergoes large librations that encompasses $L_3$, $L_4$ and $L_5$. 

As previously stated, the libration of the resonant angle $\zeta$ is slow with respect to the average mean-motion $\eta$. The fundamental libration frequency $\nu$ is proportional to $\sqrt{(m_1+m_2)/m_0}\eta$. In the neighbourhood of the $L_4/L_5$ equilibria \citep{Charlier1906}:
\begin{equation}
\nu_0 =\sqrt{\frac{27}{4}}  \sqrt{\frac{m_1+m_2}{m_0}}\eta .
\label{eq:nu_L}
\end{equation}

\subsection{Dynamics in the eccentric and inclined directions}
\label{sec:dynei}
In order to study the co-orbital dynamics for low eccentricities and inclinations, $\gH_P$ can be expanded in Taylor series in the neighbourhood of ($x_1$,$x_2$,$y_1$,$y_2$)=($0$,$0$,$0$,$0$), at 2nd order in $x_j$, $y_j$. This expansion can be written as \citep{La89,RoPo2013}:  

\begin{equation}
\begin{aligned}
\ol \gH= & \ol \gH_0(\zeta,Z,Z_2)+\ol \gH^{(2)}_x(\zeta,Z,Z_2,x_j,\xt_j)\\
 & +\ol \gH^{(2)}_y(\zeta,Z,Z_2,y_j,\yt_j)\, .
\end{aligned}
\label{eq:Hbexp}
\end{equation}
where $\ol \gH_0$ is given in appendix \ref{ap:ham}, and $\gH^{(2)}_x$ and $\gH^{(2)}_y$ are sums of quadratic monomials in ($x_j$,$\xt_j$) and ($y_j$,$\yt_j$), respectively. 

From this form we learn two things: at low eccentricity and inclination, the dynamics of the variables $\bm{x}=(x_1,x_2)$ and $\bm{y}=(y_1,y_2)$ are decoupled, and the dynamics of the variables $\zeta$ and $Z$ are independent of $x_j$ and $y_j$ at first order. In the conservative case, the variational equations of $\bm{x}$ and $\bm{y}$ are given by \cite{RoPo2013,RoNi2015}:
\begin{equation}
\dot{\bm{x}}=
M_x(\zeta)
\bm{x} \, ,   \ 
\dot{\bm{y}} =
M_y(\zeta)
\bm{y}  \, ,
   \label{eq:xy}
   \end{equation}
with
\begin{equation}
M_x(\zeta)=
   \begin{pmatrix}
  \frac{A_x(\zeta)}{m_1} &  \frac{\overline B_x(\zeta)}{\sqrt{m_1 m_2}} \\
  \frac{B_x(\zeta)}{\sqrt{m_1 m_2}}&  \frac{A_x(\zeta)}{m_2}
   \end{pmatrix}  \, ,   \ 
M_y=
   \begin{pmatrix}
  \frac{A_y(\zeta)}{m_1}&  \frac{\overline B_y(\zeta)}{\sqrt{m_1 m_2}} \\
  \frac{B_y(\zeta)}{\sqrt{m_1 m_2}}&  \frac{A_y(\zeta)}{m_2}
  \end{pmatrix} \, ,
   \label{eq:MxMy}
   \end{equation}
where $A_x$, $B_x$, $A_y$, $B_y$ are complex-valued functions of $\zeta$ and are given in Appendix \ref{ap:ham}. $\overline B$ represents the complex conjugate of $B$.

%In the coplanar circular case, 1-Dimension models  were developed \citep{Ed1977,RoPo2013}, describing the co-orbital dynamics as long as $m_1$ and $m_2$ are not too close to each other (outside of the Hill's sphere).

\subsubsection{The eccentric direction}
\label{sec:conse}

 \begin{figure}
\begin{center}
\includegraphics[width=0.49\linewidth]{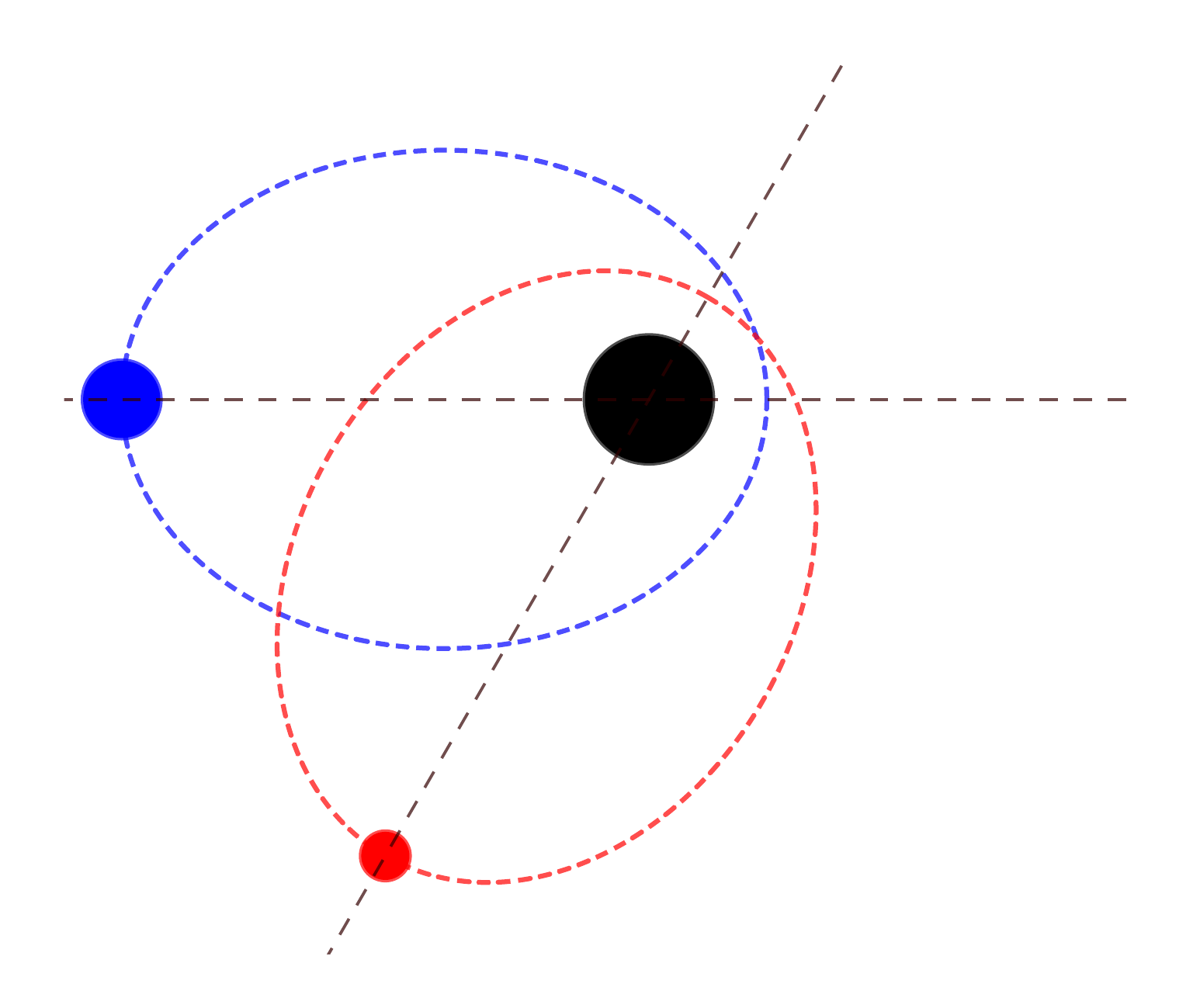}\includegraphics[width=0.49\linewidth]{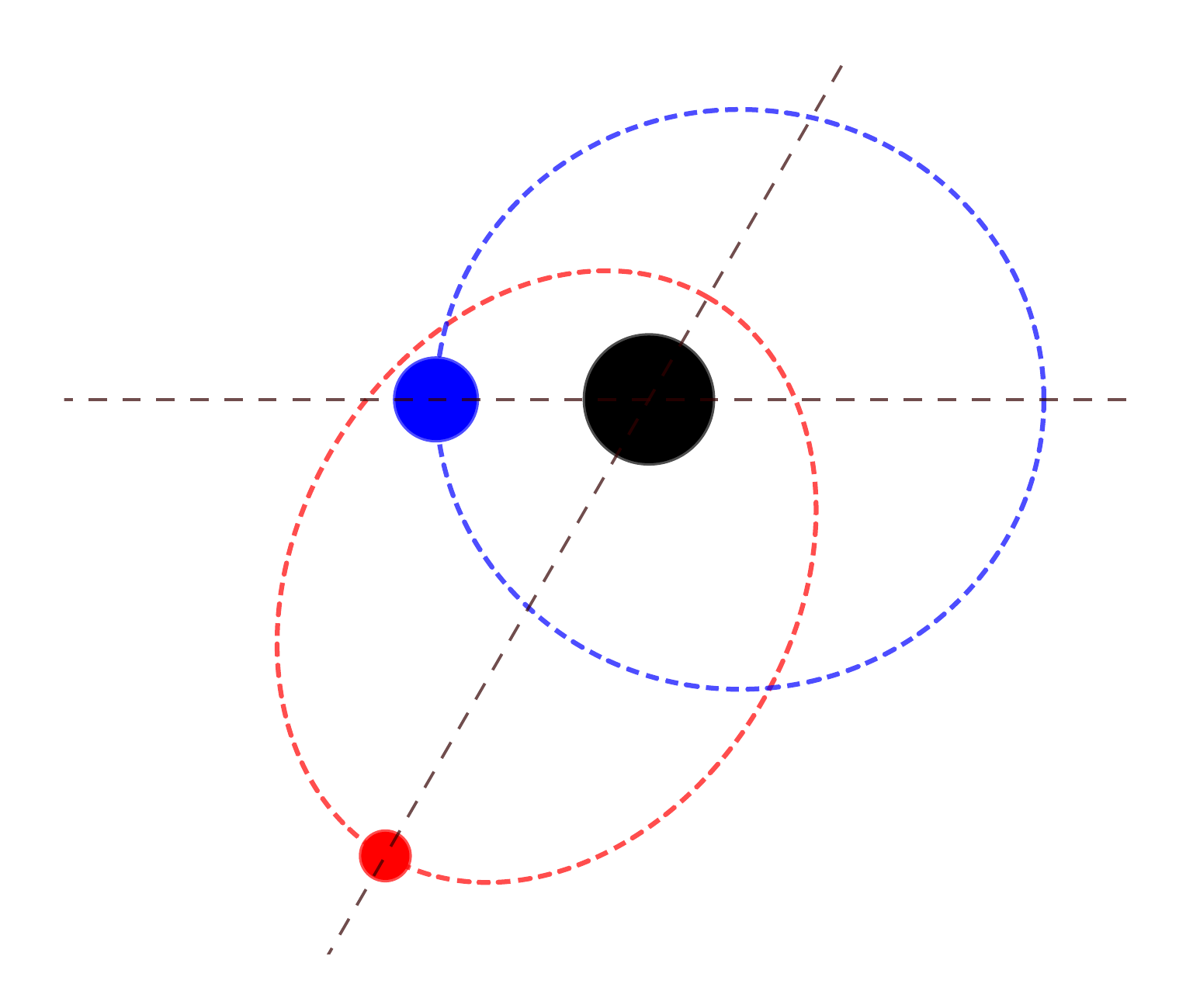}
\caption{\label{fig:AL4EL4} Left, the Eccentric Lagrangian equilibrium: $e_1=e_2$, $\varpi_1-\varpi_2=\zeta=\pm 60^\circ$ and right the Anti-Lagrangian equilibrium. At first order in the eccentricities: $m_1e_1=m_2e_2$, $\varpi_1-\varpi_2=\zeta+180^\circ$.  }
\end{center}
\end{figure}

The dynamics in the direction of the eccentricities is given by the system (\ref{eq:xy}). Although the coefficients of the matrix $M_x$ are functions of the resonant angle $\zeta$, $\zeta$ does not depend on the eccentricities at first order. One can hence evaluate $M_x$ at a fixed point of the ($\zeta$,$Z$) variables ($L_4$ will be described here, results are equivalent for $L_5$), and study the dynamics of the $x_j$ variable in the neighbourhood of the $L_4$ circular equilibria. The eigenvector of the matrix $M_x(L_4)$ gives the direction of two remarkable families of quasi-periodic orbits that are represented in Fig. \ref{fig:AL4EL4} \citep{GiuBeMiFe2010,RoPo2013}:\\

\noindent - The first eigenvector, paired with a null eigenvalue $g_-=0$, is tangent to the eccentric Lagrangian family ($EL_4$), for which $e_1=e_2$ and $\Dv=\varpi_1-\varpi_2=\zeta$.\\

\noindent - The second eigenvector, paired with the eigenvalue $g_+=27/8 (m_1+m_2)/m_0 \eta$, is tangent to the Anti-Lagrangian family ($AL_4$). For low eccentricities, this family is tangent to $m_1 e_1=m_2 e_2$ and $\Dv=\varpi_1-\varpi_2=\zeta+\pi$.\\

Description of the dynamics at larger eccentricities can be found for example in \cite{NeThoFeMo02}, \cite{GiuBeMiFe2010}, and \cite{LeRoCo2018}.

\subsubsection{The inclined direction}
\label{sec:consi}
 \begin{figure}
\begin{center}
\includegraphics[width=0.99\linewidth]{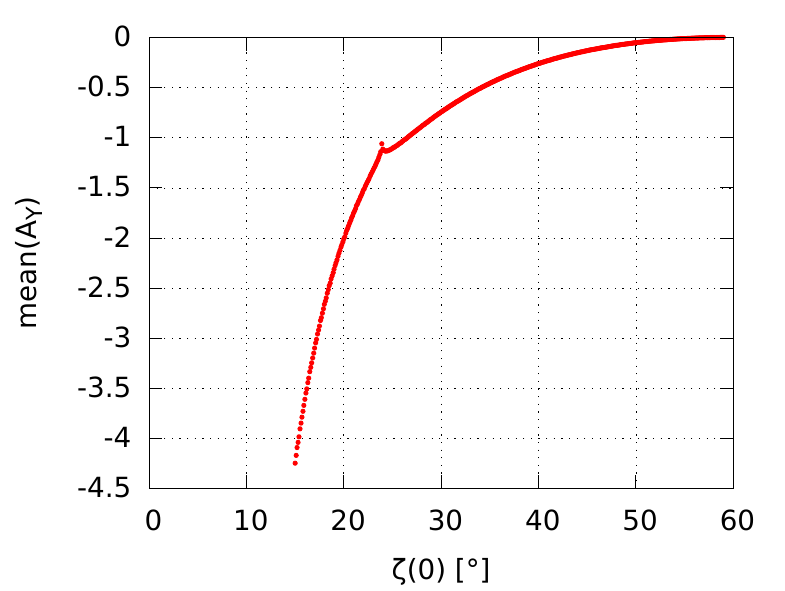}
\caption{\label{fig:Ab_vs_zet0} Evolution of $\bar A_y$ with respect to the min value of $\zeta$ on a given orbit. The Lagrangian equilibrium is located at $\zeta_0=60^\circ$, while the separatrix between the horseshoe and tadpole domains is at $\zeta_0\approx 24^\circ$. }
\end{center}
\end{figure}

The dynamics in the direction of inclination is given by the system (\ref{eq:xy}). In opposition to the eccentric direction, we do not learn anything by evaluating $M_y$ at the $L_4$ equilibria since all of its coefficients vanish for $\zeta=60^\circ$. However, since the evolution of $\zeta$ is fast with respect to the secular evolution on the $y_j$ variables, we can obtain an approximation of the secular dynamics in the  direction of the inclination for a given trajectory by averaging the expression of this system over a period $2\pi/\nu$ with respect to the time $t$. 

We note that $Im(B_y(\zeta))=-A_y(\zeta)$ (see Appendix \ref{ap:ham}). In addition, the real part of $B_y(\zeta)$ is proportional to the expression of $\ddot \zeta$ (Eq. \ref{eq:eqerdi}), which is the derivative of a periodic function of period $2\pi/\nu$. Its average value over $2\pi/\nu$ is hence null. As a result, the averaged $M_y$ can be written:
\begin{equation}
\bar M_y=-i \frac{m_1m_2}{2m_0} \eta  \bar A_y
   \begin{pmatrix}
  \frac{1}{m_1}&  \frac{-1}{\sqrt{m_1 m_2}} \\
  \frac{-1}{\sqrt{m_1 m_2}}&  \frac{1}{m_2}
  \end{pmatrix} \, ,
   \label{eq:Myav}
   \end{equation}
where $\bar A_y$ is the averaged value of $A_y(\zeta)$ over a period $2\pi/\nu$, see Fig. \ref{fig:Ab_vs_zet0}. At $\zeta_0=60^\circ$ (Lagrangian equilibrium), $\bar A_y=0$ and the system is degenerate. If $\bar A_y \neq 0$, we identify the two eigenvectors:\\

   \noindent - The first eigenvector, paired with a null eigenvalue $s_-=0$, is tangent to the direction $I_2=I_1$ and $\Omega_2=\Omega_1$: the two planets orbit in the same plane, inclined by $I_1=I_2$ with respect to the reference frame.\\

   \noindent - The second eigenvector, paired with the eigenvalue $s_+=-i (m_1+m_2) \eta  \bar A_y/(2m_0)$, is tangent to the direction $m_1 I_1=m_2 I_2$ and $\Omega_2=\Omega_1+\pi$: the inclination of both co-orbitals is constant and their lines of nodes slowly precess at the frequency $s_+$. \\
   
 We note that in the second direction the $Oxy$ plane of the reference frame is perpendicular to the total angular momentum of the system (i.e. the $Oxy$ plane is the invariant plane).

%, then we can show that, in this reference frame, the inclinations of the co-orbitals are constant. Indeed, by definition of the invariant plane we have, at first order in the inclination:
%\begin{equation}
%m_1 y_1=m_2 y_2 \exp^{i \pi}\, .
%   \label{eq:yjrelation}
%   \end{equation}
%%
%
%Combining equation (\ref{eq:Myav}) and (\ref{eq:yjrelation}), we obtain:
%%
%\begin{equation}
%\dot y_1=-im_1m_2 blabla \bar A_y (1/m_1-\sqrt m_1/(m_2\sqrt m_2)) y_1\, ,
%   \label{eq:y1ev}
%   \end{equation}
%%
%and a similar expression for $y_2$. The norms of the $y_j$ are hence constant over time. As a result, in the frame linked to the invariant plane, the inclination of co-orbital bodies are constant at first order, with a slow precession of the line of the nodes.

%
%%
%hamiltonian can be obtained using for example Laskar \& robutel. previous studies developed Integrable model. Showed that e and i direction were decoupled at the lowest order. Showed stability condtions and diferent eigenvectors. 

\section{Stability of the Lagrangian equilibria $L_4$ and $L_5$ under external forces and mass change - the coplanar circular case}
\label{sec:H3bp}

In this section we study the stability of the Lagrangian equilibria in the coplanar circular case under a generic dissipation, modelled by forces $F_1$ and $F_2$ applied on each planet. We assume that these forces are small with respect to the attraction by the central star. Using Gauss' equations, the Poincaré variables are modified in the following way: 
  
%\begin{equation}
%\begin{aligned}
%\dot\Lambda_{j,d} &=m_j a_j T_j\, , \vspace{2cm} &\dot \lambda_{j,d} &=-\frac{2}{a_jn_j} R_j \, .
%\end{aligned}
%\label{eq:poincvard}
%\end{equation}
%
\begin{equation}
\begin{aligned}
\dot\Lambda_{j,d} &=\Gamma_j = \frac{\Lambda_j^2}{ \cG m_0 m_j^2} F_{t,j} \, , \\
\dot \lambda_{j,d} &= R_j =-\frac{2 \Lambda_j}{\cG m_0 m_j^2 } F_{r,j} \, .
\end{aligned}
\label{eq:poincvard}
\end{equation}
Where $F_{r,j}$ is the radial force applied on the planet $j$, while $ \Gamma_j $ is the torque induced by the tangential force $F_{t,j}$. If the forces vary significantly over the orbital time-scale, they will also excite the eccentricities of the planets. For this work, we assume that variation of these forces over an orbital period is negligible, and that non-axisymmetric dissipative forces applied on the planets can be parametrised by $ \Lambda_1$, $\Lambda_2$, and $\zeta= \lambda_1-\lambda_2$. 
%We also consider a slow, isotropic mass change for the two planets. We assume that this mass change occur on a time-scale long with respect to the resonant motion, and so $m_1$ and $m_2$ remain adiabatic invariant of the problem. Denoting $ \dot L_{m}$ the contribution of this mass change to the evolution of the total angular momentum, we obtain:
%%
%\begin{equation}
%%\left\{ 
%\begin{aligned}
%\dot L &= \Gamma_1+  \Gamma_2 + \dot L_{m}\, ,
%\end{aligned}
%%\right.
%\label{eq:sysL}
%\end{equation}
%% %

\subsection{Model of the 1:1 MMR under dissipation and mass change}

We start by developing an analytical model for the dynamics of the co-orbital configuration in the neighbourhood of the Lagrangian equilibria $L_4$ and $L_5$ in the dissipative case. 
%, as well as in model of ring-induced migration \citep{Wisdom19??}, and tides \citep{Burns1977}. 
To develop this model, we head back to the Hamiltonian transformations that we described in Sect. \ref{sec:coorbcons}: we rewrite the equation of variation using the canonical variables $\Lambda_j$, $\lambda_j$. The equations of motion are given by the equation canonically associated with the Hamiltonian $H$, Eq. (\ref{eq:Hpoinc}), to which we add the effect of the dissipation (Eq. \ref{eq:poincvard}), and a slow, isotropic mass change for the planets parametrised by two constants $\dot m_1$ and $\dot m_2$. 
%We then close the system of equation by adding the equation for the evolution of the total angular momentum of the system:
%% 
%\begin{equation}
%%\left\{ 
%\begin{aligned}
%\dot L &= \Gamma_1+  \Gamma_2 + \frac{\dot m_1}{m_1} \Lambda_1 + \frac{\dot m_2 }{m_2} \Lambda_2 \, .
%\end{aligned}
%%\right.
%\label{eq:sysL}
%\end{equation} 
%

We then perform a change of variables to uncouple the fast (i.e. associated to the mean motion) and semi-fast (i.e. resonant) degrees of freedom \citep{RoNi2015}:
\begin{equation}
\begin{aligned}
 \begin{pmatrix}
  \zeta \\
  \varphi
   \end{pmatrix} &=
   \begin{pmatrix}
 1 & -1 \\
  \frac{m_1}{m_1+m_2} &  \frac{m_2}{m_1+m_2}
   \end{pmatrix} 
      \begin{pmatrix}
  \lambda_1 \\
  \lambda_2
   \end{pmatrix}\, , \\ 
       \begin{pmatrix}
   \hat \Delta \\
   L
   \end{pmatrix} &=
   \begin{pmatrix}
  \frac{m_2}{m_1+m_2} & -\frac{m_1}{m_1+m_2} \\
  1 & 1
   \end{pmatrix} 
      \begin{pmatrix}
 \Lambda_1 \\
  \Lambda_2
   \end{pmatrix}\, .
   \end{aligned}
   \label{eq:tZDel}
   \end{equation}
It should be noted that this change of co-ordinates is canonical in the conservative case, here we need to add the terms relative to the dissipative forces and mass changes. Averaging over the fast angle $\varphi$, we obtain the following system:
\begin{equation}
\begin{aligned}
\dot{\hat{ \Delta}}  & = \mu_0^2 \frac{m_1 m_2 (m_1+m_2)^2}{m_0 L^2 } \left( 1-\frac{1}{\delta(\zeta)^3} \right) \sin \zeta\\
&  + \frac{m_2 \Gamma_1-m_1\Gamma_2}{(m_1+m_2)} + \frac{m_1^2 \dot m_2 + m_2^2 \dot m_1}{m_1m_2(m_1+m_2)}\hat \Delta\, , \\
\dot{\zeta} & =  - 3 \mu_0^2 \frac{(m_1+m_2)^5}{m_1m_2 L^4} \hat \Delta + R_1-R_2\, ,\\
\dot L &= \Gamma_1 +  \Gamma_2+\left( \frac{\dot m_1+\dot m_2 }{m_1+m_2}\right) L   \, , \\
%&+ \frac{\dot m_1}{m_1} \left( \frac{m_1}{m_1+m_2} L + \hat\Delta \right) + \frac{\dot m_2 }{m_2}  \left( \frac{m_2}{m_1+m_2} L - \hat \Delta \right)
\dot{\varphi} & =  f(\hat \Delta, L, \zeta)  + \frac{m_2R_2+m_1R_1}{m_1+m_2}\, ,\\
\end{aligned}
\label{eq:syserdid}
\end{equation}
with $f$ a polynomial function of $\hat \Delta$ and $L$, with trigonometric terms in $\zeta$, and $\delta(\zeta) =\sqrt{2-2\cos \zeta}$. $\Gamma_j$ and $R_j$ remain unchanged as we neglected their evolution over an orbital period. For the rest of this study, we focus on the evolution of the resonant degree of freedom ($\hat \Delta$,$\zeta$). 

\subsection{Constant torques, radial forces and masses}
\label{sec:constantF}
At first we only consider the constant part of the torques and radial forces,  $\Gamma_{j0}$ and $R_{j0}$, and constant masses ($\dot m_j=0$). In this case, the evolution of the resonant degree of freedom ($\hat \Delta$,$\zeta$) is given by the equations associated to the following Hamiltonian:
\begin{equation}
\begin{aligned}
\gH_r =&  \frac{m_1m_2\eta_L L}{m_0(m_1+m_2)} (F(\zeta)  - C_\Gamma \zeta)\\
 & - \frac{3}{2} \eta_L \frac{(m_1+m_2)^2}{m_1m_2 L} \hat \Delta( \hat \Delta -2 \hat \Delta_{eq}) \, ,\\
\end{aligned}
\label{eq:Hr}
\end{equation}
where $F(\zeta)=\cos \zeta- (2-2\cos \zeta)^{-1/2}$,  
 \begin{equation}
C_\Gamma = \left( \frac{\Gamma_{10}}{m_1} - \frac{\Gamma_{20}}{m_2} \right)  \frac{m_0}{\eta_L L}\, ,
\label{eq:CGamma}
\end{equation}
and 
 \begin{equation}
 \hat \Delta_{eq} = \frac{L m_1 m_2 (R_{10} - R_{20})}{3 \eta_L (m_1 + m_2)^2 }\, .
\label{eq:Delhatexp}
\end{equation}
\subsubsection{Asymmetry of the phase space}

 \begin{figure}
\begin{center}
\includegraphics[width=0.99\linewidth]{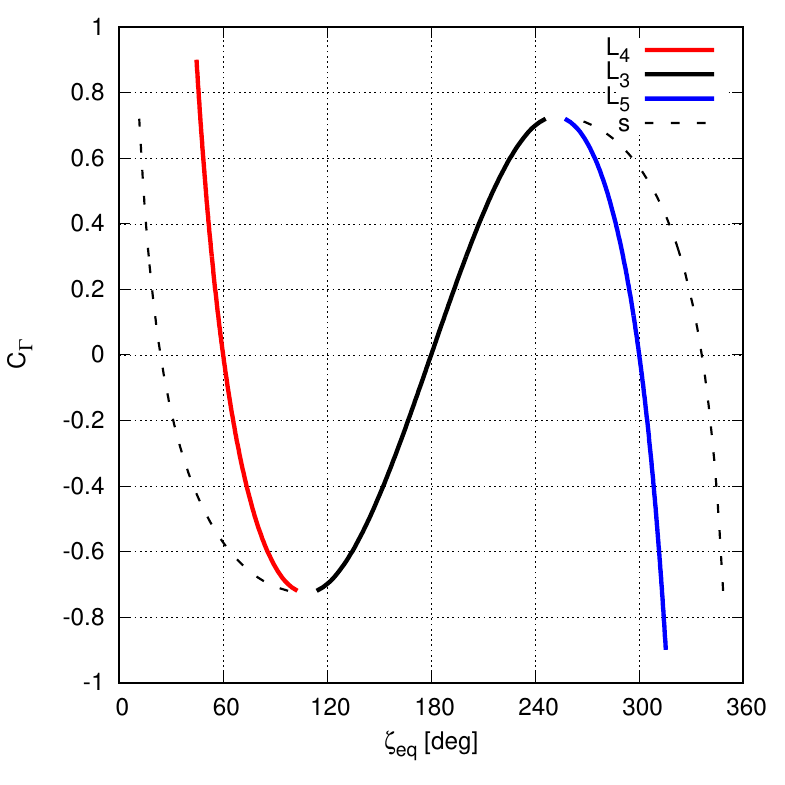}
\caption{\label{fig:L4L3L5} Evolution of the position of $L_4$, $L_3$, $L_5$, and the separatrix emanating from $L_3$, in the direction of $\zeta$ with respect to $C_\Gamma$, for $\hat \Delta= \hat \Delta_{eq}$. }
\end{center}
\end{figure}
In the restricted case ($m_1 \ll m_0$, $m_2=0$), the application of a constant torque on the co-orbitals result in a distortion of the phase space \citep{Murray1994,SiDu2003}. For a negative torque applied on the massive planet, it leads to a smaller tadpole domain for trailing massless particles than for leading ones, as the hyperbolic Lagrangian point $L_3$ gets closer to $L_5$, and further away from $L_4$. We study here the displacement of the perturbed circular coplanar Lagrangian equilibria for two massive bodies, as a function of the dimensionless quantity $C_\Gamma$, equivalent to the variable `$\alpha$' in \cite{SiDu2003}. The position of these equilibria are obtained by solving the system:
\begin{equation}
\begin{aligned}
\dot \zeta & =+\partial \gH_r / \partial \hat \Delta & = 0\, ,\\
\dot{\hat{ \Delta}}& =-\partial \gH_r / \partial \zeta &=0\, .
\end{aligned}
\label{eq:Hrpf}
\end{equation}

For small enough dissipative forces ($C_\Gamma \ll 1$), we can compute analytically the positions of the new equilibria in the neighbourhood of their position in the absence of dissipation, $\zeta=\pi/3$, and $\hat \Delta=0$. In order to keep track of the relative size of the terms in Eq. (\ref{eq:syserdid}), we introduce the small dimensionless tracer $\eps$. We make the assumption that the perturbative terms $C_\Gamma$ and $\sqrt{R_j/\eta}$ are of similar size with $m_j/m_0$, traced by $\eps$. $\eps$ is just a tool to neglect second-order perturbative terms, and we can later take $\eps=1$ for numerical estimation of the variables. Using the implicit function theorem, we look for a shift of size $\eps$ in the value of $\zeta$ and $\hat \Delta$ with respect to the non-dissipative case. We hence replace $\hat \Delta$ by $\eps \hat \Delta_{L_4}$ and $\zeta$ by $\pi/3+\eps z$ in the system (\ref{eq:Hrpf}), and solve it. At lowest order in $\eps$, the new $L_4$ equilibrium is located at:
\be
\hat \Delta_{L_4}= \eps^3 \hat \Delta_{eq}\, ,\ \ \ z_{L_4} =-  \frac{4}{9} \eps C_\Gamma\, .
\label{eq:L4eqc}
\ee
%\begin{equation}
%\left\{ 
%\begin{aligned}
%\Delta_{L_4} &= \eps^{3/2} \frac{L^3 \sqrt{m_0} (R_{01} - R_{02})}{\sqrt{6} (m_1 + m_2)^{7/2} \mu_0^2}\, , \\
%z_{L_4} & =-  \frac{4}{9} \eps C_\Gamma \, .\\
%\end{aligned}
%\right.
%\label{eq:L4d}
%\end{equation}
%
Similarly, we compute the evolution of the position of the $L_3$ equilibria by developing the system (\ref{eq:Hrpf}) in the neighbourhood of $\Delta=0$ and $z=\zeta-\pi=0$. Under the effect of the dissipative terms, the fixed point $L_3$ is shifted by:
\begin{equation}
\hat \Delta_{L_3}= \eps^3  \hat \Delta_{eq}\, ,\ \ \ z_{L_3}  = \frac{8}{7}  \eps C_\Gamma \, .\\
\label{eq:L3d}
\end{equation}
%-((2 gammar^2 m00 mu0^(-2 + k) sm1m2sgam^(-2 + 2 k) (tau1r - tau2r))/( 9 (m1 + m2)^2 tau1r tau2r))
 %
 %
%As explained in appendix \ref{ap:relative_size}, the displacement $ z_{L_3} $ and $z_{L_4}$ are not necessarily small. 
We hence have:
\begin{equation}
\begin{aligned}
\zeta_{L_3}-\zeta_{L_4} & = \frac{2 \pi}{3} + \eps(z_{L_3}- z_{L_4}) \\
& =   \frac{2 \pi}{3}+\frac{100}{63} \eps C_\Gamma \, .
\end{aligned}
\label{eq:zl3mzl4}
\end{equation}
As a result, $L_3$ and $L_4$ get closer if $ \frac{\Gamma_{10}}{m_1} <\frac{\Gamma_{20}}{m_2}$. If for example the torque per mass unit of the leading planet is lower that the torque per mass unit of the trailing planet, this results in a smaller trojan domain for the studied configuration.

% This displacement is parametrised by the dimensionless quantity $C_\Gamma$,
% %
% \begin{equation}
%C_\Gamma = \left( \frac{\Gamma_1}{m_1} - \frac{\Gamma_2}{m_2} \right)  \frac{m_0}{\eta_L L}\, ,
%\label{eq:I1exp}
%\end{equation}
%
For larger $C_\Gamma$, we solve the system (\ref{eq:Hrpf}) numerically. This system has three solutions for $\zeta \in [0^\circ,360^\circ]$ as long as $|C_\Gamma| \lesssim 0.72$. For larger absolute values of $C_\Gamma$, two of the three roots merge and vanish. The positions of the equilibria are shown in Fig. \ref{fig:L4L3L5}.
 For these 3 equilibria, the value of $\hat \Delta$ remain $\hat \Delta_{eq}$.
We also compute, for this value of $\hat \Delta$, the position of the separatrix emanating from $L_3$. To do so, we find the solutions of the equation $\gH_r (\hat \Delta_{eq} ,  \zeta_{L_3})=\gH_r ( \hat \Delta_{eq} , \zeta)$. The two separatrices were added as dashed lines to Fig. \ref{fig:L4L3L5}, and illustrate clearly the variation of the width of the trojan domain in the direction of $\zeta$ as a function of $C_\Gamma$, as the orbits librating around $L_4$ (resp. $L_5$) have to remain between the dashed and solid black lines. These results are in agreement with those of \cite{SiDu2003}, and generalise them to the case of two massive bodies.

\subsubsection{Stability of the Lagrangian equilibria}
\label{sec:stabW}

 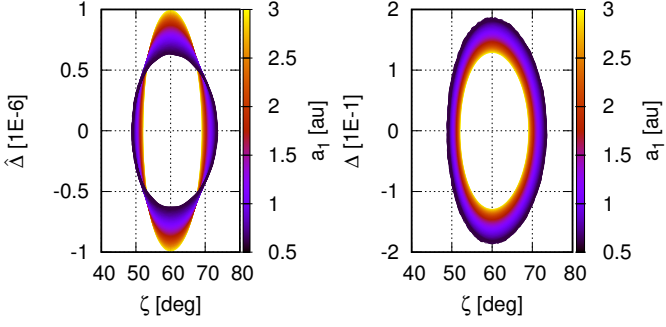
\begin{figure}
\begin{center}
\includegraphics[width=0.49\linewidth]{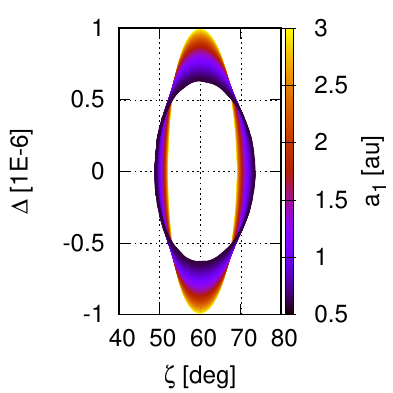}\includegraphics[width=0.49\linewidth]{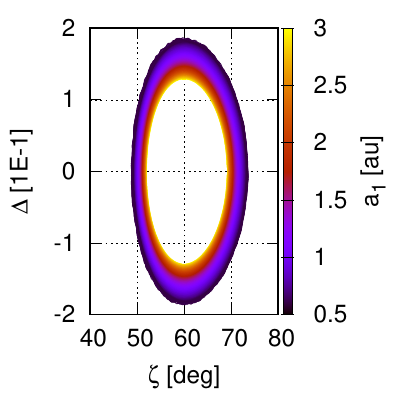}\\
  \setlength{\unitlength}{1cm}
\begin{picture}(.001,0.001)
\put(-4.35,2.45){\rotatebox{90}{$\hat{}$}}
\end{picture}
\caption{\label{fig:Wevol} Evolution of a trajectory in the ($\hat \Delta$,$\zeta$) plane (left) and ($\Delta$,$\zeta$) plane (right), of a $m_1=m_2=10^{-4}m_0$ pair of planet as their semi-major axis (colour code) decreases under the influence of constant torques applied on each planet. }
\end{center}
\end{figure}

We head back to small perturbations ($C_\Gamma \ll 1$, $ \sqrt{R_j/\eta}  \ll 1$). The torques applied on each planet slowly change the total angular momentum of the system. %
%Additionally, we consider a slow, isotropic mass change for the planets. 
Here we describe how the evolution of $L$ changes the orbit of the co-orbitals on a long time-scale with respect to the resonant motion.

Let us consider a pair of co-orbitals on a trajectory librating in the neighbourhood of the $L_4$ equilibria. We use $z=\zeta-\zeta_{L_4}$ and $d\hat \Delta=\hat \Delta-\hat \Delta_{eq}$ to describe the trajectory. On the resonant time-scale, this trajectory follow a level curve of the Hamiltonian (\ref{eq:Hr}). This level curve $\cL$ can be parametrised by $z_\cL>0$, with the energy of the curve being $H_r(\Delta_{eq},\zeta_{L_4}+z_\cL)$. For $z_\cL=0$, $\cL(z_\cL)$ goes through $d \hat \Delta = \hat \Delta_{\cL}$, estimated by solving $\gH_r (\hat \Delta_{eq} ,  \zeta_{L_4} + z_\cL)=\gH_r (\hat \Delta_{eq} + \hat \Delta_{\cL} , \zeta_{L_4})$. We obtain:
\begin{equation}
\begin{aligned}
\hat \Delta_{1\cL}=\frac{\sqrt{3}}{2} \eps^{3/2} W z_\cL+ \gO(\eps^{5/2}z_\cL)\, ,
\end{aligned}
\label{eq:IL4}
\end{equation}
where
\begin{equation}
W=\frac{m_1 m_2 L}{\sqrt{m_0}(m_1+m_2)^{3/2}}\, .
\label{eq:W}
\end{equation}
As $L$ slowly evolves, so does $W$ and $H_r$, and hence the level curve followed by the trajectory. However, the area enclosed by this level curve is an adiabatic invariant \citep{He1982}. As a result, a slow decrease of $W$ leads the trajectory to follow level curves of larger and larger $z_\cL$, while $\hat \Delta_{1\cL}$ decreases, see the left panel of Fig. \ref{fig:Wevol} for an example. We hence need to give a more precise definition of the stability we want to consider. As the position of the fixed points and separatrix are scale-free in the $\zeta$ direction, see Fig. \ref{fig:L4L3L5}, we consider a trajectory to be converging toward the Lagrangian equilibria if $z_\cL$ decreases, as the trajectory is getting further away from the boundary of the tadpole orbits.

We hence normalise the variable $\hat \Delta$ by $W$, and call this new variable $\Delta$:
\be
\Delta=\frac{\hat \Delta}{W}.
\ee
By doing so, we lose the Hamiltonian formulation of the problem as we introduce a dissipative term in the equations of variation, but we explicit the effect of the change of total angular momentum on the newly-defined stability of the system:
despite the small dissipative term, on the resonant time-scale a trajectory will remain close to a level curve of the Hamiltonian part of the system. For these new level curves, $\Delta_{\cL}$ only depends on $z_\cL$. In these new variables, the stability is hence defined as a convergence toward $L_4$ in both the $\zeta$ and $\Delta$ directions, while divergence from $L_4$ happens in both $\zeta$ and $\Delta$ directions as well, as we can see in the right panel of Fig. \ref{fig:Wevol}. The evolution of $\Delta$ is given by:
\begin{equation}
\begin{aligned}
 \dot  \Delta & = \frac{1}{W}\dot{\hat{\Delta}}   - \frac{\dot Wc}{W} \Delta \, ,
\end{aligned}
\label{eq:shrinkinframe}
\end{equation}
where
\begin{equation}
\begin{aligned}
\frac{\dot Wc}{W}= \frac{\Gamma_{10}+\Gamma_{20}}{L}  \, .
\end{aligned}
\label{eq:dWc}
\end{equation}
% \frac{\frac{\dot m_1}{m_1} (m_1+2m_2) +  \frac{\dot m_2}{m_2} (m_2+2m_1)}{2(m_1+m_2)} +
%
where $\dot W_c$ is the term of $\dot W$ coming from the constant torques applied on the planets. The resonant part of the system of equations of variation (\ref{eq:syserdid}) becomes:
\begin{equation}
\begin{aligned}
\dot \Delta   =&  \eta_L  \sqrt{ \frac{m_1+m_2 }{m_0 }} \left( 1-\frac{1}{\delta(\zeta)^3} \right) \sin \zeta\\
& - \frac{\dot Wc}{W} \Delta + \frac{m_2 \Gamma_{10}-m_1\Gamma_{20}}{(m_1+m_2)W}  \, , \\
\dot{\zeta} =&  - 3  \eta_L \sqrt{ \frac{m_1+m_2 }{m_0 }}  \Delta + R_{10}-R_{20}\, .\\
\end{aligned}
\label{eq:syserdidWc}
\end{equation}

As this is our main set of variables, we give the expression of the variable $\Delta$ with respect to the orbital elements:
\begin{equation}
\begin{aligned}
\Delta & \equiv  \frac{\sqrt{m_0} \sqrt{m_1+m_2}(m_2\Lambda_1 - m_1\Lambda_2)}{m_1 m_2L} \\
& \equiv  \sqrt{\mu_0 m_0} \frac{\sqrt{m_1+m_2}}{L}(\sqrt{a_1}-\sqrt{a_2})\, .
\end{aligned}
\label{eq:Delexp}
\end{equation}
while reciprocally, the circular angular momentum of each planet reads:
\begin{equation}
\begin{aligned}
\Lambda_1 & =\frac{m_1}{m_1+m_2}L \left(1 +  \frac{m_2}{\sqrt{m_0(m_1+m_2)}} \Delta \right)\, , \\
\Lambda_2 & =\frac{m_2}{m_1+m_2}L \left(1 -  \frac{m_1}{\sqrt{m_0(m_1+m_2)}} \Delta \right)\, .
\end{aligned}
\label{eq:LajfDel}
\end{equation}

We now (and for the rest of this paper) study the stability of the new $L_4$ equilibria ($\zeta \in [0,180^\circ]$), while the stability of $L_5$ can be studied by swapping the indices of the planets. The $L_4$ equilibria is located in:
\begin{equation}
\left\{ 
\begin{aligned}
\Delta_{L_4} &=\frac{1}{3}  \eps^{3/2} \frac{1}{\eta_L} \sqrt{ \frac{m_0}{m_1 + m_2}}  (R_{10} - R_{20})\, , \\
\zeta_{L_4} & =\frac{\pi}{3}-  \frac{4}{9} \eps C_\Gamma \, .\\
\end{aligned}
\right.
\label{eq:L4d}
\end{equation}

To do so, we linearise the resonant part of the system (\ref{eq:syserdid}) in the neighbourhood of $\Delta_{L_4}$ and $\zeta_{L_4}$: 
\begin{equation}
\dot{ \begin{pmatrix}
 \Delta'\\
  z'
   \end{pmatrix}} =
J_{L_4}
   \begin{pmatrix}
  \Delta' \\
  z'
   \end{pmatrix}\, ,
   \label{eq:syslL4}
   \end{equation}
where $\Delta'=\Delta-\Delta_{L_4}$, $z'=z-z_{L_4}$, and $J_4$ is the Jacobian matrix of the system (\ref{eq:syserdid}) computed at the equilibrium (\ref{eq:L4d}).
%%
%\begin{equation}
%  J_{L_4}= \begin{pmatrix}
%\frac{\partial \dot{\hat{Z}} }{\partial \hat{Z}} & 0 & \frac{\partial \dot{\hat{Z}} }{\partial \zeta} \\
%\frac{\partial \dot{\hat{Z}}_2 }{\partial \hat{Z}}  & \frac{\partial \dot{\hat{Z}}_2 }{\partial \hat{Z}_2} & 0 \\
%\frac{\partial \dot \zeta }{\partial \hat{Z}}  & \frac{\partial\dot \zeta }{\partial \hat{Z}_2} & 0 \\
%   \end{pmatrix}_{( \hat{Z}= \hat{Z}_2=0,\zeta=\pi/3+z_{L_4})}\, .
%     \end{equation}
%% 
% %
We make a final change of coordinates that diagonalises the system (\ref{eq:syslL4}). In this new set of variables ($z_1$,$z_2$), the equations of variation reads:

\begin{equation}
\dot{ \begin{pmatrix}
  z_1\\
  z_2
   \end{pmatrix}} =
 \begin{pmatrix}
 u_c  - i \nu  & 0 \\
0  & u_c+ i \nu  \\
   \end{pmatrix}
      \begin{pmatrix}
 z_1 \\
 z_2
     \end{pmatrix}
     \label{eq:diagc}
   \end{equation}
   where:
\begin{equation}
\begin{aligned}
\nu = \eps^{1/2} \nu_0 + \gO(\eps^{3/2})\, ,
%&\frac{\sqrt{6}}{4} \frac{(m_1 + m_2)^2}{L^3} \mu_0 \left(  \frac{f(z_{L_4} )(m_1 + m_2)^3 \mu_0^2}{m_0} \right.\\
%  &  \left. - \frac{    8 L^2 ( m_1 \Gamma_{2\zeta} - m_2 \Gamma_{1 \zeta})}{m_1 m_2} \right)^{1/2}
\end{aligned}
\label{eq:eigencircIc}
\end{equation}
and
\begin{equation}
\begin{aligned}
2u_c & =- \eps^2  \frac{\dot Wc}{W} \, .
\end{aligned}
\label{eq:eigencircpWc}
\end{equation}
%
%
%The systems (\ref{eq:syserdidL}) possess a fixed point only if  $\Gamma_1=\Gamma_2=0$. 
We hence obtain a modification of the classical resonant frequency $\nu_0$ in the neighbourhood of $L_4$ (\ref{eq:eigencircIc}), plus a hyperbolic term given in (\ref{eq:eigencircpWc}). As $u_c$ and $\nu$ are not constant due to the evolution of the masses and the total angular momentum $L$, we will study the stability of `partial' equilibria \citep[see for example][]{Vorotnikov2002} by dividing the variables into two groups: the variables with respect to which the stability is investigated ($\Delta$, $\zeta$), and the remaining variables $L$, $m_1$ and $m_2$. The linearised system with a diagonal matrix Eq. (\ref{eq:diagc}) allows for a trivial application of results on the stability of the partial equilibrium $z_1=z_2=0$ (see Appendix \ref{ap:pstab}): The partial equilibrium (\ref{eq:L4d}) is stable if $u_c$ is negative or null. From now on, we also make the assumption that a positive value induces a divergence from this equilibrium. 

%The term-by-term physical interpretation of this criterion (the sign of $u$, Eq. \ref{eq:eigencircp}) is straightforward: if $R_{\zeta 1}<0$ for the leading planet 1, then it adds an negative (positive) term for the value of $\dot z$ when $z>0$ ($z<0$). As a result, z is `pushed' toward $z=0$. Similarly, a negative $ \Gamma_{I 1}$ implies that the planet 1 migrates inward faster when $\Delta>0$ (i.e. when $a_1>a_2$), which also pushes the planet toward the exact resonance.
As previously stated, $\dot W_c/W$ represents the change of the size of the resonance as $L$ evolves. If $\dot W_c/W<0$, the width of the resonance in the previous variable $\hat \Delta$ is decreasing, leading to a slow increase of $z_l$ to retain a constant area enclosed by the level curve of the trajectory. As the system increases its amplitude of libration of the resonant angle, and gets closer to the separatrix of the tadpole domain, we consider the system to be diverging from the equilibrium. Looking at the expression of $\dot W_c/W$, eq (\ref{eq:dWc}), it appears explicitly that inward migration (negative torques) have a destabilising effect on the co-orbital resonance, while outward migration tends to stabilise it. 

%In the case of constant forces applied on the planets, it is the relative size of the torques versus the accretion rate that dictate the stability of the system. 

\subsection{Stability criteria for the Lagrangian configuration $L_4$ under non-constant forces  and mass change}
\label{sec:stabu}
In this section we study the stability of the Lagrangian equilibria $L_4$ assuming that the variation of the dissipative forces can be parametrised by $\zeta$ and $\Delta$, and that other variations are slow enough to be considered as constant with respect to the resonant time-scale. We also consider a slow, isotropic mass change for both masses. As in Sect. \ref{sec:constantF}, we make the assumption that the perturbative terms $C_\Gamma$, $\sqrt{\dot m_j/m_j}$ and $\sqrt{R_j/\eta}$ are of similar size with $m_j/m_0$, traced by $\eps$. As we did in the previous section, we normalise $\hat \Delta$ by $W$. As we now consider $\dot m_j \neq 0$, we have:
\begin{equation}
\begin{aligned}
\frac{\dot W}{W}= \frac{\Gamma_{01}+\Gamma_{02}}{L} + \frac{\frac{\dot m_1}{m_1} (m_1+2m_2) +  \frac{\dot m_2}{m_2} (m_2+2m_1)}{2(m_1+m_2)}  \, .
\end{aligned}
\label{eq:dW}
\end{equation}
%  +
%
The equations of variations (\ref{eq:syserdid}) becomes:
\begin{equation}
\begin{aligned}
\dot \Delta   =&  \eta_L  \sqrt{ \frac{m_1+m_2 }{m_0 }} \left( 1-\frac{1}{\delta(\zeta)^3} \right) \sin \zeta\\
& - \frac{\dot W}{W} \Delta + \frac{m_2 \Gamma_1-m_1\Gamma_2}{(m_1+m_2)W} + \frac{m_1^2 \dot m_2 + m_2^2 \dot m_1}{m_1m_2(m_1+m_2)} \Delta \, , \\
\dot{\zeta} =&  - 3  \eta_L \sqrt{ \frac{m_1+m_2 }{m_0 }}  \Delta + R_1-R_2\, .\\
\end{aligned}
\label{eq:syserdidW}
\end{equation}
For the variables $\Delta=\hat \Delta/W$, $\zeta$.
%\begin{equation}
%\begin{aligned}
%\dot \Delta  &=  \mu_0^2 m_1 m_2 \frac{9   (m_1 + m_2)^2 }{4 L^3 m_0} z  -\frac{\dot L}{L} \Delta  + \frac{m_2 \Gamma_1-m_1\Gamma_2}{(m_1+m_2)L}\\
%%&\phantom{=} \, , \\
%\dot{\zeta} & =  - 3 \mu_0^2 \frac{(m_1+m_2)^5}{m_1m_2 L^3} \Delta + R_1-R_2\, ,\\
%\dot L &= \Gamma_1 +  \Gamma_2\, , \\
%\end{aligned}
%\label{eq:syserdidL}
%\end{equation}
%while 
 Then we linearise these perturbations in the neighbourhood of ($\Delta_{L_4}$,$\zeta_{L_4}$) by using the reduced variables $\Delta'=\Delta-\Delta_{L_4}$, $z=\zeta-\zeta_{L_4}$. 
$\Gamma_j$ and $R_j$ become:
\begin{equation}
\begin{aligned}
\Gamma_j &= \Gamma_{j0}+ \Gamma_{j\Delta} \Delta' + \Gamma_{j\zeta} z  \, , \\ 
R_j &= R_{j0}+R_{j\Delta} \Delta' +R_{j \zeta} z  \, , \\ 
\end{aligned}
\label{eq:GamRexp}
\end{equation}
where $\Gamma_{j\Delta} = \partial \Gamma_j / \partial \Delta'$, $\Gamma_{j\zeta} = \partial \Gamma_j / \partial z$, and similar expressions for $R_j$. Despite the addition of these new terms in the equations of variation, the dominant terms of $\Delta_{L_4}$ and $\zeta_{L_4}$ remains those computed in the case of constant forces, given in (\ref{eq:L4d}).

We hence linearise the rest of the system (\ref{eq:syserdidW}) in the neighbourhood of ($\Delta_{L_4}$,$\zeta_{L_4}$), modelling the dissipative forces by the expression (\ref{eq:GamRexp}). Then, as in Sect. \ref{sec:stabW}, we diagonalise the linearised system. Since the masses are not constant in this section, this transformation adds additional terms proportional to $\dot m_1$ and $\dot m_2$. However, as these terms are of size $\gO(\eps^3)$, we can neglect them when we compute the eigenvalues of the linearised system. The dominant terms of the imaginary part remain:
\begin{equation}
\begin{aligned}
\nu = \eps^{1/2} \nu_0 + \gO(\eps^{3/2})
%&\frac{\sqrt{6}}{4} \frac{(m_1 + m_2)^2}{L^3} \mu_0 \left(  \frac{f(z_{L_4} )(m_1 + m_2)^3 \mu_0^2}{m_0} \right.\\
%  &  \left. - \frac{    8 L^2 ( m_1 \Gamma_{2\zeta} - m_2 \Gamma_{1 \zeta})}{m_1 m_2} \right)^{1/2}
\end{aligned}
\label{eq:eigencircI}
\end{equation}
However, the real part of the eigenvalue gets new terms:
\begin{equation}
\begin{aligned}
2u =\eps^2 & \left[ R_{1 \zeta} - R_{2 \zeta} + \frac{ m_2 \Gamma_{1\Delta} - m_1 \Gamma_{2\Delta} }{(m_1+m_2)W} \right. \\
 & \left. -\frac{\Gamma_{01}+\Gamma_{02}}{W} -\frac{1}{2} \frac{\dot m_1 + \dot m_2}{m_1+m_2} \right] \, .
\end{aligned}
\label{eq:eigencircp}
\end{equation}

We hence obtain a small modification of the classical resonant frequency $\nu_0$ in the neighbourhood of $L_4$ (\ref{eq:eigencircI}), plus a hyperbolic term given in (\ref{eq:eigencircp}). We note that at lowest order in $\eps$ the equilibrium point remains elliptic because of our assumptions on the relative size of the dissipative terms with respect to $m_j/m_0$.  As in Sect. \ref{sec:stabW}, we can deduce the stability of the system by estimating the sign of $u$: a negative value of $u$ induces a convergence of the system toward the exact resonance, while a positive value led to a divergence.

The term-by-term physical interpretation of this criterion (the sign of $u$, Eq. \ref{eq:eigencircp}) is straightforward: if for example $R_{1 \zeta}$ is the only non-zero term of $u$ and using Eqs. (\ref{eq:poincvard}) and (\ref{eq:GamRexp}), we have:
\be
\dot z = R_{1 \zeta} z
\ee
If $R_{1 \zeta}<0$, $z$ converges toward $0$, hence $\zeta$ converges toward $\zeta_{L_4}$. Similarly, a negative $ \Gamma_{1\Delta}$ implies that the planet `1' migrates inward faster when $\Delta>0$ (i.e. when $a_1>a_2$), which also pushes the planet towards the exact resonance.

 The terms in $\Gamma_{j0}$ and $\dot m_j$ were already described in the previous section, and take into account the evolution of the width of the resonance as the masses and total angular momentum of the configuration slowly evolve.
 % $R_{j \zeta}$ is the partial derivative of $\dot \lambda_{j,d}=R_j$ in the neighbourhood of $L_4$, the evolution of the mean longitude of the planet $j$ with respect to the dissipation. Taking for example the leading planet 1: if $R_{1\zeta}<1$, then $\dot \lambda_{1,d}<0$ when $z>0$: the leading planet 1 tend to "slow down" when $\lambda_1-\lambda_2=\zeta>\zeta_{L_4}$, helping the convergence toward $L_4$ The relative size and sign of these various contributions lead the system to either converge toward, or diverge from, the equilibria.
In the following sections, we apply the stability criterion $u$ in the case of planets evolving in a protoplanetary disc.

\section{Axi-symmetric dissipative forces: 1D protoplanetary discs}
\label{sec:type1}

Gravitational interactions between the planets and their parent disc impact on the planets' orbital parameters, typically causing them to migrate, either inwards towards the star or outwards away from the star \citep{GoldreichTremaine,1993ApJ...419..166A,PaLa2000}.
Planet eccentricities and inclinations are also affected by the interactions with the disc, typically causing them to be damped, forcing the planets to orbit their parent star on coplanar circular orbits \citep{CreNe2006,BiKle2010}.
The orbital evolution of the planets is the result of various torque components from the disc acting on them, such as the Lindblad torque, corotation torque, and horseshoe drag \citep{BaruteauPP6}. For a single planet in a protoplanetary disc, two regimes are usually considered: As long as a planet is not massive enough to perturb the disc considerably, the planet is called to be in type I migration. As it is growing, it starts to open a gap around its orbit, and once this gap is deep enough, it migrates in type II regime. In this section, we assume that the perturbation of the disc by each planet is negligible, and hence that usual type I migration formulae can be applied on each planet individually. 

We also assume that the unperturbed disc is axi-symmetric, and we study the stability of the coplanar circular co-orbital resonance in this case. Under this assumption, only the tangential forces (or torque) will play a role in the stability of the configuration (see the expression of the criterion $u$, eq \ref{eq:eigencircp}). In the literature, these torques are often modelled by migration time-scales $\tau_{aj}$, used as prescriptions for the evolution of the semi-major axes: $\dot a_{j,d} = -a_{j}/ \tau_{aj}$, which implies a dissipative term for the evolution of the angular momentum of the planet of the form: $ \dot \Lambda_{j,d}=-\Lambda_j/(2\tau_{aj})$.

\subsection{Analytic model of type 1 migration}
\label{sec:ana_tana}

\cite{TaTaWa2002} and \cite{TaWa2004} derived a linear model of the wave excitation in three-dimensional isothermal discs to obtain an analytical model for the torques induced by the Lindblad and corotation resonances. Following their notation, we consider a gaseous disc parametrised by its aspect ratio $h$ and its surface density $\Sigma(r)$, such that:
\begin{equation}
h=H/r=h_0 r^{f}
\label{eq:diskheight}
\end{equation}
where $H$ is the scale height, $h_0$ is the aspect ratio at $1$~au and $f$ is the flaring index, and
\begin{equation}
\Sigma(r)=\Sigma_0 r^{-\alpha}
\label{eq:surfdens}
\end{equation}
where $\Sigma_0$ is the surface density at $1$~au and $\alpha$ parametrises the slope of the surface density. At first order in the inclination and eccentricity, we have:
\begin{equation}
\tau_{aj}= \frac{\tau_{wj}}{2.7+1.1 \alpha}h^{-2}
\label{eq:taua}
\end{equation}
% \left(\frac{H}{r} \right)^{-2},
for the evolution of the semi-major axis \citep{TaTaWa2002}, with
\begin{equation}
\tau_{wj}= \left( \frac{m_j}{m_0} \right)^{-1} \frac{m_0}{\Sigma(r_j) r_j^2} h^4 \Omega_j^{-1}\, ,
\label{eq:twave}
\end{equation}
where $\Omega_p=	(\cG m_0/r_j^3)^{1/2}$ is the Keplerian angular velocity of the planet. 
%
%%The damping time-scale of the Inclination and eccentricity is given by:
%\begin{equation}
%\tau_I=\tau_w /0.544 \, , \ \ \tau_e=\tau_w /0.780\, . 
%\label{eq:tauItaue}
%\end{equation}
%%
%
Considering Eqs. (\ref{eq:diskheight}) and (\ref{eq:surfdens}), and neglecting the effect of the eccentricity on the star-planet distance, $\tau$ reads:
\begin{equation}
\tau_{wj}= \frac{m_0}{m_j} \sqrt{\frac{ m_0}{\cG}}\frac{h_0^4 }{\Sigma_0}\, a_j^{\alpha+4f-1/2}\, .
\label{eq:twave2}
\end{equation}%

At this stage, it becomes clear that taking $\tau_{aj}$ as a constant is a strong assumption that would force the semi-major axis to behave as an exponential decay. We hence introduce the variable $K_j$ that parametrises the local slope of the migration:
\begin{equation}
\tau_{aj}=\tau_{j}a_j^{K_j} \, .
\label{eq:taj}
\end{equation}%
%
%\begin{equation}
%\tau_{aj} \propto \frac{m_0}{m_j} \sqrt{\frac{ m_0}{\cG}}\frac{h_0^4 }{\sigma_0}\, a_j^{\alpha+2f-1/2}\, .
%\label{eq:twave2}
%\end{equation}%
%Assuming that in type I migration the effect of the planets on the disc is negligible (i.e. the planets do not open a gap), 
%
The effect of the disc on the angular momentum of each planet is hence given by:
\begin{equation}
\dot \Lambda_{j,d}= -  m_j^{2K_j} \mu_0^{K_j} \frac{\Lambda_j^{1-2K_j}}{2 \tau_{j}}\, .
\label{eq:LambdotK}
\end{equation}%
Expanding the torque applied on each planet at first order in $\Delta$, $\dot \Lambda_{j,d}=\Gamma_{j0}+\Gamma_{j\Delta} \Delta$, we have:
\begin{equation}
\Gamma_{j0}=  - \frac{m_j  }{m_1+m_2 } \frac{ L }{2 \bar a_L^{K_j}\tau_{j} } \, ,
\label{eq:Gammaj0}
\end{equation}%
and:
\begin{equation}
\Gamma_{j\Delta} = (-1)^j \frac{(1-2{K_j}) L }{2 \bar a_L^{K_j}  \tau_{j} } \, .
\label{eq:GammajI}
\end{equation}%

\begin{figure}
\begin{center}
\includegraphics[width=0.99\linewidth]{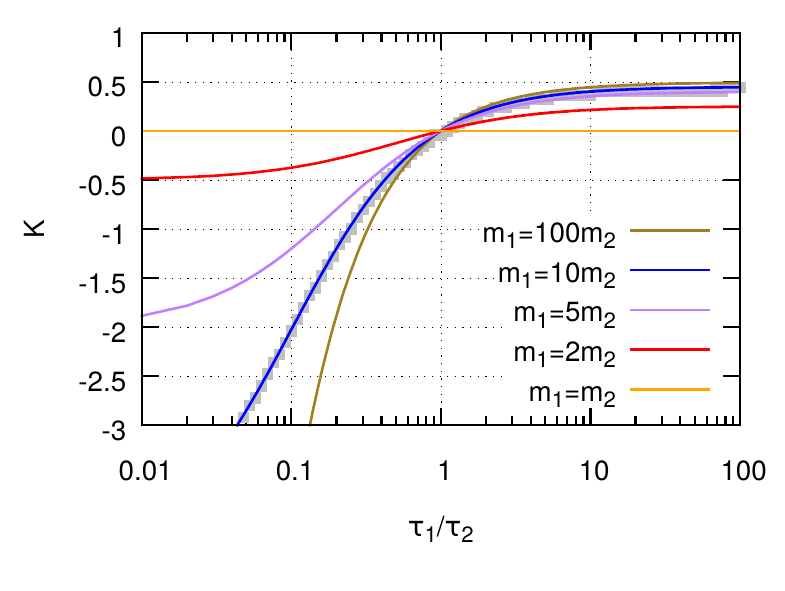}\\
  \setlength{\unitlength}{1cm}
\begin{picture}(.001,0.001)
\put(-1.8,6){divergent}
\put(1.5,5){convergent}
\end{picture}
\caption{\label{fig:tauacrit} Attraction criterion for the $L_4$ and $L_5$ equilibria in the dissipative case, for different values of $m_2/m_1$. The equilibria are attractive if $K$ is below the curve, and repulsive if $K$ is above. The grey squares represent the attraction limit for a horseshoe orbit and were derived numerically, see the text for more details. }
\end{center}
\end{figure}
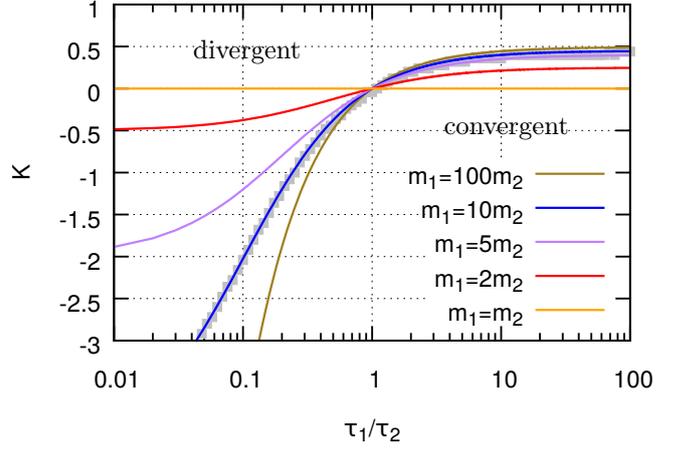

The effect of dissipative forces modelled by $\tau_{aj}$ (eq \ref{eq:taj}) on the co-orbital configuration can be deduced from the expression (\ref{eq:Gammaj0}) and (\ref{eq:GammajI}) and the criterion $u$, Eq. (\ref{eq:eigencircp}). The Lagrangian point is attractive if:
\begin{equation}
((2K_2-1)m_1+m_2)\tau_{1} +((2K_1-1)m_2+m_1)\tau_{2}  < 0  \, .
\label{eq:stabcon}
\end{equation}%
Figure \ref{fig:tauacrit} represents the $K$ values for which the equilibrium becomes repulsive (above the line of a given mass ratio) as a function of $\tau_1/\tau_2 \approx \tau_{a_1}/\tau_{a_2}$, in the special case $K=K_1=K_2$. These results apply to the neighbourhood of the $L_4$ or $L_5$ equilibria, hence to tadpole orbits with small amplitudes of libration, as they were derived from the linearisation of the system (\ref{eq:syserdidW}) in the neighbourhood of $L_4$.
We then check the validity of this criterion for different amplitudes of libration in the case $m_1=10 m_2$. We integrate the equations of the 3-body problem using the variable-step integrator DOPRI \citep{DOPRI}. In addition, the migration of the semi-major axis is modelled using 
\be
\ddot{\bm{r}}_j = - \dot{\bm{r}}_j a_j^{-K} /(2 \tau_{j})\, ,
\ee
where $\bm{r}_j$ is the position of the planet $j$ with respect to the star. The tests were made using $m_0=1$, $m_1=5 \times 10^{-5}$, $m_2=5 \times  10^{-6}$, taking as initial conditions $e_j=I_j=0$, $a_1=a_2=1\,$au and $\zeta_0=58^\circ$, $50^\circ$, $40^\circ$, $30^\circ$ and $20^\circ$. For each initial value of $\zeta$, a grid of cases were integrated for different values of $K$ and $\tau_{w_1}/\tau_{w_2}$, using $\tau_{w_1}=1/{m_1}$. The grey squares in Fig. \ref{fig:tauacrit} represent the threshold values of $K$ for which the configurations change from converging (below the curve) to diverging (above it) for $\zeta_0=20^\circ$ (horseshoe configuration). All other initial amplitudes of libration (initial values of $\zeta_0$) gave very similar results. It implies that, at least for the chosen masses, the attraction of the exact resonance does not depend significantly on the amplitude of libration in the case of axi-symmetric dissipative forces.

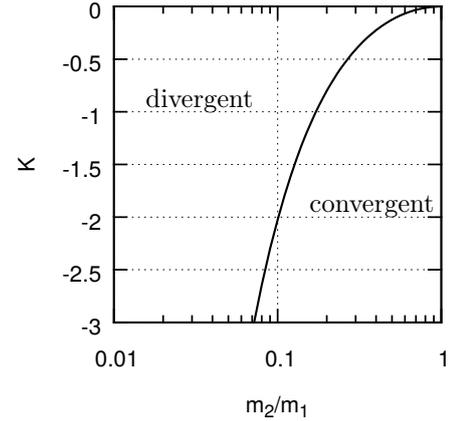
\begin{figure}
\begin{center}
\includegraphics[width=0.7\linewidth]{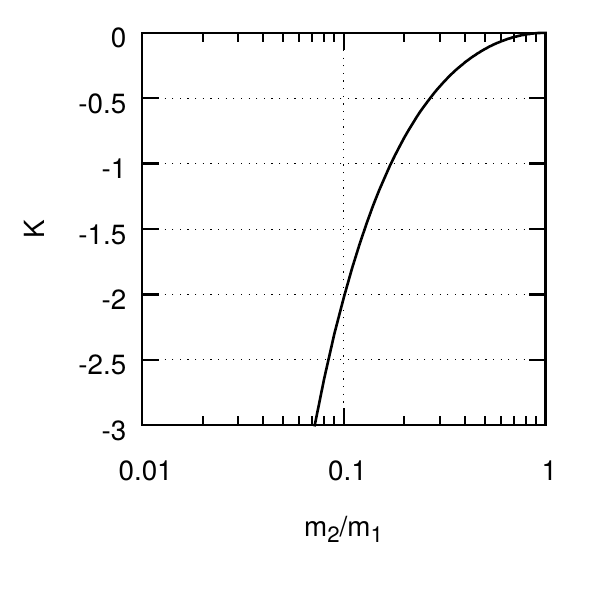}\\
  \setlength{\unitlength}{0.77cm}
\begin{picture}(.001,0.001)
\put(-1.5,6.5){divergent}
\put(1.3,4.7){convergent}
\end{picture} 
\vspace{-1cm}
\caption{\label{fig:tauacritTana}Stability threshold for the $L_4/L_5$ equilibria when the torque induced by the protoplanetary disc is modelled using Eq. (\ref{eq:taua}). The configuration diverge when $K$ is above the line and converge when it is below.}
\end{center}
\end{figure}

 In the case of the torque from \cite{TaTaWa2002}, we have
\begin{equation}
\tau_{aj}=\tau_{j}a^K_j=\frac{1}{2.7+1.1 \alpha} \frac{m_0}{m_j} \sqrt{\frac{ m_0}{\cG}}\frac{h_0^2 }{\Sigma_0}\, a_j^{\alpha+2f-1/2}\, .
\label{eq:tajtana}
\end{equation}%
Which hence verifies the relation $K_1=K_2=\alpha+2f-1/2$, in addition to $\tau_{a_1}/\tau_{a_2}= m_2/m_1$. Applying these additional constraints, the stability of the Lagrangian points depends only on $m_2/m_1$ and the parameter $K$. The critical value for $K$ is thus:
\begin{equation}
K_{0} =- \frac{(\frac{m_2}{m_1}-1)^2}{4\frac{m_2}{m_1}}
\label{eq:stabcri}
\end{equation}
which is represented in Fig. \ref{fig:tauacritTana}.

%
% We note that Eq. (\ref{eq:stabcri}) and Fig. \ref{fig:tauacrit} can be applied to other torque formula than the one described in Eq. (\ref{eq:taj}).
%
%
%
%
%From equation (\ref{eq:twave2}), we note that $\tau_{a_1}/\tau_{a_2}= m_2/m_1$. into expression \ref{eq:eigencirc}, the real component of the eigenvalue is equal to zero on the curve represented in Fig. \ref{fig:tauacritTana}. Above the curve, the configuration slowly diverges from the equilibrium. Under the curve, it converges toward the equilibrium. As we will see in section \ref{sec:gavintype1}, $K$ is generally above $-1$ in a typical protoplanetary disc, which would lead to most co-orbitals slowly getting out of resonance during type I migration.

\subsection{Stability in an evolving protoplanetary disc}
\label{sec:gavintype1}

\begin{figure*}
\includegraphics[scale=0.42]{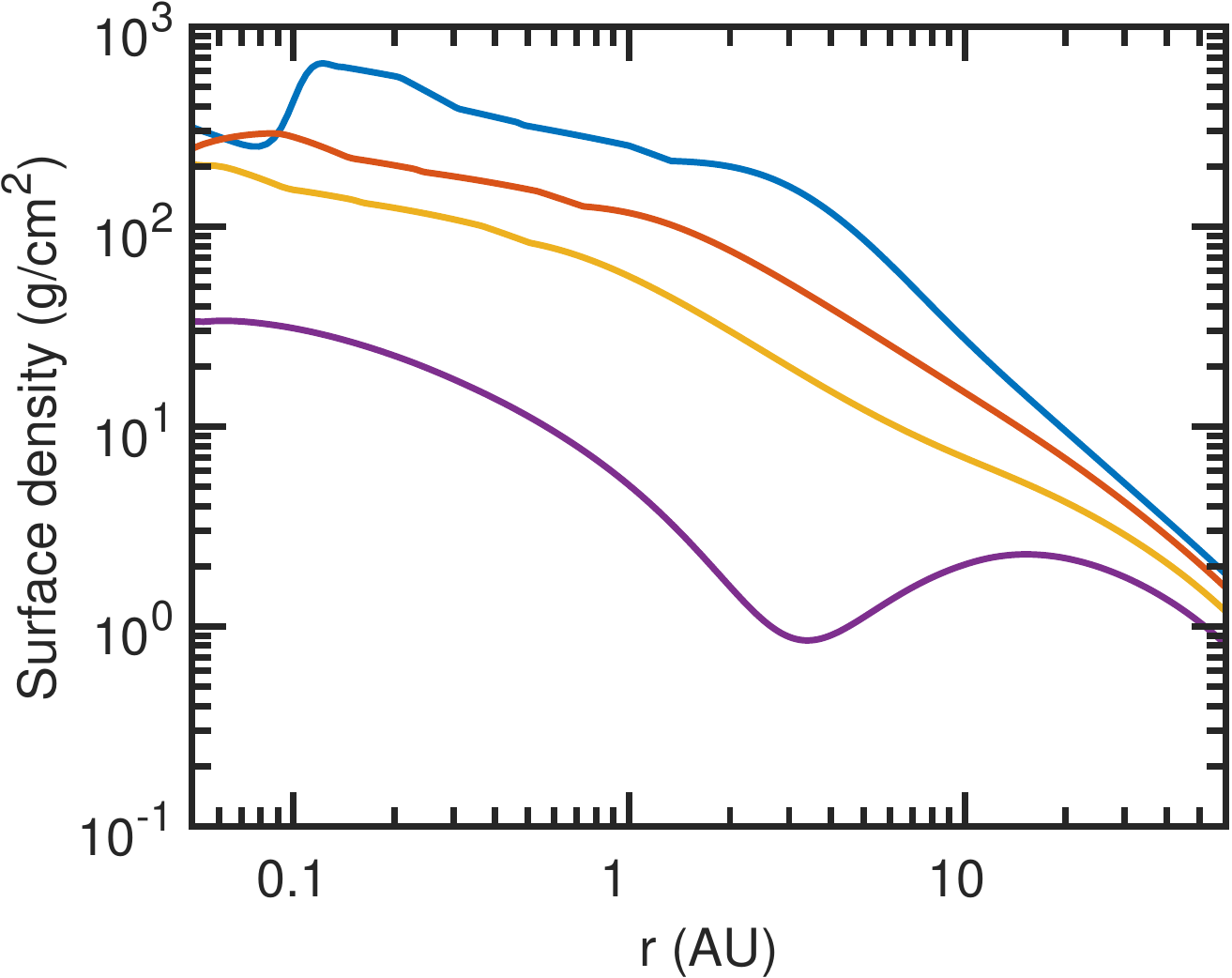}
\hspace{0.2cm}
\includegraphics[scale=0.42]{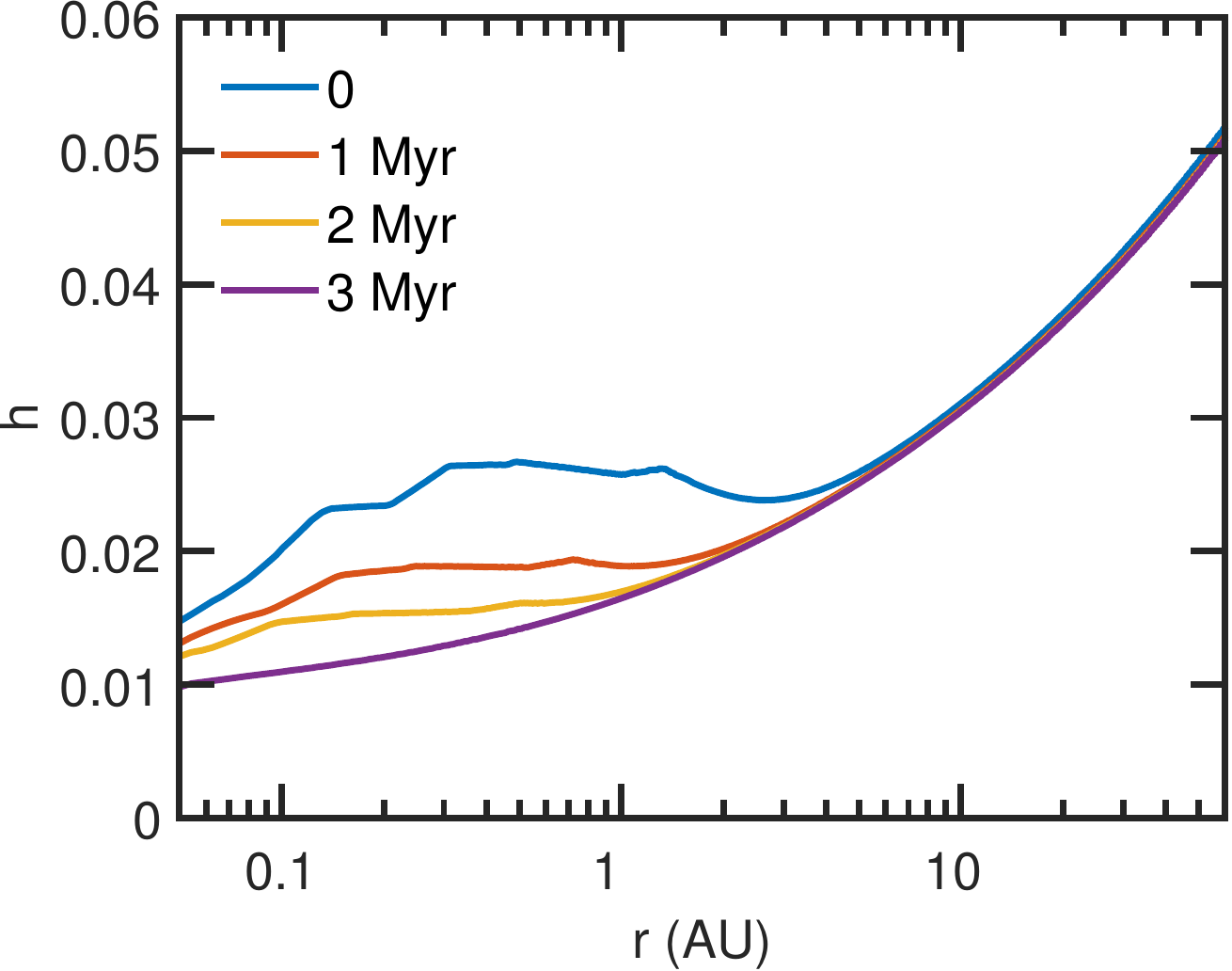}
\hspace{0.2cm}
\includegraphics[scale=0.42]{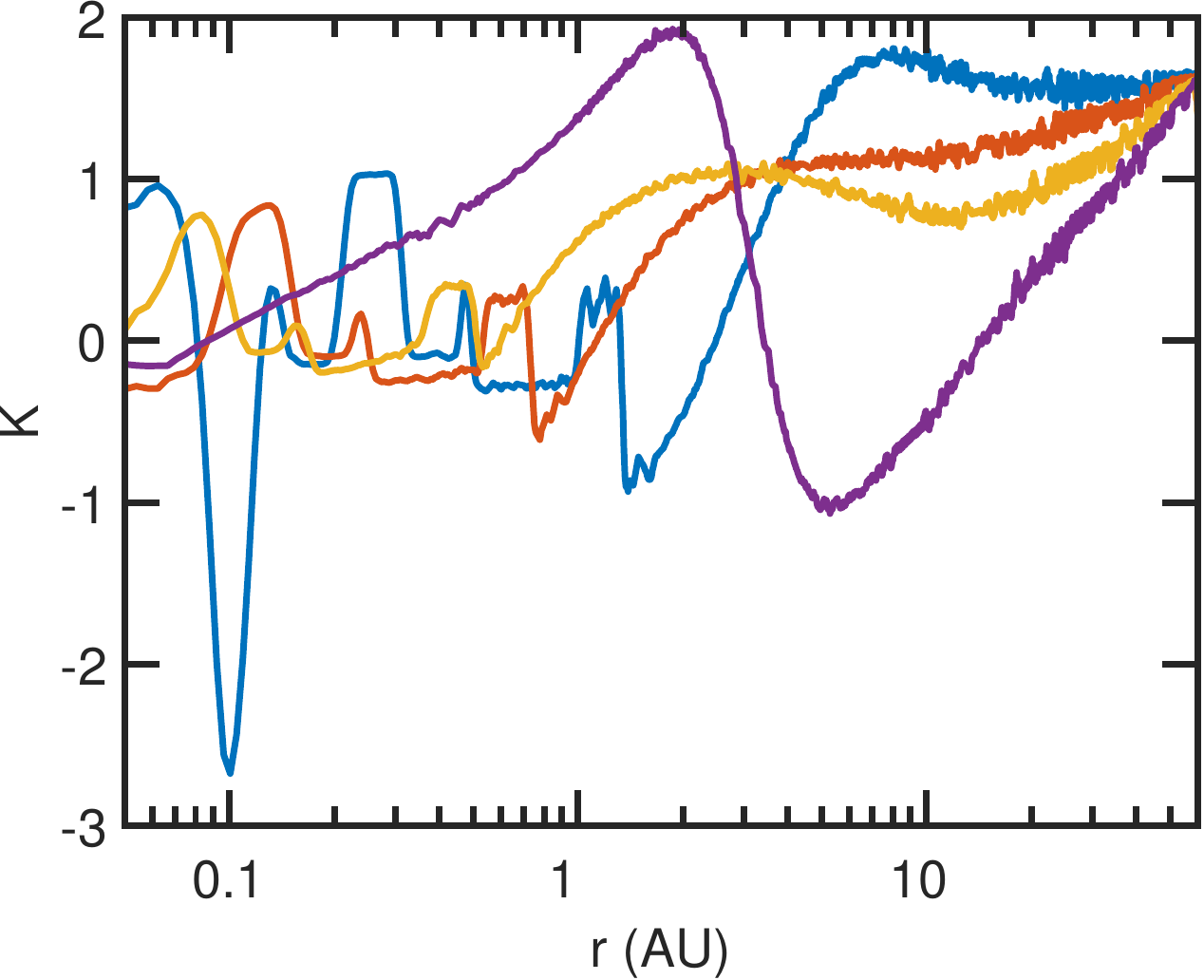}
\caption{Gas surface densities (left panel), disc aspect ratios (middle panel) and K (right panel) at t = 0 Myr (blue lines), 1 Myr (red lines), 2 Myr(yellow lines) and 3 Myr (purple lines) for a typical protoplanetary disc model.}
\label{fig:disc_profiles}
\end{figure*}

The section above describes the evolution of the co-orbital resonance in a static environment, where the migration time-scales and disc parameters are assumed to be constant.
However, protoplanetary discs do not remain static, so therefore it is important to determine how the co-orbital resonance evolves in a more global, ever-changing environment.
Thus we now examine the behaviour of the resonance as the protoplanetary disc evolves on Myr time-scales, and also as a planet migrates from one region of a disc to another where the value of $K$ can differ significantly.
We use the disc model presented in \citet{ColemanNelson16}, where the standard diffusion equation for a 1D viscous $\alpha$-disc model is solved \citep{Shak}.
Temperatures are calculated by balancing viscous heating and stellar irradiation against blackbody cooling.
We use the torque formulae from \citet{pdk10,pdk11} to calculate type I migration rates due to Lindblad and corotation torques acting on a planet.
The Lindblad torque emerges when an embedded planet perturbs the local disc material, forming spiral density waves that are launched at the Lindblad resonances in the disc.
Corotation torques arise from both local entropy and vortensity gradients in the disc, and their possible saturation is included in these simulations.
The influence of eccentricity and inclination on the migration torques and the damping of eccentricities and inclinations are also included \citep{CreNe2008,Fendyke}.
These torques exchange angular momentum between the planet and the gas disc, and depending on their strength and direction can result in a torque being exerted on the planet, either inwards or outwards.

Figure \ref{fig:disc_profiles} shows the gas surface density (left panel), disc aspect ratio $h$ (middle panel) and the calculated value for $K$ (right panel) at 4 different times throughout the disc lifetime.
The blue line shows the profiles at the beginning of the disc lifetime, with the red, yellow and purple lines showing the profiles at 1, 2 and 3 Myr respectively.
The disc lifetime here was $\sim3.6$ Myr. $K=\alpha+2f-1/2$, as defined in Sect. \ref{sec:ana_tana}.

\begin{figure}
\centering
\includegraphics[scale=0.6]{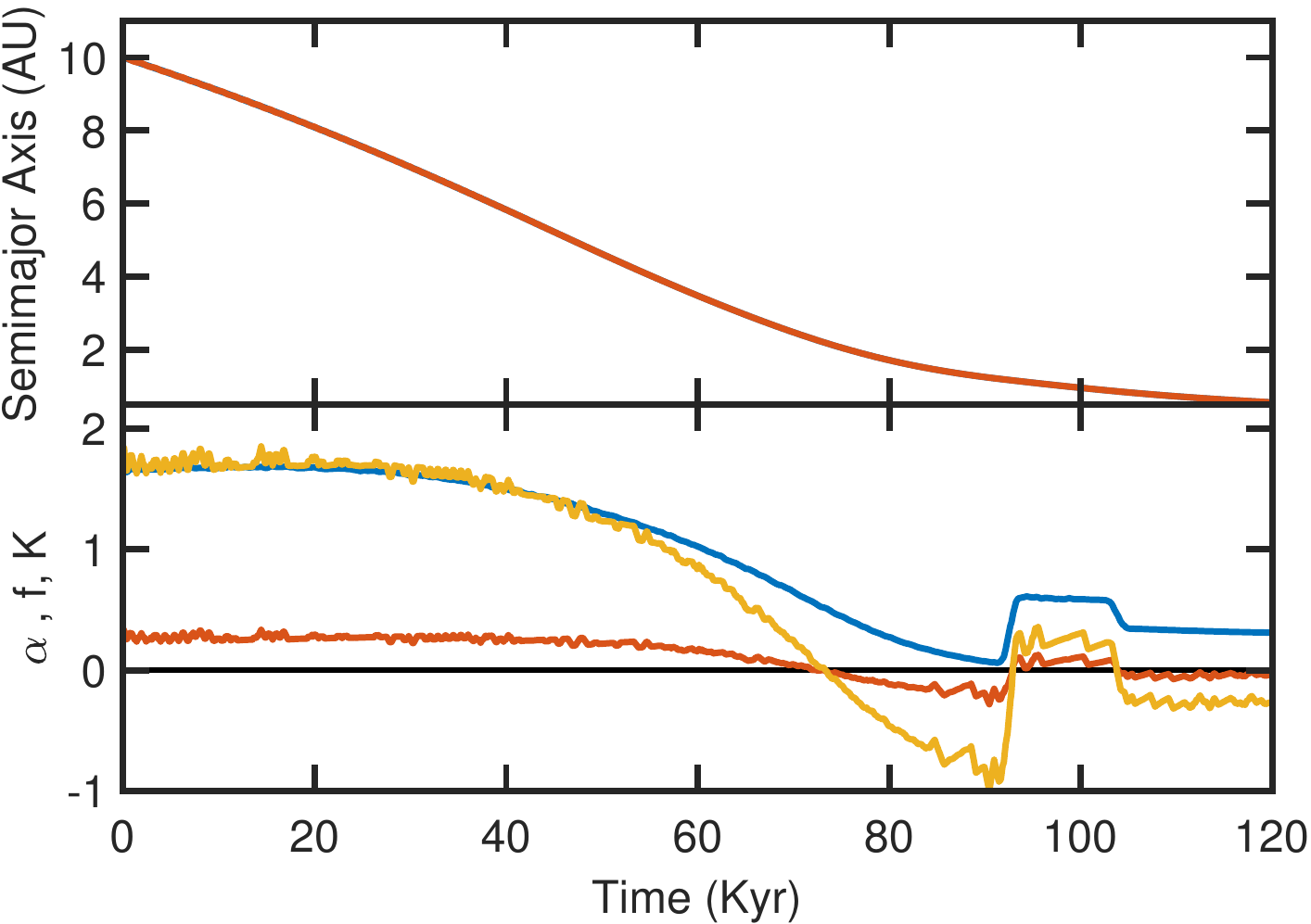}
\caption{{\it Top panel:} temporal evolution of semimajor axis for a pair of co-orbital planets.
{\it Bottom panel:} Surface density index $\alpha$ (blue line), aspect ratio index $f$ (red line), and corresponding $K$ (yellow line) at the planets' location over time.}
\label{fig:sma_zeta}
\end{figure}

\begin{figure}
\centering
\includegraphics[scale=0.6]{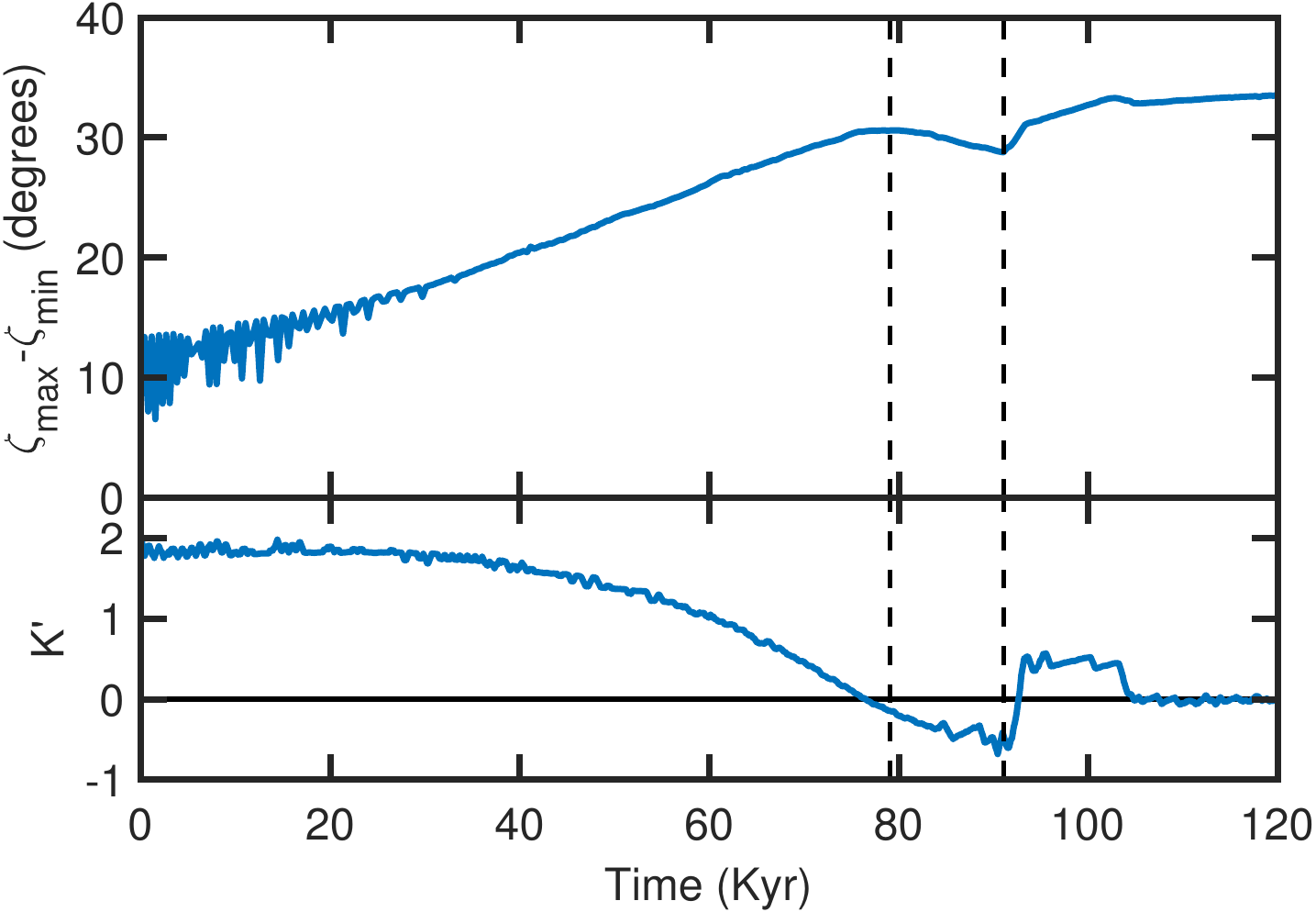}
\caption{$\zeta_{\rm max}-\zeta_{\rm min}$ and $K$' over time.
When $K$' is positive, resonance is diverging, and when $K$' is negative, the resonance is converging.
The dashed lines show there $K$' changes sign, indicating the change from convergence to divergence or vice versa.}
\label{fig:libtime_k}
\end{figure}

%Whilst Fig. \ref{fig:disc_profiles} showed how the protoplanetary disc evolves, and how that evolution impacts on $K$ across the disc both radially and temporally, it is interesting to 
We then examine the evolution of a pair of co-orbital planets that are undergoing type I migration in the protoplanetary disc represented in Fig. \ref{fig:disc_profiles}. The masses of the planets are 10 and 5 $\me$ for the primary and secondary respectively, giving a mass ratio of 2.
The top panel of Fig. \ref{fig:sma_zeta} shows the evolution of semimajor axis as the planets migrate through the disc, showing that they remain in co-orbital resonance, even when migrating from 10 $\au$ down to the inner edge of the disc.
The bottom panel shows the values for the surface density gradient $\alpha$ (blue line), the aspect ratio gradient $f$ (red line), and the corresponding value of $K$ (yellow line) at the planet's location.
As the planets originate in the outer irradiation dominated region of the disc, $K$ has a value $\sim 1.75$.
But as the planets migrate in closer to the central star, they enter the viscous dominated region of the disc where the transition in opacities significantly impacts $\alpha$ and $f$, causing $K$ to drop to below 0 for a time before rising back to just above 0.
As the planets near the inner edge of the disc, $K$ settles to just below 0.
Whilst the planets are migrating through the different regions of the disc, they will either be converging towards the resonance or diverging away from the resonance, depending on the local disc conditions.
Figure \ref{fig:libtime_k} shows the amplitude of libration (top panel) and the corresponding $K'=K-K_0$ for the two migrating planets in Fig. \ref{fig:sma_zeta}.

When looking at Fig. \ref{fig:libtime_k}, it can be seen that the amplitude of libration is increasing at the start of the simulation whilst $K$' $\sim 2$.
As the planets migrate into the inner regions of the disc, $K$' drops to begin fluctuating around 0.
The first dashed line shows where the amplitude of libration begins to converge, and this lines up just after $K$' reaches negative values.
This is expected from the analytical model in sect. \ref{sec:ana_tana}.
However, the analytical model uses the simplified type I migration torque formulae from \citet{TaTaWa2002}, whereas the evolving disc model here uses torque formulae from \citet{pdk10,pdk11} that includes a more accurate treatment of the corotation torque. Despite these differences, it is interesting to see that the behaviour of the co-orbital resonance roughly matches what is expected from the Tanaka formulae, i.e. converging when $K$' is negative and diverging when $K$' is positive.
%Looking at the second dashed line in Fig. 	\ref{fig:libtime_k}, after $\sim 91$ Kyr, the amplitude of libration has again begun to diverge.
%This happens just before $K$' returns to a positive value, again showing the impact that the complex corotation torque can have on the resonance, slightly deviating away from what is expected from the \citet{TaTaWa2002} formula.

%Whilst figs. \ref{fig:sma_zeta} and \ref{fig:libtime_k} showed how the co-orbital resonance evolves as a pair of planets migrate, it is important to examine whether the resonance survives upto and after the end of the disc lifetime.
The planets migrating in figs. \ref{fig:sma_zeta} and \ref{fig:libtime_k} migrated until they reached the disc inner edge, close to the central star.
Whilst doing so, they moved away from the co-orbital resonance, eventually breaking out of the resonance.
However global simulations of planet formation have shown that planets in co-orbital configurations cover a wide range of semimajor axes at the end of the disc lifetime \citep{ColemanNelson16,ColemanNelson16b}.

\subsubsection{The effects of changing planet mass over time}
\begin{figure}
\centering
\includegraphics[scale=0.6]{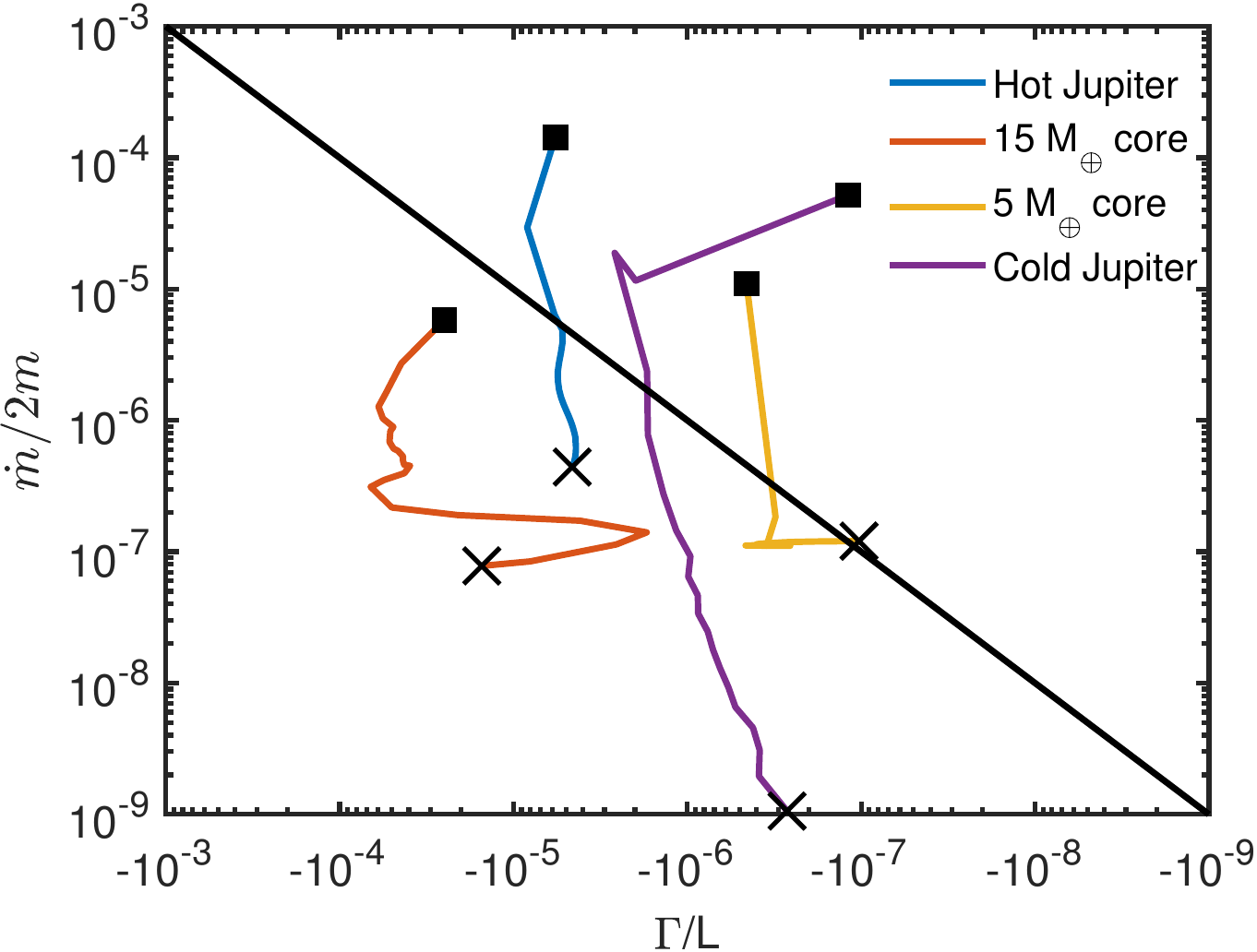}
\caption{Evolution of the relative change in mass against the relative change in angular momentum for four different planets over time.
Black squares denote the starting point for the planets, whilst black crosses denote where the planets finish.
Regions above the black line act to stabilise the co-orbital resonance, whilst below the line destabilises.}
\label{fig:mdot_torque}
\end{figure}
We now consider the case where the mass of a planet in the co-orbital resonance is changing over time in addition to its migration in the disc,meaning that it is accreting gas or planetesimals.
As show in Sect. \ref{sec:stabu}, both slow mass accretion and migration in the disc changes the width of the resonance in the way that leads to a divergence from the equilibrium (if the size of the resonance decreases due to inward migration), or to a convergence towards it (mass accretion or outward migration).

Figure \ref{fig:mdot_torque} shows the relative change in mass against the relative change in angular momentum for a number of evolving planets at different locations of the protoplanetary disc and of different masses. We note that these planets are not in co-orbital configuration but are used to probe the torque felt and their accretion rate during their evolution in the disc.
The planets shown in fig. \ref{fig:mdot_torque} were placed in a nominal protoplanetary disc similar to those shown in \citet{ColemanNelson16}.
The black line shows where the sum relative changes in mass and angular momentum equate to zero, indicating no change in the stability of the co-orbital resonance.
Planets that are evolving above this line will be accreting mass at a faster rate than they are migrating inwards and as such will be introducing a stabilising effect to their co-orbital region.
For planets evolving below the line, then the opposite will occur, and they will diverge from the equilibrium.
It is interesting to note, but not shown in the figure, that if a planet is undergoing outward migration, then the co-orbital resonance will be always be stabilising, so long as the planet is not losing mass at a significant rate.
Looking at the regimes that specific planets operate in, we see that when a planet is small and growing through planetesimal or pebble accretion (yellow line), it is typically above the black line, since the planet is increasing in mass faster than it is migrating, stabilising the co-orbital region.
For more massive planets, of mass between 10--20 $\rm M_{\oplus}$, that are accreting few pebbles and/or planetesimals, but are accreting gas slowly (red line), they will tend to sit below the black line, destabilising the resonance.
This is due to them migrating faster than they are accreting, however this is also highly dependant on the local disc profiles, which could allow planets to become trapped and migrate inwards slowly, reducing the magnitude of the destabilisation.
For example, the red line shows the evolution of a $\sim15\rm M_{\oplus}$ planet initially orbiting at 5 au.
The location of the planet in the protoplanetary disc, will also affect how close they appear to the black line in fig. \ref{fig:mdot_torque}, since the migration torques are dependant on the local disc conditions, for example the $15 \rm M_{\oplus}$ planet shown by the red line would be shifted to the right of the plot if it was initially orbiting at 20 au, instead of 5 au as is shown in the plot.
For planets that grow into giant planets through runaway gas accretion (blue and purple lines), initially they are in a regime of significant stabilisation as they accrete gas extremely quickly.
Once the runaway gas accretion phase ends, they accrete gas at a slower rate, but migrate at a similar rate, reducing the effects of the stabilisation, before ultimately making it destabilising.
Again, given the location in the disc that these planets occupy, this will mainly affect the rate of change of angular momentum, possibly making the co-orbital region stabilise or destabilise at a faster rate.
For the giant planet that survives migration (purple line), meaning it does not migrate into the central star, the period of slow type-II migration (at the bottom of Fig. \ref{fig:mdot_torque}) is destabilising for the co-orbital region.
This is again due to the planet migrating at a faster relative rate than it is able to accrete gas.
%Given that giant planets that survive migration typically form late in the protoplanetary disc lifetime \citep{ColemanNelson16b}, the location of the planet in the disc should have little effect on the evolution of the co-orbital region of these planets. 

Both mass accretion and migration hence have significant impacts on the evolution of the co-orbital resonance in the disc and have to be taken into account. 
In Sect. \ref{sec:type2}, we will estimate how the perturbations of the disc induced by the pair of planets perturbs the torque that is applied to each of them.

\subsection{Population outcome}
\label{sec:pop_outcome}
\begin{figure}
\begin{center}
\includegraphics[width=1\linewidth]{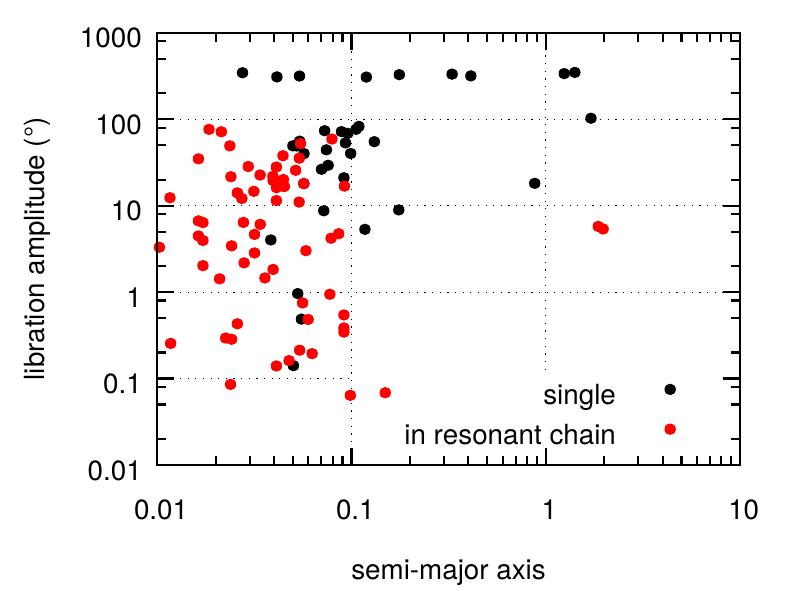}
\caption{\label{fig:pop_outcome} Final amplitude of libration and semi-major axis of the co-orbitals formed in 880 synthetic systems around a 0.1 Solar mass star. 12\% of the system had co-orbitals at the end of the run. Isolated co-orbital pairs are displayed in black, while those that are in resonance with another planet are displayed in red. Horseshoe configurations have an amplitude of libration above $180^\circ$ degree. Other configurations are trojan. }
\end{center}
\end{figure}

While sect. \ref{sec:gavintype1} examined the evolution of the co-orbital resonance in an evolving protoplanetary disc, it is interesting to see how often these co-orbital resonances actually occur in a much larger suite of simulations.
To do this, we searched for co-orbital resonances in a recent population of simulations which studied planet formation around low mass stars.
The simulations initially began with a realistic range of initial conditions, such as disc mass and solid mass, and used a similar disc model to that described in sect. \ref{sec:gavintype1}, but which had been adapted to being suitable for a low mass star  \citep[see Sect. 3 of][]{ColemanProxima17}.
The main aim of these simulations was to examine the formation of planetary systems, through pebble or planetesimal accretion, around low-mass stars of mass $0.1 {\rm M}_{\bigodot}$, similar to Trappist-1 and Proxima Centauri \citep{Coleman2019}.
%This population of systems around low-mass stars seemed most suitable as most of the generated planets remain in type I migration throughout the disc lifetime.
Initially in these simulations, a number of low mass planetary embryos ($m_{\rm p}<0.1\me$) would be scattered throughout the disc, and would be able to either accrete pebbles or planetesimals, and also undergo type I migration.
% (this population of systems around low-mass stars seemed most suitable as most of the generated planets remain in type I migration throughout the disc lifetime). 
As the planets migrate they become trapped in resonant chains, typically involving first-order resonances, but could also become trapped in co-orbital resonances.
As the systems evolve, the planets migrate to their final locations, sometimes maintaining their resonant chains and also their co-orbital configurations. 
The planetary systems are integrated for 3 to 5 million years after disc dispersal to allow the planetary systems to continue to evolve in an undamped environment.

We analysed the outcome of the 880 systems generated in the pebble accretion scenario and found co-orbitals in $\approx 12\%$ of the final systems.
Figure \ref{fig:pop_outcome} displays all of these co-orbitals as a function of their semi-major axis and amplitude of libration at the end of the simulation. As expected, co-orbitals that migrated on their own and survived until the end of the disc lifetime tend to have a large amplitude of libration. However, the subgroup of co-orbitals that were trapped into a resonant chain with other planets seemed to be able to migrate close to the inner edge of the disc while retaining a small amplitude of libration. 

Indeed, we show in appendix \ref{sec:reschain} that co-orbital configurations that would be unstable on their own can be stabilised during the protoplanetary disc phase by the presence of another planet trapped in first order mean motion resonances either inside, or outside, of the co-orbital pair.

\section{Evolution in a protoplanetary disc: 2D hydrodynamical simulations}
\label{sec:type2}

The torques applied on the planets in the previous sections were obtained for a single planet embedded in a disc. When there is more than one planet in the disc, the surface density perturbations by the other planets can alter the torque on each individual planet \citep{BaruteauPapaloizou2013,PiRa2014,Broz2018}. On the other hand, moderate-mass planets might open a partial gap around their orbits that can also vary the torque from its pure type-I value. In this section we take these effects into account by running two-dimensional (2D) locally isothermal hydrodynamical simulations using \texttt{FARGO} \footnote{http://fargo.in2p3.fr/-Legacy-archive-} code \citep{2000A&AS..141..165M} for a system with two co-orbital planets in different mass regimes.

\subsection{Disc and planets setups}
\label{hydro:setup}

The disc in our simulations is extended radially from 0.3 to 2.5~au and azimuthally over the whole $\rm 2\pi$. It is gridded into $N_{r}\times N_{\phi} = 873 \times 1326$ cells with logarithmic radial segments. The resolution is chosen such that the half horseshoe width of a $3M_{\oplus}$ planet can be resolved by about 6 cells. The surface density profile is $\Sigma=\Sigma_{0} r^{-\alpha}= 2\times 10^{-4} r^{-0.85}$ in code units. This corresponds to $1777\ g/cm^2$ when the radial unit is 1~au and the mass unit is 1$M_{\bigodot}$. The disc viscosity follows the alpha prescription of viscosity $\nu_{\rm visc} = \alpha_{\rm visc} c_{s} H$ where $c_s$ is the sound speed and $H$ is the disc scale height. These two quantities are related to the aspect ratio $h$ as $h=H/r=c_s/v_{k}$, $v_{k}$ being the Keplerian velocity. The disc is flared with the aspect ratio of $h=h_{0} r^{f} = 0.05 r^{0.175}$. The reason of such choices for surface density and aspect ratio profiles will be explained in Sec.~\ref{hydro:disc1d2d}.

Planets are initiated radially at $r_{1}=r_{2}=1$~au and azimuthally at $\lambda_{1}=0$ (more massive one) and $\lambda_{2}=+50$ or $-50$ degrees. The mass of the planets are increased gradually during the first 50 years to avoid abrupt perturbations in the disc while they are kept on circular orbits until $t=100$~yr. The total time of the simulations is 2000 or 3000~years depending on the migration rate of the planets. In cases that the planets migrate faster, we had to stop the simulations earlier to avoid the effect of the inner boundary. In order to have the consistency with the 1D simulations, we used $\epsilon_{p}= 0.4h$ for smoothing the planet's potential. The planets do not accrete gas during these simulations and their masses stay constant after $t=50$~yrs.
%The initial eccentricity of the planets are zero.
% and mass of the planets range from $1M_{\oplus}$ to $1M_{\rm Jup}$.

\begin{table*}[]
\setlength{\extrarowheight}{7pt}
\caption{Value of the different terms of the stability criterion $u$ for different hydro simulations with identical initial conditions except for the masses of the co-orbitals, given in the first two columns. The upper block shows the three models with low-mass planets which do not alter the surface density profile. In the second block, the planets are more massive and perturb the disc by creating a shallow gap around their orbits. In the two models of the last block, we used a full gap-opening planet as one of the co-orbitals in order to eliminate the co-rotation torque on the low-mass one.}
\label{table:typeI}
\centering
\small
\tabcolsep=0.11cm
\begin{tabular}{ c c | c c c c c c || c c || c | c l }
\hline\hline
$m_{1}$ & $m_{2}$ & -$\frac{\Gamma_{10}}{W}$ & $\frac{m_2 \Gamma_{1\Delta}}{(m_1+m_2)W}$ & $R_{1\zeta}$ &  -$\frac{\Gamma_{20}}{W}$ &  $-\frac{m_1 \Gamma_{2\Delta}}{(m_1+m_2)W}$ & $-R_{2\zeta}$ & $2u$ &  $\frac{d \zeta_{max}-\zeta_{min}}{dt}$ & $C_\Gamma$   & $K_1$ & $K_2$ \\
\hline
[$M_\oplus$] &[$M_\oplus$] & [yr$^{-1}$] & [yr$^{-1}$]& [yr$^{-1}$]& [yr$^{-1}$]& [yr$^{-1}$]& [yr$^{-1}$]& [yr$^{-1}$]& [rad.yr$^{-1}$]& - &- &- \\
\hline
3.0&5.0&3.9e-6&-2.0e-5&1.4e-5&1.3e-5&2.2e-5&4.2e-6&3.6e-5&5.9e-6&6.5e-2&-1&2	 \\
3.0&3.0&5.1e-6&-3.1e-5&7.7e-6&6.4e-6&-1.8e-5&4.1e-6&-2.6e-5&-4.1e-6&2.2e-2&-3&-0.9	\\
5.0&3.0&1.1e-5&-7.4e-5&7.4e-6&5.4e-6&-5.8e-5&7.3e-6&-1.0e-4&-1.9e-5&-1.9e-2&-5&-3	\\
\hline
3.0&15.0&5.4e-7&-6.2e-5&5.7e-5&5.1e-5&2.0e-4&1.7e-6&2.5e-4&3.2e-5&1.7e-1&-1e1&1e1\\
6.0&12.0&5.8e-6&-2.3e-5&3.4e-5&3.6e-5&1.9e-4&4.0e-6&2.5e-4&4.1e-5&1.1e-1&-0.5&6\\
10.0&10.0&1.6e-5&1.8e-4&2.7e-5&2.5e-5&-2.6e-4&6.8e-6&-9.1e-6&-2.8e-6&4.4e-2&6&-5	\\
12.0&6.0&3.1e-5&3.1e-5&1.4e-5&9.3e-6&-2.2e-4&2.4e-5&-1.1e-4&-1.8e-5&-5.5e-2&2&-5	\\
15.0&3.0&4.8e-5&-2.6e-4&7.1e-6&1.9e-6&-8.2e-5&1.8e-5&-2.7e-4&-4.5e-5&-1.4e-1&-1e1&-4	\\	
%1$M_{J}$ & 1$M_{E}$ & 2.1E-5 & 1.6E-5 & -1.4E-6 & -1.2E-7 & 2.9 E-4 & -2.8E-4 & -  &  - \\
%1.0&333.3&1e-7&0.00027&-0.00037&2.3e-5&1.8e-5&-1.6e-6&4.5&1.3e+02&-6.1e-5&1.3e-6\\
%333.3&1.0&2.3e-5&1.8e-5&-2e-6&-1.3e-7&0.00027&-0.00032&1.3e+02&-2.5&-1.4e-5&-4.5e-6\\	
%10.0&333.3&1.1e-6&0.00017&-0.00035&1.1e-5&0.00012&-1.4e-5&3&1.8e+02&-&-\\
%333.3&1.0&2.3e-5&1.8e-5&-1.9e-6&-1.3e-7&0.00027&-0.00032&1.3e+02&-2.6&-9.7e-6&-3.4e-6	\\
%333.3&1.0&2.1e-5&1.8e-5&-1.6e-6&-1.2e-7&0.00029&-0.00029&1.4e+02&-3&4e-5&1.5e-6\\
%1.0&333.3&8.7e-8&0.00023&-0.00032&2.1e-5&1.8e-5&-1.3e-6&4.5&1.4e+02&-5.1e-5&4.4e-6	
%12$M_{E}$ & 6$M_{E}$ & 3.1E-5 & -9.4E-6 & 1.3E-5 & 9.3E-6 & -2.7 E-4 & 2.2E-5  & 0.8  & -6.8\\
%6$M_{E}$ & 12$M_{E}$ & 5.8E-6 & -8.7E-6 & 3.4E-5 & 3.6E-5 & 1.9 E-4 & 1.6E-6  & 0.12  & 5.9\\
%3.0&15.0&5.5e-7&-6.9e-5&6e-5&5.1e-5&0.00019&1.9e-6&-12&9.8	\\
%15.0&3.0&4.9e-5&-0.00025&6.9e-6&1.9e-6&-7.9e-5&1.2e-5&-12&-3.6	 \\
\hline
1.0&333&8.5e-8&2.3e-4&-3.2e-4&2.1e-5&1.7e-5&-1.3e-6&-4.9e-5&4.6e-6&-1.1e-3&5&1e2	\\
333&1.0&2.1e-5&1.7e-5&-1.6e-6&-1.2e-7&2.9e-4&-2.8e-4&4.3e-5&2.1e-6&-9.9e-3&1e2&-3	
\end{tabular}
\end{table*}

We used various planet mass pairs as given in the left columns of Table \ref{table:typeI}, where $m_1$ is the mass of the leading planet and $m_2$ the mass of the trailing one.
% $m_{1}, m_{2}= [10,10]M_{\oplus}$, $m_{1}, m_{2}= [12,6]M_{\oplus}$, and $m_{1}, m_{2}= [15,3]M_{\oplus}$. The depth of the gap created by the highest mass in these simulations is not more than 20\% and therefore the migration can be considered in type I. For each model, we ran two simulations, one with $K>0$ and one for $K<0$. The $K$ values are chosen such that based on Fig.~\ref{fig:tauacrit}, we expect to have divergence for the positive values and convergence for negative values. The details of the disc parameters are given in table~\ref{table:typeI}. 
%In all of these simulations, both planet are initially set $50^\circ$ apart, with the same semi-major axis $a_1=a_2=1\,$au, with zero eccentricity. 
We saved the forces and torque on each planet 20 times per year, allowing us to estimate the value of the different terms of the stability criterion $u$, given by Eq. (\ref{eq:eigencircp}). The method to compute these different terms is given in Appendix \ref{ap:Forcespd}. Columns 3 to 8 of Table \ref{table:typeI} give the averaged values of each quantity over the whole simulation time, and the 9th column is the quantity $2u$, the sum of these contributions.

\subsection{Comparison of 1D to 2D discs}
\label{hydro:disc1d2d}
First, we compare the evolved disc in the hydro-models to our 1D discs. In Sect. \ref{sec:gavintype1}, we saw that the criterion (\ref{eq:stabcon}), based on the analytical torques from \cite{TaTaWa2002}, was a good approximation to estimate the stability of the system. This criterion is a function of the local flaring index and surface density slope $f$ and $\alpha$ through the parameter $K$ which parametrises the local slope of the torque felt by both planets. In all of the simulations presented in this section, the flaring index and initial surface density slopes are $f=0.175$, $\alpha=0.85$ implying that the disc is viscously in equilibrium. These values correspond to $K=0.7$, for which the co-orbital planets in type-I migration should always be diverging (see Fig. \ref{fig:tauacritTana}). In our hydrodynamical runs, we estimate $K_j$ --the value of $K$ which is felt by the $j$th planet-- that can be linked to the quantities $\Gamma_{j0}$ and $\Gamma_{j\Delta}$ by the following relation (see Eq. \ref{eq:GammajI}):
\be
K_j=\left(1+\frac{m_j \Gamma_{j\Delta}}{(m_1+m_2)\Gamma_{j0}}\right)/2
\label{eq:Kj}
\ee 
The $K_j$s for the hydro simulations are hence computed and given in the last two columns of Table \ref{table:typeI}. It appears clearly that the local slope of the torque felt by each planet is different from what is expected from \cite{TaTaWa2002}. The upper block of the table (the first three rows) contains the models with low-mass planets such that they do not perturb the disc greatly. The maximum surface density perturbation $\delta\Sigma/\Sigma_{0}$ in these models is only about 3\%. Therefore the difference of $K_j$ with respect to the pure type-I migration which is expected for this type of planet is due to the presence of the co-orbital companion. For the models in the second block of the table, $\delta\Sigma/\Sigma_{0}$ is at most 35\%. In these models, the torques are modified both by the presence of the other planet and the partial gap. Hence, the stability of co-orbitals is expected to be different from those obtained using a 1D disc model. However, we note that if one of the co-orbitals is significantly more massive than the other, the constant part of the torque that applies on it remains the same regardless of the position of the lower mass planet. For large mass discrepancies, the first two terms of $u$ (Eq. \ref{eq:eigencircp}) might hence be properly estimated by the 1D model, see for example Fig. \ref{fig:mdot_torque}. However, hydrodynamical simulations are needed to estimates the $\Gamma_{j\Delta}$ and $R_{j\zeta}$ terms.

\subsection{Low to moderate mass planets: Super-Earth to mini-Neptune}
\label{hydro:hydrytypeI}

%\subsubsection{Outcome of the simulation}
%TTTTTTTTTTTTTTTTTTTTTTT
%

%5.9E-6*57;
%                 0.0003363
%> 57*-4.1E-6;
%                -0.0002337
%> 57*-1.9E-5;
%                 -0.001083
%> 57*2.5E-4;
%                   0.01425
%> 57*2.5E-4;
%                   0.01425
%> 57*-9.1E-6;
%                -0.0005187
%> 57*-1.1E64;
%                 -6.27E+65
%> 57*-4.5E-5;
%                 -0.002565

%FFFFFFFFFFFFFFFFFFFFFFF   typeI_analysis_5_3.pdf
\begin{figure}
\begin{center}
\includegraphics[width=\linewidth]{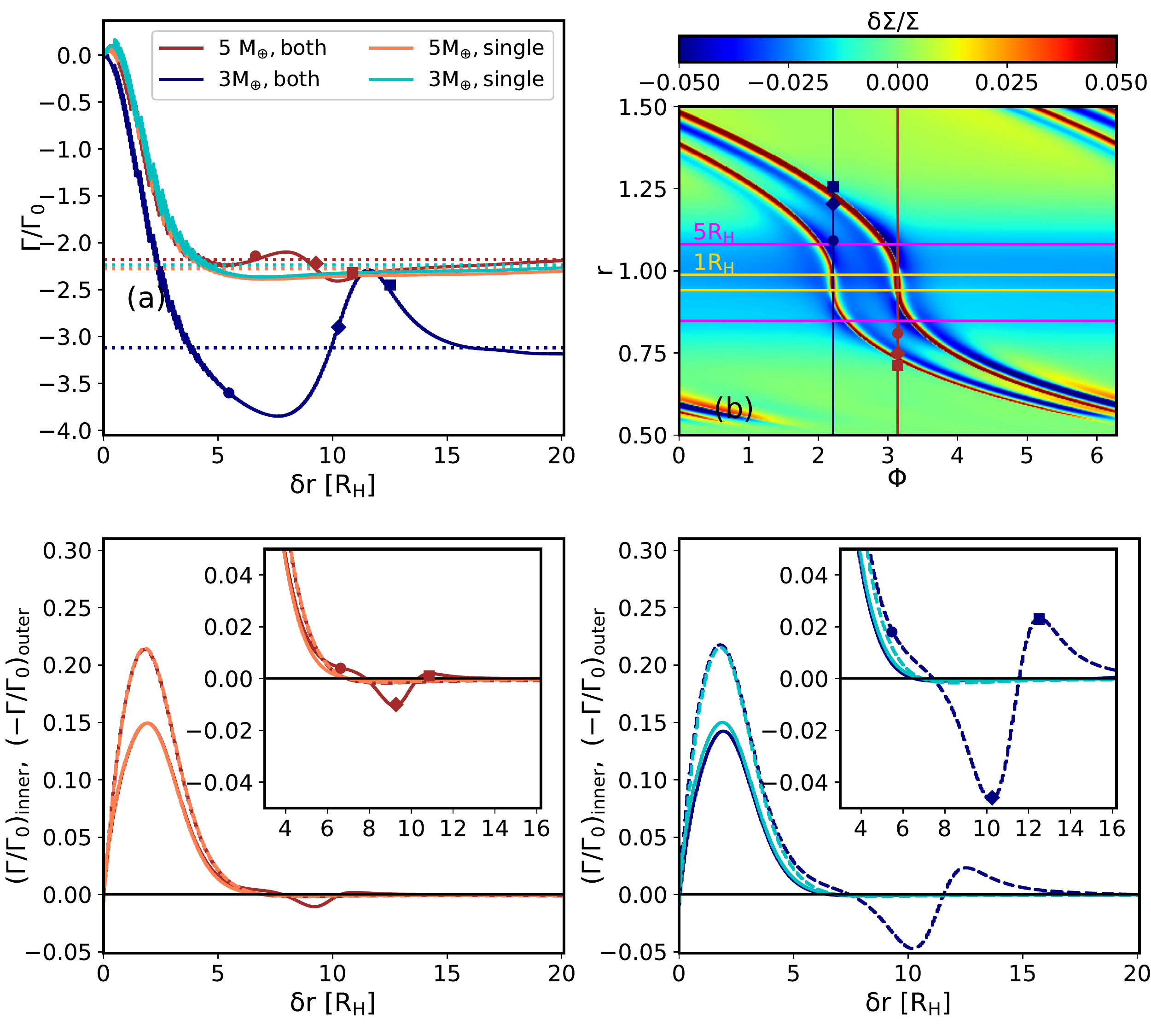}
\caption{\label{fig:hydrotypeI53} \textit{(a)}: Comparing the torque from the disc between $a-\delta r$ and $a+\delta r$  (solid lines) on $m_{1}=5M_{\oplus}$ planet (red lines) and $m_{2}=3M_{\oplus}$ (blue lines). The lighter colours represent the torques from the simulations with a single planet and the darker ones for the co-orbital simulation. The same colour dotted lines mark the torque from the whole disc. The $y$-axis is the scaled torque and $x$-axis the distance from the planets' orbit in unit of their mutual Hill radius $R_{H}$. \textit{(b)}: Perturbed surface density. Two horizontal lines are drawn at $1$ and $5R_{H}$ from the planets' orbits to guide the eye. \textit{(c)}: The torque on $m_{2}$ as a function of distance from the planet. Panel~(a) is the cumulative torque but this panel and panel~(d) show the torque only from the grid cells at a given distance from the planet. The dashed and solid lines belongs to the outer and inner disc, respectively. The colour code is the same as in panel~(a). To ease the comparison, we plot the negative of the torque from the outer disc. The symbols mark where the torque on the planets change due to the presence of the second planet. \textit{(d)}: The same as panel~(c) but for $m_{1}$.}
\end{center}
\end{figure}

%FFFFFFFFFFFFFFFFFFFFFFF   typeI_analysis_12_6.pdf
\begin{figure}
\begin{center}
\includegraphics[width=0.99\linewidth]{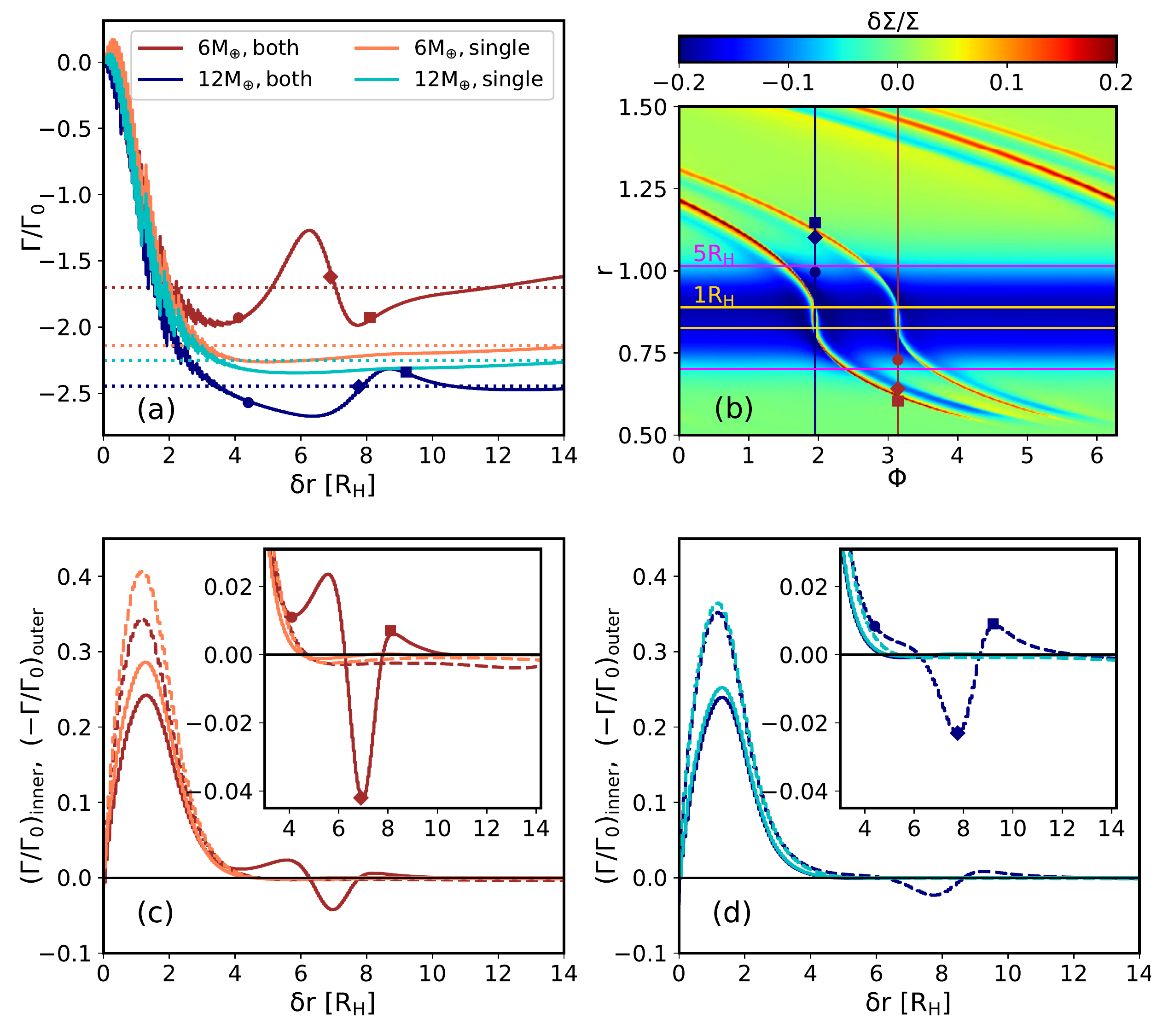}
\caption{\label{fig:hydrotypeI126} The same as Fig.\ref{fig:hydrotypeI53} but for $m_{1}=6M_{\oplus}$ and $m_{2}=12M_{\oplus}$.}
\end{center}
\end{figure}

%\begin{figure}
%\begin{center}
%\includegraphics[width=.9\linewidth]{torqueex_lm.pdf}\\
%\caption{\label{fig:torqueex_lm} Evolution of the torque from the disc applied on the small planet $m_{2}$ over a single libration period in the case $m_1,m_2=[12,6]M_{\oplus},\ K>0$. The negative torque becomes smaller when the planets get close to each other and  larger when are farther.}
%\end{center}
%\end{figure}
%FFFFFFFFFFFFFFFFFFFFFFFFF

%The last two columns of Table \ref{table:typeI} gives the values $K_j$, which parametrise the local slope of the evolution of the angular momentum of each planet (Eq. \ref{eq:Kj}). For all simulation, we set the disc initial conditions such that the torques from \cite{TaTaWa2002} should be $K_j=0.7$ for all planets.

In this section we study the evolution of co-orbitals in the super-Earth to mini-Neptune mass regime which are the planets that do not open a full gap in the disc. For comparison with the stability criterion developed in Sect. \ref{sec:stabu} (attractive equilibrium for negative $u$, repulsive for positive $u$), we give in Table~\ref{table:typeI} the quantity $d (\zeta_{max}-\zeta_{min})/dt$, averaged over the simulation time. For the first two blocks of the table, where the no-gap and partial-gap opening planets are listed, the stability of the Lagrangian equilibrium is indeed correctly predicted by the criterion $u$ as it has the same sign as $d (\zeta_{max}-\zeta_{min})/dt$. On these sets of simulations, we can see a trend that was already remarked by \cite{PiRa2014}: a more massive leading planet ($m_1>m_2$) tends to stabilise the co-orbital configuration, while if the more massive planet is trailing, the system slowly evolves away from the equilibrium.

%In Fig.~\ref{fig:hydrotypeI}, we plot the rate of change in the libration angle $\rm d \zeta/dt$, averaged over the final 1000~years of the simulations, for each of the models. In spite of our prediction, both models for $[10,10]M_{\oplus}$ show convergence and in the rest of them with larger $m_{1}$, the planets diverge from the Lagrangian point. 

%
%
To see how much the presence of the second planet can affect the torque, we present the torque analysis for the models $(m_{1},m_{2}) = (5,3)M_{\oplus}$ in Fig.~\ref{fig:hydrotypeI53}. In panel~(a), we plot the scaled torque $\Gamma_{j}/\Gamma_{0}$, with $\Gamma_{0}=(\dfrac{m_{j}/m_{0}}{h})^2 \Sigma_{p} a^4 \Omega_{p}^2$, versus distance from the planet's orbit $\delta r$, which is scaled to the planet's mutual Hill radius $R_{H} = \sqrt[3]{(m_{1}+m_{2})/3m_{0}}$. $\Gamma_{j}/\Gamma_{0}$ is the torque that is exerted on the $j$th planet by the material within $\pm\delta r$ of the planet's orbit. In this figure, we compare the torque on each planet in the co-orbital simulation with the corresponding single-planet model. The scaled torques on single planet models are almost identical as expected for type-I migration. The torque on the co-orbital $5 M_{\oplus}$ follows the single planets up to about 5~$R_{H}$. It means that the torque from the co-rotation region and the its own spiral in this area is identical to the single planet model. As we move further out, the torque levels up until 8~$R_{H}$, decreases until 10~$R_{H}$, and increases again until it reaches the value of the total torque (dotted line). The cause of this variation can be found in panel~(b), where the surface density perturbation is shown, and in panel (c), in which we plot the torque from each location in the disc on the planet. In panel~(c), the solid line demonstrates the torque from the inner disc and the dashed line is the \textit{negative} of the torque from the outer disc. The advantage of this plot is that it shows where the torque from the disc in the co-orbital model differs from the single one. We see in panel~(c) that the torque from the inner and outer disc is identical until about 5~$R_{H}$, where the positive torque from the inner disc increases in the co-orbital model. This area of the positive torque, marked by a red dot, corresponds to the area between the two inner arms of the planets. The presence of the second planet creates a slight depression in the surface density at the left side of the planet compared to the right. It makes the torque from this area more positive. As we move further the torque drops strongly due to the fact that we get close to the over-dense part of the second planet's spiral arm which is located on the left side of the planet and exerts a negative torque on the $5 M_{\oplus}$. Then, this over-dense arm moves to the right side of the planet and its effect turns to a positive torque. The sum of all these components which arise from the \textit{depletion between the planets' arms} and \textit{other planet's spiral} is responsible for the deviation of the torque from the single planet models. The torque on the $3 M_{\oplus}$ is very similar except that the effect of the other planet is stronger due to its stronger arm (panel~d): the diamond and square symbols mark the passage of the $5 M_{\oplus}$ planet's spiral arms by the azimuthal position of the $3 M_{\oplus}$ planet. 

As the spiral arms follow the planets in their libration around the Lagrangian equilibrium, both the torques and radial forces applied on each planet evolve over the libration time-scale, which is responsible for the non-negligible terms in columns 4, 5, 7 and 8 of Table \ref{table:typeI}.
%However, in this model, the torque from the inner disc at about 2~$R_{H}$ is slightly smaller than the single planet model that also contribute in the total torque. 

In Fig.\ref{fig:hydrotypeI126}, we present the same torque analysis for the model $(m_{1},m_{2}) = (6,12)M_{\oplus}$, which has two partial gap opening planets. As in the low-mass co-orbital model, the main cause of the deviation from the single-planet models is the depression of the surface density between the two planets' arms and the presence of the other planet's arm. In addition, the partial gap between the two planets ($\phi \in [\pi,4\pi/3]$) is deeper than the rest of the gap, which creates an additional offset for the torques 	applied on each planet.  \\
%One should note that the difference between the scaled torque of the single models, coral and cyan dotted lines, is because of the deeper partial gap around $12 M_{\oplus}$ than $6 M_{\oplus}$ that suppresses the torque from its spirals. 

The torques on the co-orbital planets is hence very different from those that would apply on a single planet, and the difference originates from the suppression of the surface density between the planets and the effect of the other planet's spiral arm. According to our current knowledge, there is no extensive study that tell us how the depletion of mass between the co-orbitals changes by the disc parameters or the mass of the planets. On the other hand, the torque from the other planet's spiral arm depends on the strength of the arm which depends on the planet's mass, and opening angle of the spiral which is a function of the disc aspect ratio. Hence, we expect the stability of the co-orbitals depends on these two parameters because as the planets librate around their equilibrium point, their distance from each other and consequently from each other's spiral would also change. In the following section, we remove the complexity of the partial gap and the co-rotation torque by replacing one of the co-orbitals with a gap-opening planet.

%As the lower-mass planet librates around the $L_4$ point, the distance between the planets' spirals changes and it causes some oscillations in the value of the torque. This is clearly visible in Fig.~\ref{fig:torqueex_lm} which shows the torque on the less massive planet during one libration period for the model $[12,6]M_{\oplus}$. As a result, for co-orbital planets not only the average torque on each planet differs from the torque that would be applied on single planets, but also this torque depends on the value of the resonant angle $\zeta$.

\subsection{Gap-opening planets: A Jupiter and an Earth}
\label{hydro:hydrytypeII}

%FFFFFFFFFFFFFFFFFFFFFFFFF

%\begin{figure}
%\begin{center}
%\includegraphics[width=1\linewidth]{hydro_typeII_dzdt-eps-converted-to.pdf}\\
%\caption{\label{fig:hydrotypeII} The same as Fig.~\ref{fig:hydrotypeI} but for systems with at least one gap-opening planet.}
%\end{center}
%\end{figure}

\begin{figure}
\begin{center}
\includegraphics[width=.9\linewidth]{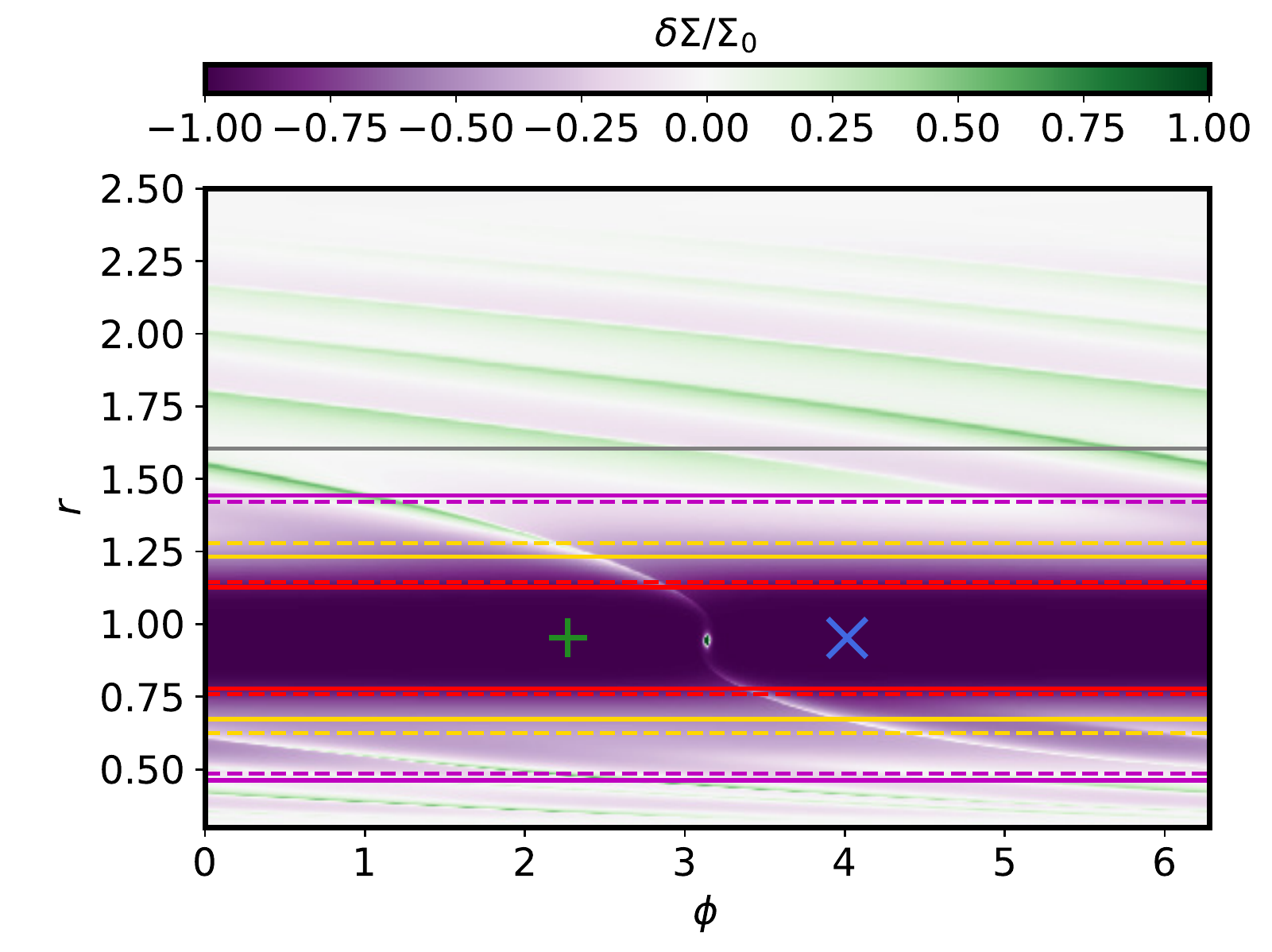}\\
\includegraphics[width=.9\linewidth]{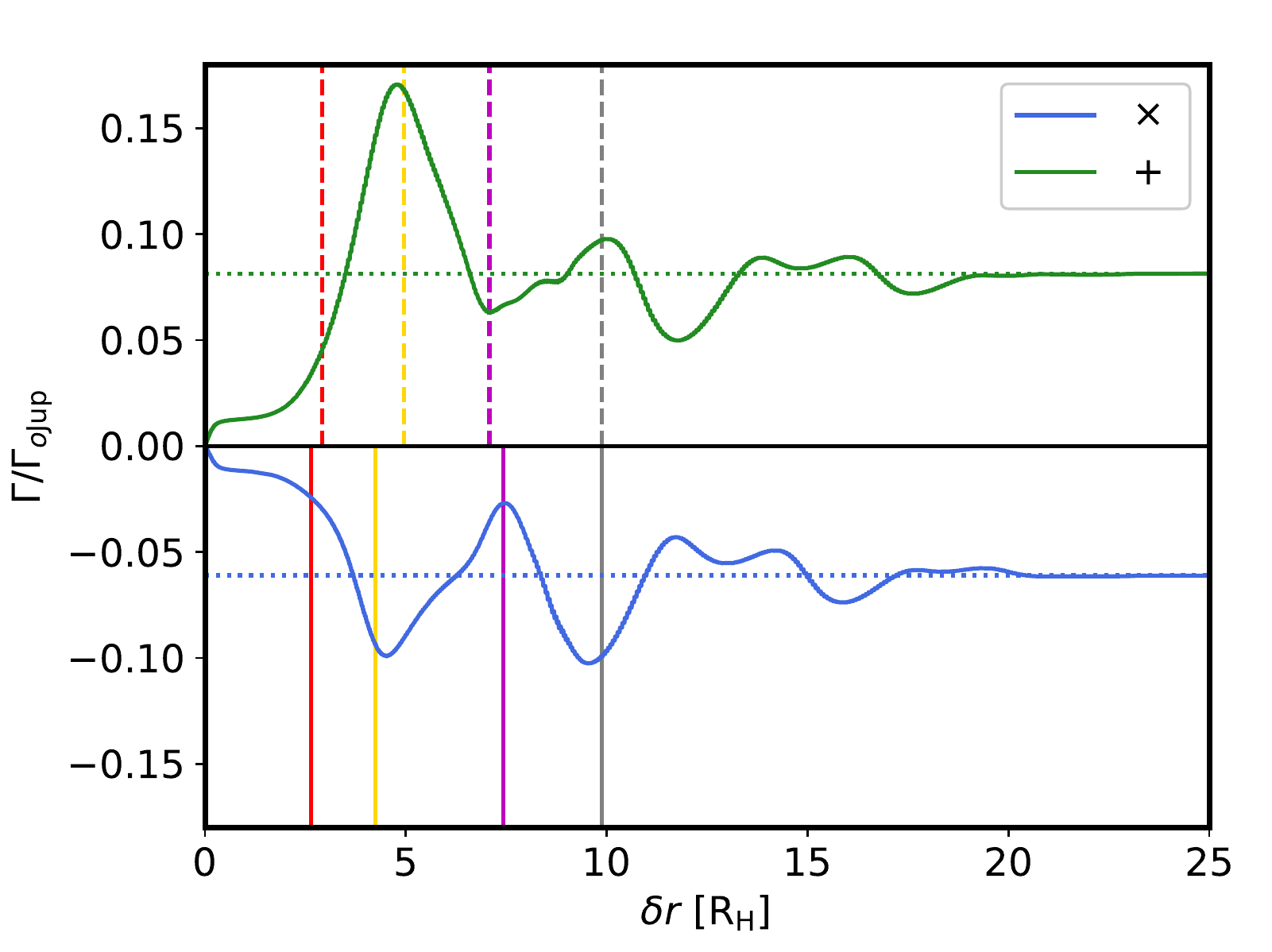}\\
\caption{\label{fig:hydrotypeIItorque} \textit{Upper}: The perturbed surface density  for the models with a Jupiter and an Earth. The location of the low-mass planets are marked with $+$ and $\times$ signs. Because the surface density perturbation is identical in both cases, we only show the surface density map of one of them but marked the location of both planets for comparison. The red, yellow, magenta, and grey lines mark different distances from the planets' orbit. The dashed lines belong to the model with the low-mass planet at $\times$ and the solid lines are for the the one with planet at $+$. These lines are also added to the lower panel to denote the effect of the Jupiter's spirals. \textit{Lower}: The same as panel~(a) of Fig.~\ref{fig:hydrotypeI53}. The green and blue curves show the torque on the low-mass planet initiated close to $L_4$ and $L_5$ equilibria, respectively. The scaling is the Jupiter's Hill radius $R_{H}$ for the $x$-axis and $\Gamma_{0}$ for the $y$-axis that is calculated using the Jupiter's mass.}
\end{center}
\end{figure}

\begin{figure}
\begin{center}
\includegraphics[width=0.49\linewidth]{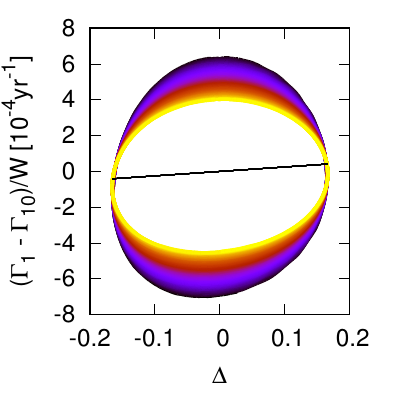}\includegraphics[width=0.49\linewidth]{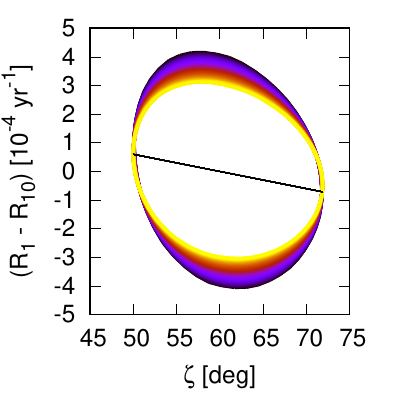}
\caption{\label{fig:hydrotorqueJupiter} Evolution of torque (left) and radial force (right) on the Earth-mass planet in the leading case `$\times$'. The black lines show the linear approximation for the evolution of these quantities with respect to $\Delta$ and $\zeta-\zeta_{eq}$, respectively.}
\end{center}
\end{figure}

\begin{figure}
\begin{center}
\includegraphics[width=1\linewidth]{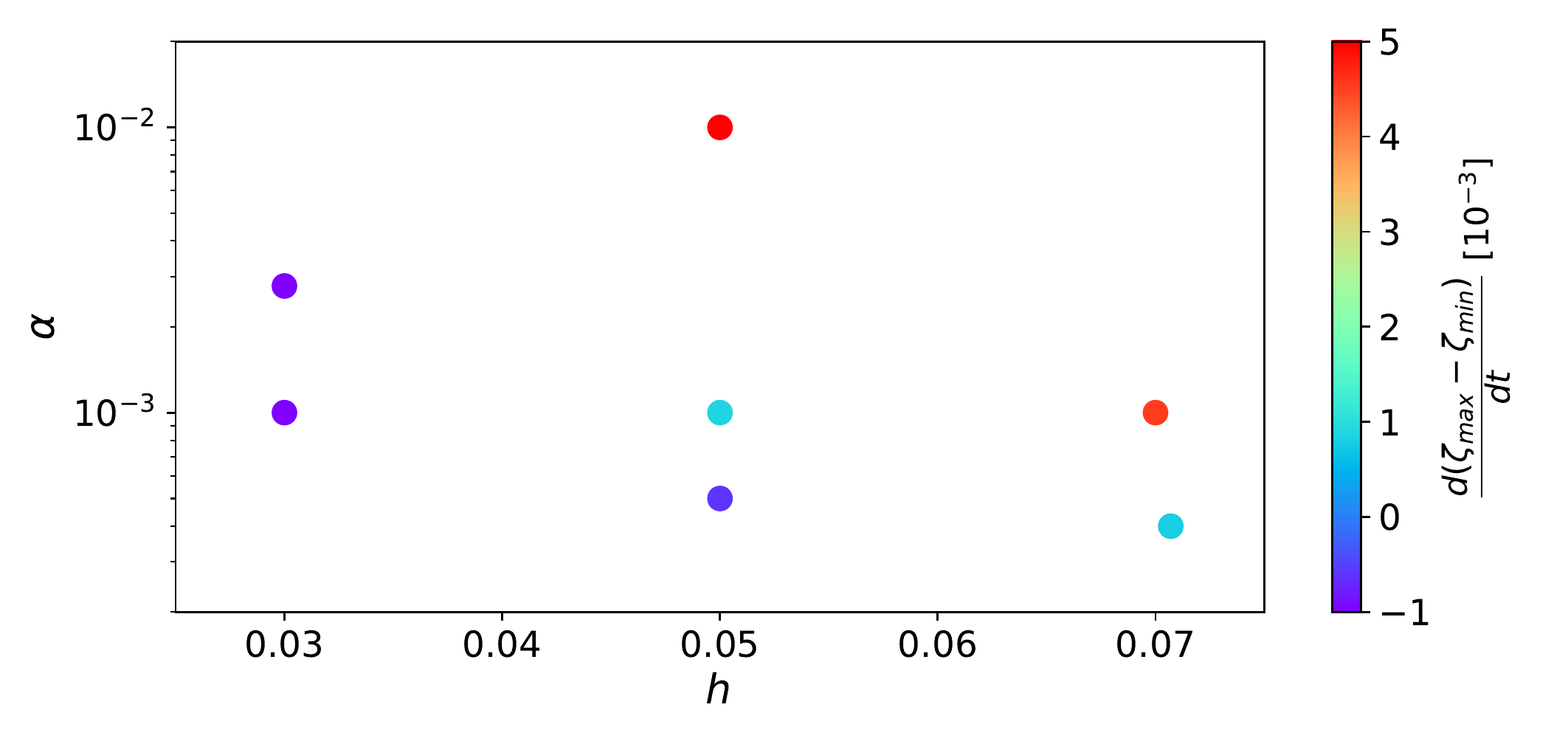}\\
\caption{\label{fig:hydrodiscpara} Evolution of the libration angle for models with $m_{1}=10M_{\oplus}$, $m_{2}= 1M_{\rm Jup}$ and in discs with different values for $\alpha_{\nu}$ and aspect ratio $h$. }
\end{center}
\end{figure}

%FFFFFFFFFFFFFFFFFFFFFFFFFFFFFFFFFFFFFFFFFFF

%Low-mass planets, which do not open a deep gap, feel the torque from their own and also from the spirals of the another planet (hereafter ex-spirals). 
In order to isolate the effect of the spiral arm of one planet on the other, we ran two simulations with a Jupiter-mass planet and an Earth-mass planet. In one of the simulations the Earth-mass planet is leading, and in the other it is trailing.
% As the last block of table~\ref{table:typeI} shows, both of these models diverge regardless of the sign of $u$ parameter. The reason might be that the linear calculation cannot be applied on this case. However, having a closer look at the torque variation at these two points helps us understand the results.

In Fig.~\ref{fig:hydrotypeIItorque} we present the disc surface density perturbation and the torque analysis for these models. The torque on the Earth-mass planet in these models only comes from the Jupiter-mass planet, either from the material accumulated in its Hill radius or spirals. As the upper panel of Fig.~\ref{fig:hydrotypeIItorque} shows, the spirals of the Earth-mass planet are so weak because there is little material in the gap in the disc to form them. The lower panel shows that the sign of the torque on the Earth-mass planet depends on its location compared to the Jupiter-mass planet. Here we explain the torque analysis for the planet on the right side (indicated by a cross sign) and the opposite argument is applied for the planet on the left (marked by a plus sign). Following the solid blue line in the lower panel, we see the (negative) torque slowly increases until the red line, where the gap edge is located. This indicates that the main torque (see the dashed blue line) does not come from the material around the Jupiter-mass planet. As we add the contribution of the material from the red to the yellow line, a large negative torque is exerted on the planet by the inner spiral which is located to the left side of the planet. As we get further, the continuation of the inner spiral adds a positive torque but since it is farther than the section on the left, it cannot change the torque considerably. The grey line marked where we reached the inner edge of the disc, and therefore, the oscillations after the grey line only originate from the outer disc.

Based on the calculations in Sec.~\ref{sec:stabu}, the partial derivatives $\Gamma_{j\Delta}=\partial \Gamma_j/\partial \Delta$ and $R_{j\zeta}=\partial R_j/\partial (\zeta-\zeta_{eq})$ are key parameters the for stability of the co-orbitals, see columns 4, 5, 7 \& 8 of Table \ref{table:typeI}. Figure \ref{fig:hydrotorqueJupiter} represents the evolution over time of the torque and radial forces for the leading-Earth-mass planet case `$\times$'. On the left panel, we can see that $\Gamma_{1\Delta}>0$, which leads to a positive term in the expression of $u$ hence destabilising the configuration. On the other hand, the left panel shows that $R_{1\zeta}<0$, which induces a stabilising term. These two effects oppose one another and determine the attractiveness of the equilibrium. We note that in this particular case, our estimation of $u$ did not match the evolution of the system: we computed a negative $u$, while the system is diverging from the equilibria. As the sum of the two dominant terms is an order of magnitude lower than each of these terms, this might be due to higher-order effects in the expansion of the torques and forces that were neglected when we computed the expression of $u$ (in our runs, the semi-amplitude of libration is of $\sim 0.17$ radians which is at the limit of the validity of the linear model).

%Nonetheless, the critical parameters for the stability of the configuration clearly appear in table \ref{table:typeI}: $\Gamma_{j\Delta}$ and $R_{j\zeta}$ for the low mass planet.
 The partial derivatives $\Gamma_{j\Delta}$ and $R_{j\zeta}$ applied on the Earth-mass planet come from the spiral arm of the Jupiter-mass planet, see Fig. \ref{fig:hydrotypeIItorque}. As these effects are of opposite sign for leading and trailing planets (bottom panel of Fig. \ref{fig:hydrotypeIItorque}), they lead to a qualitatively similar behaviour for the stability around the $L_4$ and $L_5$ equilibria of the giant planet. To confirm this trend, we ran a set of simulations with a 10$M_\oplus$ planet either leading or trailing a Jupiter-mass planet for different disc parameters, varying the $\alpha$ parameter of the viscosity and the aspect ratio. Viscosity affects the gap depth and width, and the aspect ratio widens or tightens the spirals. For all the tested disc parameters, either both leading and trailing 10$M_\oplus$ planets converged toward the equilibria, or they both diverged away from it.

In Fig. \ref{fig:hydrodiscpara} we show the results for these 7 different disc profiles that we tested.
%, changing the alpha-parameter of the viscosity and the aspect ratio.  
Seemingly, the stability of the planets inside the gap of a massive planet is a delicate trade-off between the disc parameters, although based on this small set, lower viscosity and smaller aspect ratios (that result in deeper and wider gaps) seem to stabilise co-orbital configurations. 
%In addition, these parameters affect the strength and opening angle of the spirals that will change the torques and radial forces that apply on the low-mass planet.
 As this dependency is key to estimating the probability of the existence of Earth to super-Earth mass trojan companions to giant planets, we will investigate this topic more thoroughly in a future study.

\section{Stability in the direction of the eccentricity and the inclinations}
\label{sec:stabei}
In this section we study the effect of dissipation on the evolution of the eccentricities and inclinations of the co-orbitals, for low values of $e_j$ and $I_j$ ($\lesssim 0.1$). At first order, the Poincaré variables (Eq. \ref{eq:poincvar}) read:
\begin{equation}
\begin{aligned}
x_j =\frac{\sqrt{\Lambda_j}}{\sqrt{2}}e_j\operatorname{e}^{i\varpi_j} \ \, ,\text{and}\ \ y_j =\frac{\sqrt{\Lambda_j}}{\sqrt{2}}I_j\operatorname{e}^{i\Omega_j} \, .
\end{aligned}
\label{eq:xydef}
\end{equation}
In the absence of dissipation, these variables follow the equations of variation given by the system (\ref{eq:xy}).

We assume that the evolution of the orbital elements induced by dissipative forces can be modelled by migration and damping time-scales:
\begin{equation}
\begin{aligned}
\dot a_j =-a_j/\tau_{a_j}\, ,\ \dot e_j =-e_j/\tau_{e_j}\, ,\text{and}\ \dot I_j & =-I_j/\tau_{I_j} \, .
\end{aligned}
\label{eq:dampmig}
\end{equation}
We note that modelling the migration by such a law is equivalent to taking $K=0$ in Sect. \ref{sec:ana_tana}. It can be shown that the results of this section remain valid for any value of $K$, as the local variations of the torques over the resonant time-scale have a negligible effect on the evolution of the variable $x_j$ and $y_j$. %We also assume that the mass of the planets can vary slowly. 
We hence consider the following non-conservative terms:
\begin{equation}
\begin{aligned}
\dot x_{j,d} & =x_j \left( -\frac{1}{4\tau_{aj}} -\frac{1}{\tau_{ej}} \right)\, ,\\
\dot y_{j,d} & =y_j \left( -\frac{1}{4\tau_{aj}} - \frac{1}{\tau_{I,j}} \right)\, ,\\
\end{aligned}
 \label{eq:Lambdajevol}
\end{equation}
%\frac{\dot m_j}{2m_j}
for the evolution of the $x_j$ and $y_j$. The equation of variations hence read:
\begin{equation}
\dot{\bm{x}}=
M_x
\bm{x}
   +  \dot{\bm{x}}_d,   \ 
\dot{\bm{y}}=
M_y
\bm{y}
   +  \dot{\bm{y}}_d \, ,
   \label{eq:ty1d}
   \end{equation}
where the $M_x$, $M_y$ can be found in appendix \ref{ap:ham},  $\dot{\bm{x}}_d=(\dot x_{1,d} ,\dot x_{2,d} )$ and $\dot{\bm{y}}_d=(\dot y_{1,d} ,\dot y_{2,d} )$ (Eq. \ref{eq:Lambdajevol}). As we will be primarly interested in the evolution of the orbital elements $e_j$ and $I_j$, we normalise the variables $x_j$: $X_j=e_j \operatorname{e}^{i\varpi_j} = x_j/\sqrt{\Lambda_j/2}$ and $Y_j=I_j \operatorname{e}^{i\Omega_j} = y_j/\sqrt{\Lambda_j/2}$. The evolution of these new variables reads:
\begin{equation}
\begin{aligned}
\dot{\bm{X}} & =M_X(\zeta)\bm{X}  +  \dot{\bm{X}}_d\, , \\
\dot{\bm{Y}} & =M_Y(\zeta)\bm{X}  +  \dot{\bm{Y}}_d\, ,
% \dot{X}_j & = \frac{\sqrt{2}}{\sqrt{\Lambda_j}} \dot x_j(X_1,X_2,\zeta)  - \frac{\dot \Lambda}{2\Lambda} X_j  \, ,\\
% \dot{Y}_j & = \frac{\sqrt{2}}{\sqrt{\Lambda_j}} \dot y_j(Y_1,Y_2,\zeta)  - \frac{\dot \Lambda}{2\Lambda} Y_j  \, ,\\
\end{aligned}
\label{eq:eqmotsfxy}
\end{equation}
%
%where the $M_X(\zeta)$ and $M_Y(\zeta)$ are given in appendix \ref{ap:ham}, 
and the dissipative terms read:
\begin{equation}
\begin{aligned}
 \dot{X}_{j,d} &= \left(  \frac{\dot \Lambda_j}{2\Lambda_j} -\frac{1}{4\tau_{aj}} -\frac{1}{\tau_{ej}}  \right) X_j \, , \\
 \dot{Y}_{j,d} & = \left(  \frac{\dot \Lambda_j}{2\Lambda_j} -\frac{1}{4\tau_{aj}} -\frac{1}{\tau_{Ij}}  \right) Y_j\, .
% \dot{X}_j & = \frac{\sqrt{2}}{\sqrt{\Lambda_j}} \dot x_j(X_1,X_2,\zeta)  - \frac{\dot \Lambda}{2\Lambda} X_j  \, ,\\
% \dot{Y}_j & = \frac{\sqrt{2}}{\sqrt{\Lambda_j}} \dot y_j(Y_1,Y_2,\zeta)  - \frac{\dot \Lambda}{2\Lambda} Y_j  \, ,\\
\end{aligned}
\label{eq:xydis}
\end{equation}
%

%
%by the total angular momentum $\Gamma$:
%\begin{equation}
%\begin{aligned}
%\hat \cZ & = \Gamma  \cZ \, , \  &\hat \cZ_2  &= \Gamma  \cZ_2 \, ,\\
%x_j & = \Gamma^{1/2} X_j\, , \ &y_j  &= \Gamma^{1/2} Y_j\, . 
%\end{aligned}
%\label{eq:shrinkinframe}
%\end{equation}
%
%%
%in the plane, and 
%
%in the $X_j$ and $Y_j$ directions.
%%

\subsection{Stability in the direction of the eccentricities}

\label{sec:Xstab}
\subsubsection{Constant masses}

 \begin{figure}
\begin{center}
\includegraphics[width=0.99\linewidth]{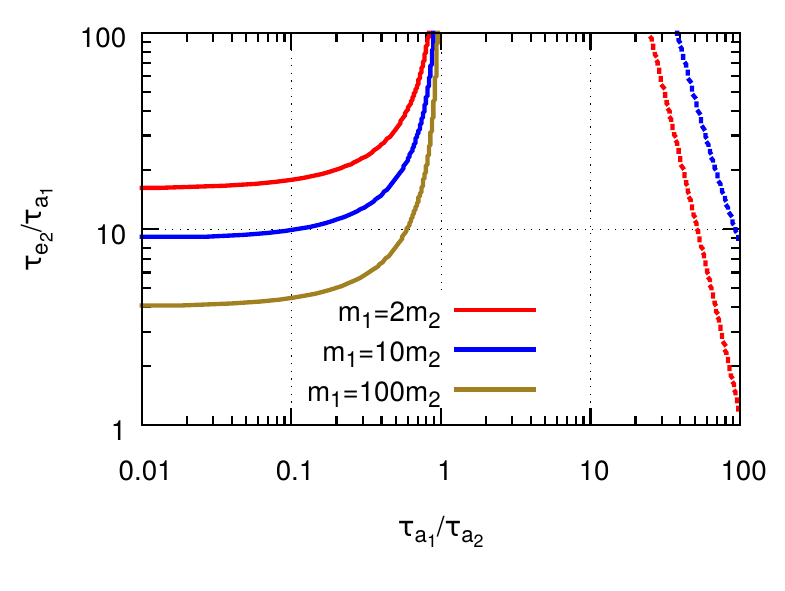}
\caption{\label{fig:tauecritg} Example of attraction criteria in the eccentric directions in the dissipative case, for different values of $m_2/m_1$. For this example, the relations $m_1=10^{-4}m_0$, $\tau_{e_1}=\tau_{e_2} m_2/m_1$, and $\tau_{a_1}=10/m_1$ were chosen. Orbits in the neighbourhood of $L_4$ will tend toward $e_1=e_2=0$ if $\tau_{e_2}/\tau_{a_1}$ is chosen below both curves of a given colour. The solid lines represent the stability limit in the anti-Lagrangian direction, while the dashed one is the limit in the eccentric Lagrangian direction, see the text for more details.}
\end{center}
\end{figure}
 \begin{figure}
\begin{center}
\includegraphics[width=0.49\linewidth]{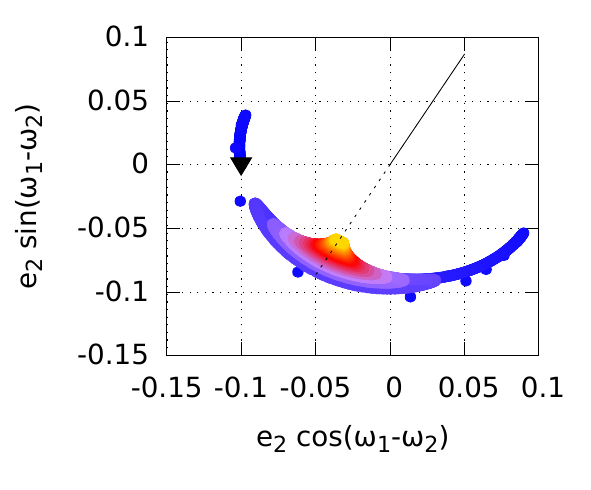}
\includegraphics[width=0.49\linewidth]{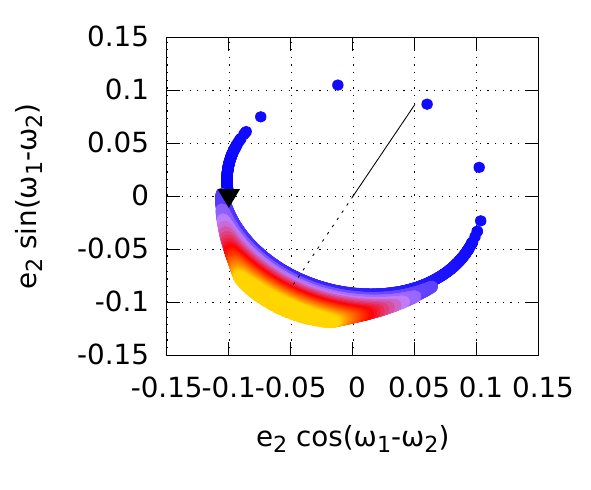}
\caption{\label{fig:Execce} Examples of the evolution of the quantities $e_2$ and $\Dv$ for different initial conditions. The black dashed lines represent the direction of the anti-Lagrangian configuration while the black solid line is the direction of the Eccentric Lagragian equilibria. In both cases, the initial conditions are $\omega_2=\omega_1+\pi$,  $I_1=I_2=0$, $m_1=10 m_2= 1\times 10^{-4} m_0$, $a_1=a2=1$ au, $e_1=0.1$, $e_2=m_1 e_1/m_2$ and $\zeta=60^\circ$. Each trajectory is integrated for $20$ Kyr, with the initial conditions represented by the black triangle, and the colour code representing the time (blue at $t=0$, yellow at $20$ Kyr). On the left panel, $\tau_{e_2}/\tau_{a_1}=5$, while on the right $\tau_{e_2}/\tau_{a_1}=20$. In both cases, $\tau_{a_1}/\tau_{a_2}=0.01$. }
\end{center}
\end{figure}
We study the stability in the direction $x_j$, related to the eccentricity and the argument of periastron, in the neighbourhood of the $L_4$ circular equilibrium for constant masses. In this section we consider dissipation time-scales that are not necessarily small with respect to $\frac{m_j}{m_0}\eta_L$. The equation of variation of the variables $x_j$ is given by 
\be 
\dot{\bm{X}}=M_X(L_4)\bm{X}  +  \dot{\bm{X}}_d\, ,
\ee
 where $M_X(L_4)$ is obtained by estimating the terms of $M_X$ at the circular $L_4$ equilibria, given by Eq. (\ref{eq:L4eqc}). 
%
% We hence obtain:
% %
% \be 
%\dot{\bm{x}}=M_x(L_4)\bm{x}  +  \dot{\bm{x}}_d\, ,
%\ee
%%
% 
% 
%\begin{equation}
%M_X(L_4)=
%   \begin{pmatrix}
%  \frac{A_X}{m_1} -X_{1,d}/X_1 &  \frac{\overline B_X}{\sqrt{m_1 m_2}} \\
%  \frac{B_X}{\sqrt{m_1 m_2}}&  \frac{A_X}{m_2}-\dot X_{2,d}/X_1
%   \end{pmatrix} 
%   \end{equation}
%   %
%   where $A_x$ and $B_x$ can be found in Appendix (\ref{ap:ham}). At first order in $\eps$, $M_X(L_4)$ can be diagonalised, with the diagonal elements being:
%   %
%
The system of equations (\ref{eq:eqmotsfxy}) have two eigenvalues. At first order in $\eps$:
 \begin{equation}
 \begin{aligned}
%g^{(h)}_0 & =0\, , \\
%g^{(h)} & = -i \eps 3\frac{-18+29\sqrt{3}z_{L_4}}{16} \frac{m_1+m_2}{m_0} \eta_\Gamma\\
%      &  -\eps \left( \frac{(m_1-m_2)(\tau_{a_1}-\tau_{a_2})}{4(m_1+m_2)\tau_{a_1}\tau_{a_2}}+ \frac{\tau_{e_1}+\tau_{e_2}}{\tau_{e_1}\tau_{e_2}} \right)
g_{X\pm} & = i \eps \frac{g_{L_4} }{2}  - \frac{\eps}{8}\left( \frac{m_1-m_2}{m_1+m_2}\frac{1}{\tau_{a-}}+\frac{4}{\tau_{e+}} \right)\\
  & \pm \eps \sqrt{ \left(\frac{1}{8\tau_{X-}}\right)^2-  i  \frac{1}{2} \frac{m_1-m_2}{m_1+m_2} \frac{g_{L_4}}{\tau_{X-} }+ \left(i  \frac{g_{L_4} }{2} \right)^2}\, ,
    \end{aligned}
   \label{eq:vape}
   \end{equation}
   where 
    \begin{equation}
    1/\tau_{a\pm}= \frac{\tau_{a_1}\pm\tau_{a_2}}{\tau_{a_1}\tau_{a_2}} \ \ \,   1/\tau_{e\pm}= \frac{\tau_{e_1}\pm\tau_{e_2}}{\tau_{e_1}\tau_{e_2}}\, ,
 %  \tau_{X-}=1/ \left( \frac{\tau_{a,1}-\tau_{a,2}}{\tau_{a,1}\tau_{a,2}}+4\frac{\tau_{e,1}- \tau_{e,2}}{\tau_{e,1}\tau_{e,2}} \right)\, ,
   \label{eq:eigene}
   \end{equation}
   $1/\tau_{X-}=1/\tau_{a-}+4/\tau_{e-}$, and
     \begin{equation}
   g_{L_4}=\eps (-\frac{27}{8} + \frac{87}{16}  \sqrt{3} z_{L_4}) \frac{m_1 + m_2}{m_0}  \eta_\Gamma\, .
   \label{eq:vapeL4}
   \end{equation}
In the conservative case, $1/\tau_{e_\pm}=1/\tau_{a_\pm}=z_{L_4}=0$, and we obtain $g_-=0$ and $g_+=i 27/8(m_1+m_2)/m_0 \eta$. The direction associated to these eigenvalues were described in Sect. \ref{sec:conse}: the eccentric Lagrangian equilibrium, where $e_1=e_2$ and $\varpi_1-\varpi_2=\zeta=\pm \pi/3$; and the anti-Lagrangian equilibrium, where $m_1e_1=m_2e_2$ and $\varpi_1-\varpi_2=\zeta+\pi=\mp 2\pi/3$.

Figure \ref{fig:tauecritg} shows the values of $\tau_{e_2}/\tau_{a_1}$ for which the real component of the eigenvalues (\ref{eq:vape}) vanishes, with respect to $\tau_{a_1}/\tau_{a_2}$. These plots were made using $m_1=10^{-4}m_0$, $\tau_{e_1}=\tau_{e_2} m_2/m_1$, and $\tau_{a_1}=10/m_1$. For a given mass ratio, the manifold $e_1=e_2=0$ is attractive below the two curves of the given colour. Above the solid line, the system diverges following the anti-Lagrangian direction; while systems above the dashed curve diverge following the eccentric Lagrangian direction. These curves were obtained using several assumptions on the relations between the masses and the damping and migration time-scales, and that the stability of the $X_j$ directions in the $\tau_{a_1}/\tau_{a_2}$, $\tau_{e_2}/\tau_{a_1}$ plane depends greatly on these assumptions.

Figure \ref{fig:Execce} represents the evolution of two configurations taken on the left border of Fig. \ref{fig:tauecritg}: $\tau_{a_1}/\tau_{a_2}=0.01$. In the left panel, $\tau_{e_2}/\tau_{a_1}=5$, while on the right $\tau_{e_2}/\tau_{a_1}=20$. The stability in the $X_j$ directions is given by the position of the configurations relative to the blue curves of Fig. \ref{fig:tauecritg}. In both cases, the motion relative to the direction of the eccentric Lagrangian equilibria (black solid lines in Fig. \ref{fig:Execce}) is quickly damped as we are far below the dashed lines in both cases. As the quantity $\varpi_1-\varpi_2$ converges toward $240^\circ$($=\zeta+180^\circ$), which is the direction of the anti-Lagrangian equilibria, the eccentricity either decreases as this direction is stable (the left case is below the solid blue curve of Fig. \ref{fig:tauecritg}), or increases if the anti-Lagrangian direction is unstable (the right case is above the solid blue curve). 

%We can also describe extreme cases for the stability in the $X_j$ directions: without loss of generality, we assume that $m_1\geq m_2$ (if not, we swap the indexes and study the neighbourhood of $L_5$ instead). For $m_1 \gg m_2$, equations (\ref{eq:vape}) become:
%   %
% \begin{equation}
% \begin{aligned}
%g_- & = - \frac{1}{\tau_{e_1}}\, , \\
%g_+ & =i g_{L_4} -   \frac{\tau_{a,1}-\tau_{a,2}}{2\tau_{a,1}\tau_{a,2}} -  \frac{1}{\tau_{e_2}}\, .
%    \end{aligned}
%   \label{eq:vapem1ggm2}
%   \end{equation}
%   %
%In this case, the circular Lagrangian $L_4$ is always attractive in the direction of the eccentric Lagrangian equilibrium for $\tau_{e_1}>0$, and the direction of the anti-Lagrangian equilibrium is repulsive only when:
%
% \begin{equation}
% \begin{aligned}
% \tau_{a,2} & >\tau_{a,1} \, , \hspace{.5cm} \text{and} \\
% \tau_{e,2} & >2\frac{\tau_{a,1}\tau_{a,2}}{\tau_{a_2}-\tau_{a,1}} \, .
%    \end{aligned}
%   \label{eq:AL4rep}
%   \end{equation}
%   %
%
%We can also study the special case $m_1=m_2$. In this case, the sign of the real part of the eigenvalues depends on the value of the square root in equations (\ref{eq:vape}), which is either pure imaginary ($1/(4 \tau_-)<g_{L_4}$), or pure real ($1/(4 \tau_-)> g_{L_4}$). In the former case, both the eccentric Lagrangian equilibrium direction and the anti-Lagrangian one are attractive, while in the latter, different regimes are possible.% Note that these limit cases do not necessarily shows on figure (\ref{fig:tauecrit}), as assumption were made for the figure, for example $\tau_{e_1}=\tau_{e_2} m_2/m_1$.

\subsubsection{Effect of mass change}

We now consider the effect of a slow, isotropic mass change on the eccentric Lagrangian equilibria and anti Lagrangian equilibria previously discussed. For this mass change to impact the evolution of the configuration, it has to be comparable to the migration and damping time-scales. We hence assume that the perturbative terms $1/\tau_{aj}$, $1/\tau_{ej}$ and $\dot{m_j}/m_j$ are of size $\eps^2$. The details of the computations can be found in appendix \ref{ap:emdot}.

\noindent \textit{- Evolution of the eccentricities along the anti-Lagrangian equilibria:} taking ${\cal X}_2=0$, Eq. (\ref{eq:edot}) yields, at second order in $\eps$:
\be
\begin{aligned}
\dot e_j=\eps^2 \frac{m_k {\cal X}_1 \bar{\cal X}_1}{4e_j(m_1+m_2)}\left( 4 m_j\dot m_k  - 4m_k\dot m_j + \frac{m_k}{T_{AL4}} \right)\\
%\dot e_2=\eps^2 \frac{m_1 {\cal X}_1 \bar{\cal X}_1}{4e_2m_2(m_1+m_2)}\left( 4 m_2^2\dot m_1  - 4m_2m_1\dot m_2 + \frac{1}{T_{AL4}} \right)
\end{aligned}
\label{eq:edotAL4}
\ee
where ${\cal X}_1 \bar{\cal X}_1$ is a positive real quantity and
\be
  \frac{1}{T_{AL4}}=m_2 \left(-\frac{1}{\tau_{a1}} +\frac{1}{\tau_{a2}}-\frac{4}{\tau_{e1}}\right) +m_1 \left(-\frac{1}{\tau_{a2}} +\frac{1}{\tau_{a1}}-\frac{4}{\tau_{e2}}\right) 
\label{eq:TAL4}
\ee

\noindent \textit{- Evolution of the eccentricities along the Eccentric-Lagrangian equilibria:} taking ${\cal X}_1=0$, Eq. (\ref{eq:edot}) yields, at second order in $\eps$:
\be
\begin{aligned}
\dot e_j=-\eps^2 \frac{ {\cal X}_2 \bar{\cal X}_2}{e_j(m_1+m_2)}\left( \frac{m_1}{\tau_{e1}}+\frac{m_2}{\tau_{e2}}  \right)\\
%\dot e_2=\eps^2 \frac{m_1 {\cal X}_1 \bar{\cal X}_1}{4e_2m_2(m_1+m_2)}\left( 4 m_2^2\dot m_1  - 4m_2m_1\dot m_2 + \frac{1}{T_{AL4}} \right)
\end{aligned}
\label{eq:edotEL4}
\ee
where ${\cal X}_2 \bar{\cal X}_2$ is a positive real quantity.\\

If the eccentricities are damped by the disc ($\tau_{ej}>0$) then the mode associated to the Eccentric Lagrangian equilibrium will always be damped toward $e_1=e_2=0$. However the eccentricities can increase along the Anti-Lagrangian equilibria if the more massive of the two planets migrate inward (or accrete gas) fast enough (see Eq. \ref{eq:edotAL4}).

\subsection{Stability in the direction of the inclinations}
\label{sec:Ystab}

\subsubsection{Constant masses}

% \begin{figure}
%\begin{center}
%\includegraphics[width=0.99\linewidth]{evolI_coorb.pdf}
%\caption{\label{fig:tauIcrit} Attraction in the eccentric directions in the dissipative case, for different values of $m_2/m_1$. Orbit in the neighbourhood of $L_4$ will tend toward $e_1=e_2=0$ if for $\tau_{e_2}/\tau_{a_1}$ is chosen bellow both curves of a given colour. The solid lines represent the stability limit in the anti-Lagrangian direction, while the dashed one is the limit in the eccentric Lagrangian direction, see the text for more details.}
%\end{center}
%\end{figure}
%%
%
 \begin{figure}
\begin{center}
\includegraphics[width=0.49\linewidth]{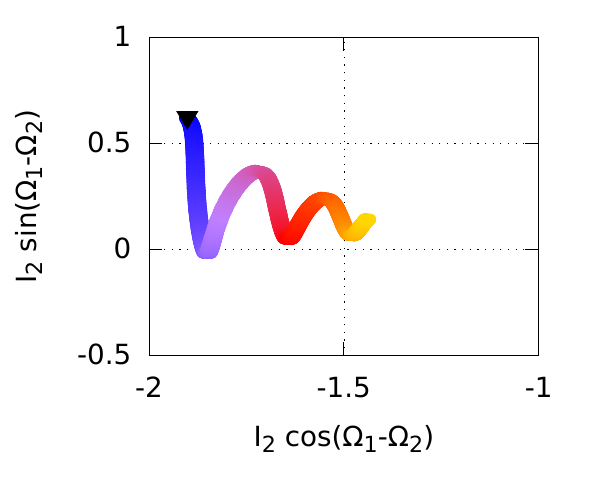}\includegraphics[width=0.49\linewidth]{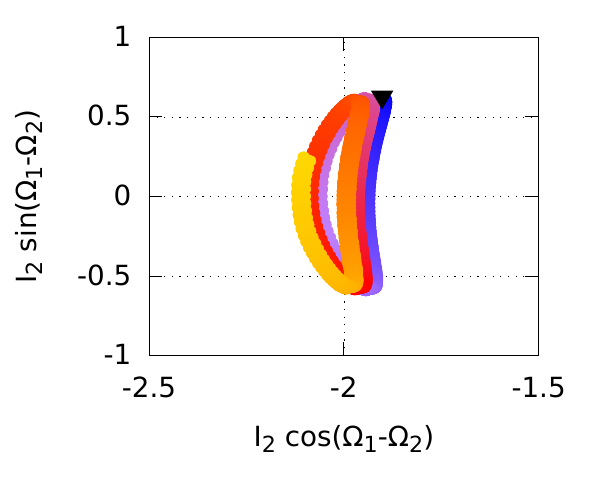}\\
\caption{\label{fig:ExeccI} Examples for the evolution of $I_2$ and $\Omega_1-\Omega_2$. In both cases, the initial conditions are $\Omega_2=\Omega_1+\pi+.3$,  $I_2=2^\circ$, $m_1=10 m_2= 1\times 10^{-4} m_0$, $a_1=a2=1$ au, $e_1=e_2=0$ and $\zeta=60^\circ$. Each trajectory is integrated for $20$ Kyr, with the initial conditions represented by the black triangle, and the colour code representing the time (blue at $t=0$). In the left panel, $\tau_{I_2}/\tau_{a_1}=1$, while on the right $\tau_{I_2}/\tau_{a_1}=30$. In both cases, $\tau_{a_1}/\tau_{a_2}=0.01$. }
\end{center}
\end{figure}
We study the stability in the direction $Y_j$, related to the inclinations and the ascending nodes of the co-orbitals, for any amplitude of libration of the resonant angle. Eq. (\ref{eq:ty1}) becomes:
% \begin{equation}
%M_Y=
%   \begin{pmatrix}
%  \frac{A_Y}{m_1} -\dot Y_{1,d}/Y_1 &  \frac{\overline B_Y}{\sqrt{m_1 m_2}} \\
%  \frac{B_Y}{\sqrt{m_1 m_2}}&  \frac{A_Y}{m_2}-\dot Y_{2,d}/Y_2
%   \end{pmatrix} 
%   \label{eq:tyd}
%   \end{equation}
%%
%Where $A_Y$ and $B_Y$ depend on the semi-fast angle $\zeta$. For a given trajectory, since the evolution of $\zeta$ is fast with respect to the secular evolution on the $Y_j$, we can obtain an approximation of the secular dynamics in the vertical direction by averaging the expression of this system over a period $2\pi/\nu_\Gamma$ with respect to the time $t$. Moreover, we note that $Im(B_Y)=-A_Y$, and the real part of $B_Y$ is proportional to the expression of $\dot Z$ in the conservative case: $\propto (1-1/\delta^3)^{1/2} \sin \zeta$ which is the derivative of a periodic function of period $2\pi/\nu_\Gamma$. It's average value over $2\pi/\nu_\Gamma$ is hence null. The system (\ref{eq:tyd}) hence becomes:
%
 \begin{equation}
M_Y=
   \begin{pmatrix}
  \frac{\overline A_Y}{m_1} -Y_{1,d}/Y_1 &  \frac{-\overline A_Y}{\sqrt{m_1 m_2}} \\
  \frac{-\overline A_Y}{\sqrt{m_1 m_2}}&  \frac{\overline A_Y}{m_2}-\dot Y_{2,d}/Y_2
   \end{pmatrix} 
   \label{eq:tyd2}
   \end{equation}
At first order in $\eps$, $M_Y(L_4)$ can be diagonalised, with the diagonal elements being:
 \begin{equation}
 \begin{aligned}
s_{Y\pm} & = i\frac{g_{Y} }{2} - \frac{1}{8}\left( \frac{m_1-m_2}{m_1+m_2}\frac{1}{\tau_{a-}}+\frac{4}{\tau_{I+}} \right)\\
  & \pm \sqrt{\left(\frac{1}{8\tau_{Y-}}\right)^2-i \frac{1}{2} \frac{m_1-m_2}{m_1+m_2} \frac{g_{L_4}}{\tau_{Y-} }+ \left(i \frac{g_{L_4} }{2} \right)^2}\, ,
    \end{aligned}
   \label{eq:eigeni}
   \end{equation}
where 
    \begin{equation}
    1/\tau_{a\pm}= \frac{\tau_{a_1}\pm\tau_{a_2}}{\tau_{a_1}\tau_{a_2}} \ \ \   1/\tau_{I\pm}= \frac{\tau_{I_1}\pm\tau_{I_2}}{\tau_{I_1}\tau_{I_2}}\, ,
 %  \tau_{X-}=1/ \left( \frac{\tau_{a,1}-\tau_{a,2}}{\tau_{a,1}\tau_{a,2}}+4\frac{\tau_{e,1}- \tau_{e,2}}{\tau_{e,1}\tau_{e,2}} \right)\, ,
   \label{eq:vapi}
   \end{equation}
   $1/\tau_{X-}=1/\tau_{a-}+4/\tau_{I-}$, and
     \begin{equation}
   g_{Y}= \frac{\bar A_Y}{2} \frac{m_1+m_2}{ m_0}  \eta_\Gamma \, .
   \label{eq:vapI}
   \end{equation}

   We note that the eigenvalues (\ref{eq:eigeni}) have a similar expression to that in the direction of the eccentricity, Eq. (\ref{eq:vape}), but here the results are valid for any amplitude of libration in the trojan and horseshoe domains. The amplitude of libration affects the stability in the $Y_j$ direction through the value of $\bar A_Y$ (see Fig. \ref{fig:Ab_vs_zet0}).
   %, to be compared with the $(-\frac{27}{8} + \frac{87}{16}\sqrt{3} z_{L_4})$ of the eccentric case. 
   %Figure \ref{fig:tauIcrit} shows the values of $\tau_{I_2}/\tau_{a_1}$ for which the real part of the eigenvalues (\ref{eq:eigeni}) vanishes, with respect to $\tau_{a_1}/\tau_{a_2}$, for $\overline A_Y/2=0.05$ (near the $L_4$ equilibria) and $1.5$ (in the horseshoe domains). 
%   Similarly to the eccentric case, these plots were made using $m_1=10^{-4}m_0$, $\tau_{I_1}=\tau_{I_2}*m_2/m_1$, and $\tau_{a_1}=10/m_10$. For a given mass ratio, the manifold $I_1=I_2=0$ is attractive bellow the two curves of the given color. Above the solid line, the system diverge following the direction $m_1 I_1=m_2 I_2$ and $\Omega_2=\Omega_1+\pi$; while systems above the dashed curve diverge following the direction $I_2=I_1$ and $\Omega_2=\Omega_1$, see section (\ref{sec:consi}).
   The orbits can either be attracted toward $I_1=I_2=0$, or diverge following $m_1 I_1=m_2 I_2$ and $\Omega_2=\Omega_1+\pi$, or $I_2=I_1$ and $\Omega_2=\Omega_1$, see Sect. \ref{sec:consi}. Examples of convergence and divergence along $m_1 I_1=m_2 I_2$, $\Omega_2=\Omega_1+\pi$ are shown in Fig. \ref{fig:ExeccI}.
 
 \subsubsection{Effect of mass change}

We now consider the effect of a slow, isotropic mass change on the inclination of quasi-circular co-orbitals (up to first order in eccentricities). As we did for the study of the evolution of the eccentricities, we assume that the perturbative terms $1/\tau_{aj}$, $1/\tau_{Ij}$ and $\dot{m_j}/m_j$ are of size $\eps^2$, and that the mass evolution is isotropic. The details of the computations are identical to the eccentric case and can be found in appendix \ref{ap:emdot}.

\noindent \textit{- Evolution of the inclinations along $m_1 I_1=m_2 I_2$, $\Omega_2=\Omega_1+\pi$:} taking ${\cal Y}_1=0$, we obtain, at second order in $\eps$:
\be
\begin{aligned}
\dot I_j=\eps^2 \frac{m_k {\cal Y}_1 \bar{\cal Y}_1}{4I_j(m_1+m_2)}\left( 4 m_j\dot m_k  - 4m_k\dot m_j + \frac{m_k}{T_{I}} \right)\\
%\dot e_2=\eps^2 \frac{m_1 {\cal X}_1 \bar{\cal X}_1}{4e_2m_2(m_1+m_2)}\left( 4 m_2^2\dot m_1  - 4m_2m_1\dot m_2 + \frac{1}{T_{AL4}} \right)
\end{aligned}
\label{eq:idotAL4}
\ee
where ${\cal Y}_1 \bar{\cal Y}_1$ is a positive real quantity and
\be
  \frac{1}{T_{I}}=m_2 \left(-\frac{1}{\tau_{a1}} +\frac{1}{\tau_{a2}}-\frac{4}{\tau_{I1}}\right) +m_1 \left(-\frac{1}{\tau_{a2}} +\frac{1}{\tau_{a1}}-\frac{4}{\tau_{I2}}\right) 
\label{eq:TI}
\ee

\noindent \textit{- Evolution of the inclination along $I_1=I_2$, $\Omega_1=\Omega_2$:} taking ${\cal Y}_2=0$, we obtain, at second order in $\eps$:
\be
\begin{aligned}
\dot I_j=-\eps^2 \frac{ {\cal Y}_2 \bar{\cal Y}_2}{I_j(m_1+m_2)}\left( \frac{m_1}{\tau_{I1}}+\frac{m_2}{\tau_{I2}}  \right)\\
%\dot e_2=\eps^2 \frac{m_1 {\cal X}_1 \bar{\cal X}_1}{4e_2m_2(m_1+m_2)}\left( 4 m_2^2\dot m_1  - 4m_2m_1\dot m_2 + \frac{1}{T_{AL4}} \right)
\end{aligned}
\label{eq:idotEL4}
\ee
where ${\cal Y}_2 \bar{\cal Y}_2$ is a positive real quantity.\\

If the inclination are damped by the disc ($\tau_{Ij}>0$) then the mode associated to $I_1=I_2$, $\Omega_1=\Omega_2$ will always be damped toward $I_1=I_2=0$. However the inclinations can increase along $m_1 I_1=m_2 I_2$, $\Omega_2=\Omega_1+\pi$ if the more massive of the two planets migrate inward (or accrete gas) fast enough (see Eq. \ref{eq:idotAL4}). We remind the reader that these results on the inclinations are valid for any amplitude of libration of the resonant angle, up to horseshoe orbits.

\section{Summary and conclusions}
\label{sec:conclusions}
\subsection{Summary}

In this paper we have studied the stability of the co-orbital resonance under dissipation in the planetary case ($(m_1,m_2) \ll m_0$). In Sect. \ref{sec:H3bp} we developed an integrable model of the 1:1 MMR perturbed by a generic dissipation and derived the stability conditions of the $L_4$ and $L_5$ equilibria. 
%These results were then compared to N-body simulations with 1-D protoplanetary disc models in Sect. \ref{sec:type1}, then to hydrodynamic simulations in Sect. \ref{sec:type2}.\\

In Sect. \ref{sec:constantF} we showed that under the effect of a constant torque applied on each planet, the phase space of the resonance becomes asymmetric, as the position of the Lagrangian equilibria $L_3$, $L_4$ and $L_5$ change. The tadpole (trojan) area is larger if the torque per mass unit applied on the leading planet is greater than the torque per mass unit applied on the trailing one. We also saw that if the difference between these two torques is too large, two out of the three equilibrium points could merge and vanish, leading to a phase space with a single equilibrium point. These results are in agreement with those of \cite{SiDu2003}, obtained in the restricted case ($m_1\ll m_0,\, m_2=0$). This effect can also contribute to the instability observed by \cite{PiRa2014}, where they showed that similar mass co-orbitals were unstable during the partial gap-opening regime, due to the opposite torques induced by a higher gas depletion between the two planet than everywhere else in the gap.

In Sect. \ref{sec:stabW} and \ref{sec:stabu}, we then studied the stability of the Lagrangian equilibria $L_4$ and $L_5$ as a function of the forces applied on each planet, their masses $m_j$, and the evolution of their mass $\dot m_j$. This study can be split into two parts: \\
-First, the evolution of the masses of the co-orbitals, along with the constant torques that are applied on them, change the width of the co-orbital resonances. It can lead to either a convergence toward the Lagrangian equilibria in the case of outward migration (positive total torque, $\Gamma_{10}+\Gamma_{20}>0$) or overall mass increase ($\dot{m}_1+\dot m_2>0$), while inward migration (negative total torque) and mass loss induces a slow divergence from the Lagrangian equilibria. These results are in agreement with those of \cite{FleHa2000}, which were obtained in the restricted case ($m_1\ll m_0,\, m_2=0$). \\
-Second, if the forces applied on each planet vary over the resonant time-scale, we show that the dependency of the torques on the semi-major axis, and the dependency of the radial component of the perturbative forces on the value of the resonant angle, impact significantly the stability of the system \citep[the effect of a radial dependency of the torque was discussed by][in the restricted case]{SiDu2003}. \\
These two effects were considered to derive the stability criterion $u$ for the Lagrangian equilibria (Eq. \ref{eq:eigencircp}).\\

Sect. \ref{sec:type1} and \ref{sec:type2} were dedicated to comparing these results to N-body simulations in 1D disc models, and hydrodynamic simulations. In Sect. \ref{sec:type1}, we applied type-I migration prescriptions on a pair of planets in an evolving protoplanetary disc. The stability criterion successfully predicts the stability of the system, as a function of their masses, their migration time-scale $\tau_{aj}$, and the slope of that migration parametrised by $K_j$. In addition, running planetary system evolution through the disc lifetime allowed us to study the balance between the destabilising effect of inward migration and the stabilising effect of mass accretion: First, planets tend to grow in mass significantly faster than they migrate, which leads to a convergence toward the exact equilibrium. However, in the later stages of the disc lifetime, the planets migrate quickly, leading to a divergence from the equilibrium. In addition, we showed that co-orbitals that belong to a resonant chain with other planets can be stabilised during the migration phase.

However, the comparison to hydrodynamics simulations show the limits of the 1D models: despite having similar initial conditions for the disc, the forces that are applied on each planet in the hydrodynamical simulation are totally different from those given by type-I prescriptions. Indeed, as the two planets evolve around the same semi-major axis, the disc is significantly perturbed both radially and azimuthally \citep[Fig. \ref{fig:hydrotypeI53}, see also][]{Broz2018}. It creates structures whose effects cannot be azimuthally averaged, as they follow the position of the planets. Notably, as both planets librate around the Lagrangian equilibria, they move relatively to one another's spiral arms. The additional torques and radial forces applied on each planet hence evolve over the libration time-scale, that can either have a stabilising effect, or destabilising one, see Table \ref{table:typeI}. It is the sum of all these terms that dictates the evolution of the system.   

In the super-Earth range (3-5$M_\oplus$) we note a trend that was observed by \cite{PiRa2014}: more massive leading planets tend to stabilise the system. We show here that this stabilisation is due to the variations of the torques felt by each planet over the resonant libration, as they are successively closer to, and then further away from, one another's spiral arm. This trend is also present in the mini-Neptune regime (up to 15$M_\oplus$) with the apparition of other structures, such as a partial gap that is deeper between the co-orbitals.

%, makes this kind of analysis more complex and will be developed in a subsequent study. 
Finally, in the case where the gap is totally open, we ran a set of simulations with a Jupiter-mass planet trailed or preceded by Earth or super-Earth mass planets. Here the dominant effect for the stability was the variation of the radial forces and torques applied on the Earth-mass planet by the Jupiter-mass planet's spiral arms, during each libration period. The symmetry of the spiral arms with respect to the Jupiter-mass planet led to a similar behaviour for leading and trailing smaller mass companions: for all tested disc profiles, both leading and trailing companions behaved in a similar way (both diverging from or both converging toward $L_4/L_5$). However, as shown in Fig. \ref{fig:hydrodiscpara}, different disc parameters change the stability of such configurations. The effect of the disc parameters on the shape and strength of the Jupiter-mass planet's spiral arms will be the subject of a future study.\\

In Sect. \ref{sec:stabei}, we studied the stability of the Lagrangian equilibrium in the direction of the eccentricities (at first order), and the stability of the whole tadpole and horseshoe domain in the direction of the inclinations (at first order as well). We have shown that even in the case were the dissipative forces tend to damp the eccentricities and inclination of the planets, those could increase along a particular family of orbits.

\subsection{Conclusions}

\subsubsection{On the limitation of 1D disc models}

We have shown that disc-planet coupling generates structures in the disc that cannot be azimuthally averaged, leading to variations over time of the torques and radial forces that applies on each planet. Using similar disc profiles in 1D and hydrodynamical simulations, these differences lead to opposite results on the stability of the Lagrangian equilibria. Similar observations were made by \cite{Broz2018} in a more general context.

\subsubsection{On the evolution of co-orbitals}

{\bf Trojan swarms:} the asymmetry between the $L_4$ and $L_5$ domains induced by the difference of torque per mass unit (Fig \ref{fig:L4L3L5}) can be used to explain the potential asymmetry between the leading and trailing Jupiter's and Neptune's trojan swarm
% \citep[see for example][]{LyHoJoMu2009,Pirani2019}. 
However it requires one to properly estimate the torques that are felt by each of the asteroids: we showed in section \ref{sec:type2} that the torque per mass unit applied by the protoplanetary disc on the leading and trailing Trojans are not negligible, and comes mainly from the Jupiter-mass planet's spiral arms (see Fig \ref{fig:hydrotypeIItorque}). These torques will hence strongly depend on the disc parameters, and are of opposite sign for $L_4$ and $L_5$ Trojans. As a result, $L_4$ and $L5$ domains would be more symmetric than if we apply the same torque on all asteroids.\\

{\bf Co-orbital exoplanets:} We have shown that the attractiveness of the Lagrangian equilibria depends on the mass distribution between the planet, the total mass, the accretion rate, the constant torques and radial forces that apply on each planet, but also on how these quantities evolve on the resonant time-scale. We have shown that long inward migration destabilises co-orbitals, while outward migration and mass accretion tend to stabilise them. Figure \ref{fig:mdot_torque} shows that the stabilising terms coming from mass accretion is comparable to the destabilising terms coming from inward migration, and hence both have to be taken into account to properly estimate the stability of a system. However, this stabilising effect comes into play mainly in the earlier phase of the planet's evolution. While in the later stages, its evolution is dominated by the migration. 

As in \cite{PiRa2014}, we also found that leading massive trojans tend to stabilise the configuration. In their study, these authors also showed that equal mass co-orbitals can be disrupted during the gap opening stages. We have shown that the stability of Earth-mass planets as trojan companions of a Jupiter-mass planet depend on the disc parameters, but that both $L_4$ and $L_5$ configurations tend to be stable or unstable for a given set of disc parameters. 

We have shown that unstable co-orbital configurations could be stabilised by being trapped in first order mean motion resonance with a third planet, although in this part of the study we neglected the perturbation coming from the different planet's spiral arms \citep{Broz2018}.

It is worth noting that the Lagrangian equilibria being repulsive does not necessarily imply that no co-orbital configurations can remain, it only implies that the amplitude of libration around the Lagrangian equilibria slowly increases over the migration time-scale, although that can lead to trojan orbits becoming horseshoe orbits, or even exiting the resonance. Similarly, attractive Lagrangian equilibria only implies a slow convergence toward it, but the configuration can still be disrupted on shorter time-scales for example through N-body interaction with other planets \citep{RoBo2009,Leleu2019}.

\subsubsection{On the detectability of co-orbitals exoplanets}

In our hydro-simulation runs, in the $[3,15]M_\oplus$ range and for a given disc profile, all configurations with a leading more massive planet were attracted toward the Lagrangian equilibria for planets. On the contrary, for the Jupiter-mass planet's Earth sized trojan the stability seemed to depend very little on who is leading in the orbit, but we showed that different disc parameters can change the attractiveness of the Lagrangian equilibria. In addition, \cite{CreNe2009} found that during the co-orbital's evolution in the disc, the mass discrepancy between the two planets keep increasing because the more massive planet starves-off the other.

Our study of the stability of the Lagrangian equilibria in the inclined direction also leads to important conclusions regarding the detectability of co-orbitals. We have shown that as long as the disc tend to damp inclinations, the system can evolve toward two directions: either coplanar co-orbitals, or mutually inclined co-orbitals following the $m_1I_1=m_2I_2$, $\Omega_1 = \Omega_2+\pi$ direction. This later direction is favoured if the proper migration of the more massive of the two planets, or its mass accretion rate, is faster than the inclination damping of the smaller planet. As the inclination damping of the smaller planet is reduced by the deeper partial or full gap created by the more massive planet, that could significantly reduce the transit probability of both co-orbitals. Similarly, even when the disc damps the eccentricities of the two planets, these eccentricities can increase following the anti-Lagrangian equilibria $m_1e_1=m_2e_2$ $\omega_1 - \omega_2= \zeta+\pi$ \citep{GiuBeMiFe2010,LeRoCo2018}.

Mutually inclined co-orbitals can still be detected using transit timing variations \citep[TTVs, ][]{FoHo2007,VoNe2014,Leleu2019} or radial velocities \citep{LauCha2002,LeRoCo2015}, however, these methods require that the co-orbitals librate with a significant amplitude around the Lagrangian equilibrium, and that the observations baseline is at least comparable with the libration time-scale. In addition, the planets have to be of comparable masses for the radial velocity method, as well as for TTVs if it is the larger of the two planets that is transiting. Finally, even in the absence of libration, the combination of transit and radial velocity measurements can be used to detect co-orbital configurations \citep{FoGa2006,LeRoCoLi2017}, although this requires good constraints on the eccentricity of the transiting planet.

\begin{acknowledgements}
The authors acknowledge support from the Swiss NCCR PlanetS and the Swiss National Science Foundation. S.Ataiee acknowledges the support of the DFG priority program SPP 1992 "Exploring the Diversity of Extrasolar Planets (KL 650/27-1)"
\end{acknowledgements}

\bibliographystyle{aa}
\bibliography{biblio.bib}

\appendix
%
%
%\section{relative size of the terms}
%\label{ap:relative_size}
%To keep track of the relevant terms in the following expressions, we introduce the small parameter $\eps=\gO(m_j/m_0)$, and we replace the planetary masses $m_j$ by $\eps m_j$, $L$ by $\eps L$ (the angular momentum is proportional to the planetary masses), $R_j$ by $\eps R_j$, and $\Gamma_j$ by $\eps^2 \Gamma_j$. These two last substitution implies that the accelerations introduced in Eq. (\ref{eq:poincvard}) 
%In order to identify the relevant terms in this system, we estimate their relative size by multiplying by $\eps$ every planetary masses, as well as the small acceleration introduced by the equation (\ref{eq:poincvard}). We obtain a system of the form:
%  %
%\begin{equation}
%\begin{aligned}
%\dot I_1  &= \eps M_{1\eps} I_1  +  \eps M_{2\eps} z
%\dot{\zeta} & = (M_{3}+\eps M_{3\eps}) I_1 +  \eps M_{4\eps} z\, ,\\
%\end{aligned}
%\label{eq:syserdidLe}
%\end{equation}
%%
% Assuming a small departure from these equilibria, we inject the expansions $z=\epsilon z_{L_4}$, $I_1=\epsilon I_{L_4}$ in the system (\ref{eq:syserdidL}). 
%
% 
% 
%   Solving $\dot z = \dot I_1=0$ at first order in $\epsilon$, we obtain:
%   
\section{Equations of the coorbital resonance at first order in $e$ and $I$}
\label{ap:ham}

The averaged Hamiltonian of the circular coplanar coorbital resonance is \citep{RoPo2013}: 
\begin{equation}
\begin{aligned}
 \ol \gH_0= & \mu_0 \left( \frac{m_1^3}{Z^2}+ \frac{m_2^3}{(Z_2-Z)^2} \right. \\
 				& \left.  +\cG m_1 m_2 \left[ \frac{m_1 m_2}{\Lambda_1\Lambda_2} \cos \zeta - \left(\frac{Z^4}{m_1^4} \right. \right.  \right. \\
 	&\left.	\left.	\left.	 +\frac{(Z_2-Z)^4}{m_2^4} - 2\frac{Z^2(Z_2-Z)^2}{m_1^2m_2^2}\cos\,\zeta \right)^{-1/2} \right] \right)\,.
\end{aligned}
\label{eq:Hbropo}
\end{equation}
While the equation of variations of the $\bm{x}$ and $\bm{y}$ variables are given by \citep{RoPo2013,RoNi2015}:
\begin{equation}
\dot{\bm{x}}=
M_x(\zeta)
\bm{x} \, ,   \ 
\dot{\bm{y}} =
M_y(\zeta)
\bm{y}  \, ,
   \label{eq:ty1}
   \end{equation}
with
\begin{equation}
M_x(\zeta)=
   \begin{pmatrix}
  \frac{A_x(\zeta)}{m_1} &  \frac{\overline B_x(\zeta)}{\sqrt{m_1 m_2}} \\
  \frac{B_x(\zeta)}{\sqrt{m_1 m_2}}&  \frac{A_x(\zeta)}{m_2}
   \end{pmatrix}  \, ,   \ 
M_y=
   \begin{pmatrix}
  \frac{A_y(\zeta)}{m_1}&  \frac{\overline B_y(\zeta)}{\sqrt{m_1 m_2}} \\
  \frac{B_y(\zeta)}{\sqrt{m_1 m_2}}&  \frac{A_y(\zeta)}{m_2}
  \end{pmatrix} \, ,
   \label{eq:MxMyr}
   \end{equation}
with
\begin{equation}
\begin{aligned}
A^{(v)} =&- i  \frac{m_1m_2 }{2m_0} \eta \left(1-\frac{1}{\delta(\zeta)^3}\right) \cos \zeta \, ,\\
B^{(v)} =&i \frac{m_1m_2 }{2m_0}  \eta \left(1-\frac{1}{\delta(\zeta)^3}\right) \exp^{i\zeta} \, ,\\
A^{(h)} = &\frac{1}{4\delta(\zeta)^5}(5\cos 2\zeta - 13 +8 \cos \zeta) - \cos \zeta\, ,\\
B^{(h)} = &\exp^{-2i\zeta}-\frac{1}{8\delta(\zeta)^5}(\exp^{-3i\zeta}+16\exp^{-2i\zeta}\\
&  -26\exp^{-i\zeta}+9\exp^{i\zeta})\, . 
\end{aligned}
\label{eq:coefxy}
\end{equation}

\section{Stability of partial equilibria}
\label{ap:pstab}
We can study the stability of the Lagrangian points even if the equations of variation (\ref{eq:diagc}) are not constant over time by studying the stability of partial equilibria \citep{Vorotnikov2002}. To do so, we divide the variables two groups: the variables with respect to which the stability is investigated $\bm{z}=$($z_1$, $z_2$), and the remaining variable $\bm{\gamma}=$($L$, $m_1$,$m_2$). Their equations of variation is given by the system:
\begin{equation}
\left\{ 
\begin{aligned}
% \dot \Zt & =  \nu_0(\Gamma)  \Zt\, , \\
% \dot  \Zt_2 &= \nu_+(\Gamma)  \Zt_2\, , \\
%  \dot  \zt  &= \nu_-(\Gamma)    \zt\, , \\
%      \dot \Gamma &= - \alpha \frac{\Gamma^{1-2k}}{\tG}\, ,
\frac{\dot z_j}{z_j} &= u(L,m_1,m_2)+ (-1)^j \nu (L,m_1,m_2)\, ,\\
\dot L &=\Gamma_1+\Gamma_2 + \frac{\dot m_1 + \dot m_2}{m_1+m_2} L\, , \\
\frac{m_j}{m_j}&=Acc_j(L,m_1,m_2) \, .
\end{aligned}
\right.
\label{eq:dagcdis}
   \end{equation}
where $Acc_j$ is the accretion rate of the planet $j$. The system (\ref{eq:dagcdis}) can be rewritten: 
\begin{equation}
\begin{aligned}
   \dot{\bm{z}} =  \bm{Z}(t, \bm{z}, \bm{\gamma})\, , \  \ \dot{\bm{\gamma}} = F(t,  \bm{z} ,\bm{\gamma}) \, , \ \ \bm{Z}(t,\mathbf 0,\bm{\gamma})= 0\, .
\end{aligned}
\label{eq:dagcdis2}
   \end{equation}
We note that the equations of variations of each component of the vector $\bm{z}$ are uncoupled (Eq. \ref{eq:dagcdis}). We hence study the stability in the direction of each component $\bm{z}_j$ separately. \\
Following \citep{Vorotnikov2002}, let $a(r)$ and $b(r)$ be arbitrary continuous, monotone increasing functions for $r \in [0,h]$, where $h$ is a positive real number, and such as $a(0)=b(0)=0$. If for the system (\ref{eq:dagcdis2}) a scalar function V exist such that
\begin{equation} 
\begin{aligned}
a(|| \bm{z}_j||) \leq V(t,\bm{z}_j,\gamma) \leq b(|| \bm{z}_j||) \ \ (a) \, , \\
 \dot{V} \leq 0 \ \ (b) \ \ \text{and} \ \ V(t,\bm{0},\bm{0}) \equiv  0 \ \ (c) \, ,
\end{aligned}
   \end{equation}
%For $t \geq 0$, $|| \bm{z}|| \leq h$, and $|\Gamma| < \infty$, then the set $\bm{z}=0$ is uniformly stable.\\
then the set $\bm{z}_j=0$ is uniformly stable. To verify these conditions, we simply take:
\begin{equation}
V(t,\bm{z}_j,\Gamma)=a(||\bm{z}_j||)=b(||\bm{z}_j||)=||\bm{z}_j||^2\, .
\label{eq:sol_theorem}
   \end{equation}
%
%with $||y||^2=y1 \bar y1+y2 \bar y2$.
(a) and (c) are automatically verified, and as $\dot V=2 Re(u \pm i\nu) \bm{z}_j  \bar{\bm{z}}_j$, (b) is verified if $u$ is negative or null. The Lagrangian point $L_4$ is hence uniformly stable if $u$ is negative or null.

\section{Effect of resonant chains}
\label{sec:reschain}
 \begin{figure}
\begin{center}
\includegraphics[width=1\linewidth]{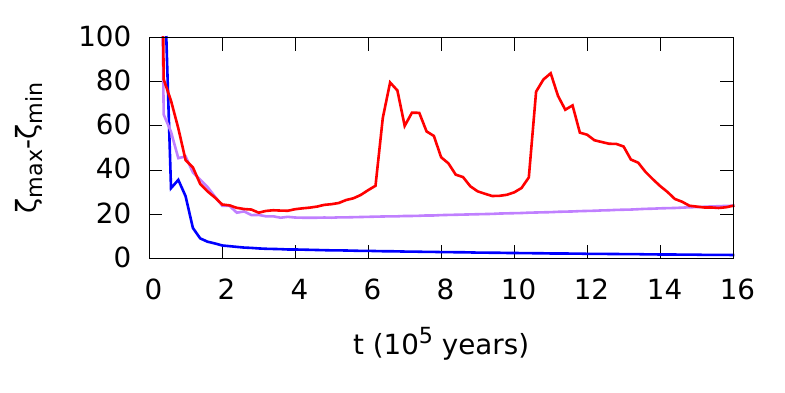}\\
\includegraphics[width=1\linewidth]{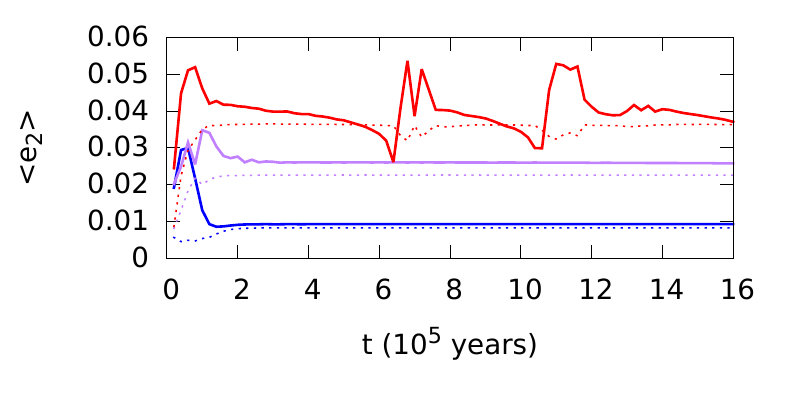}
\caption{\label{fig:rchain_ex} Evolution of the amplitude of libration of the resonant angle $\zeta_{\rm max}-\zeta_{\rm min}$ (top) and mean eccentricities (bottom) for co-orbitals ($m_1/m_2=6$) captured in a 4:3 MMR with an outer planet such that $m_3/m_1=1$ (blue), $m_3/m_1=1.6$ (purple) and $m_3/m_1=2.5$ (red). The solid lines show the eccentricity of $m_2$, while the dashed ones show the eccentricity of $m_1$. Both amplitudes of libration and mean eccentricities are taken over a single libration period. These trajectories correspond to 3 cases of the top-right panels of figs. \ref{fig:rchain_zetf} and \ref{fig:rchain_zete}.  
%{\bf Can you make the dashed lines clearer - looks like it dot-dot lines instead of dashed.}
}
\end{center}
\end{figure}
 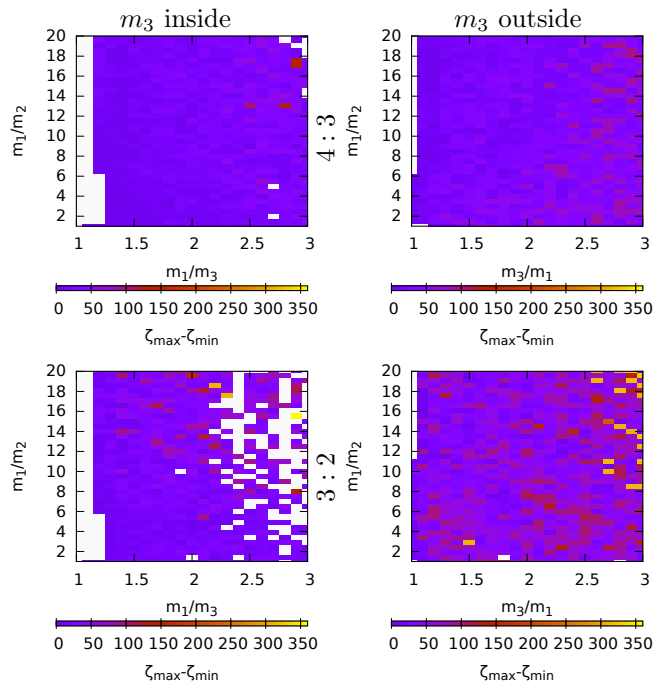
\begin{figure}
\begin{center}
\includegraphics[width=0.49\linewidth]{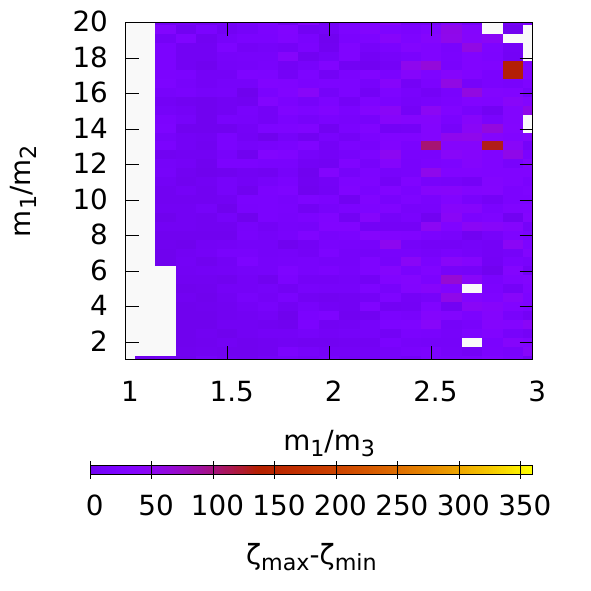}\includegraphics[width=0.49\linewidth]{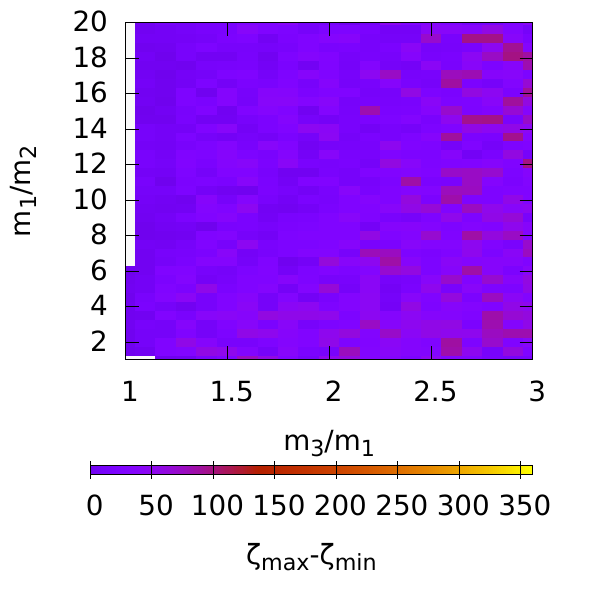}\\
\includegraphics[width=0.49\linewidth]{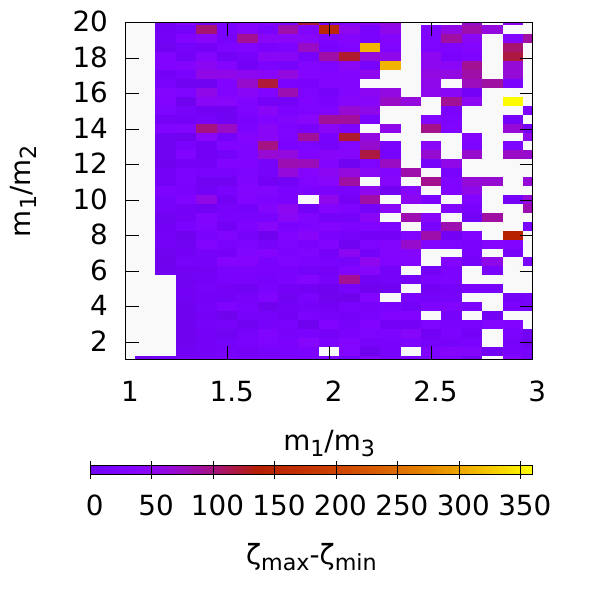}\includegraphics[width=0.49\linewidth]{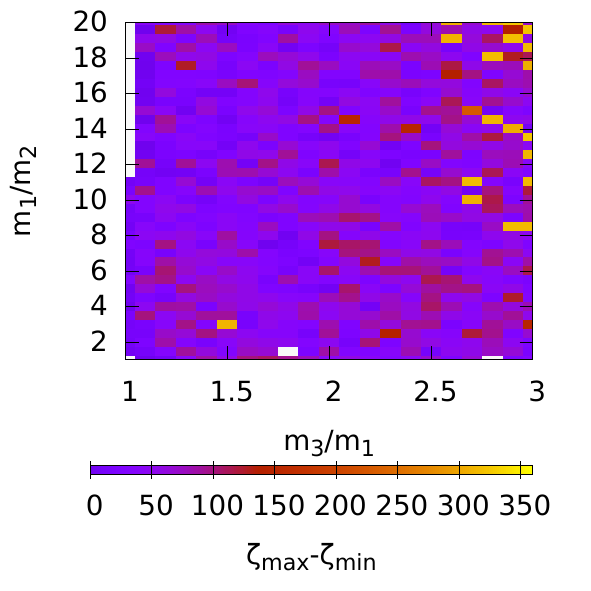}
  \setlength{\unitlength}{1cm}
\begin{picture}(.001,0.001)
\put(-0.3,2.9){\rotatebox{90}{$3:2$}}
\put(-0.3,7.4){\rotatebox{90}{$4:3$}}
\put(-2.9,9.2){$m_3$ inside}
\put(1.5,9.2){$m_3$ outside}
\end{picture}
\caption{\label{fig:rchain_zetf} Effect of the capture into the $3:4$ and $2:3$ mean motion resonance for a pair of co-orbital planets $m_1$ and $m_2$ initially in a horseshoe configuration. The co-orbitals start with an amplitude of libration of $\zeta_{max}-\zeta_{min}=320^\circ$. The colour code represents the amplitude at the end of the simulation. In most cases, the capture in MMR with another planet tends to greatly reduce the amplitude of libration of the co-orbitals' resonant angle. White pixels represent the systems for which the co-orbitals were not in the intended resonance with $m_3$ at the end of the simulation. See the text for more details. }
\end{center}
\end{figure}
 \begin{figure}
\begin{center}
\includegraphics[width=0.49\linewidth]{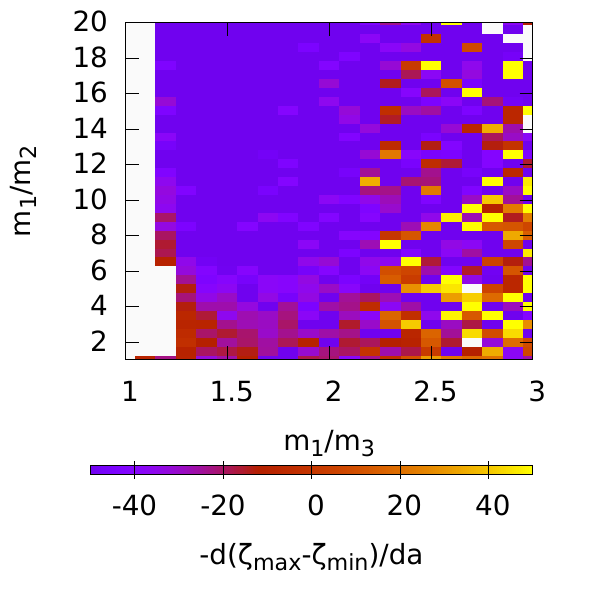}\includegraphics[width=0.49\linewidth]{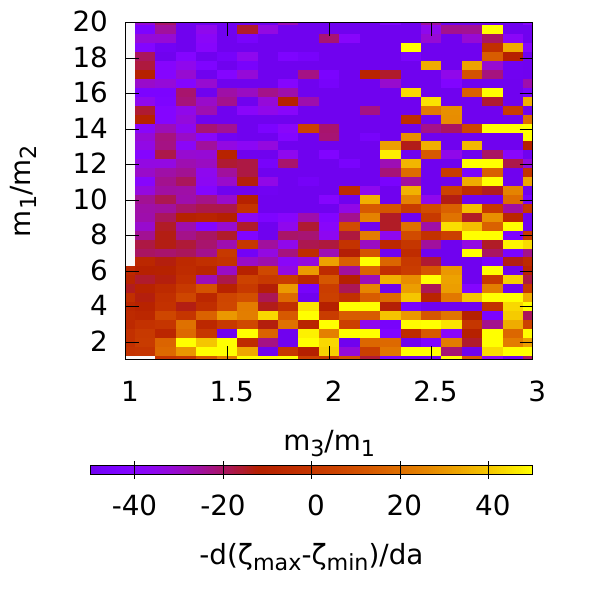}\\
\includegraphics[width=0.49\linewidth]{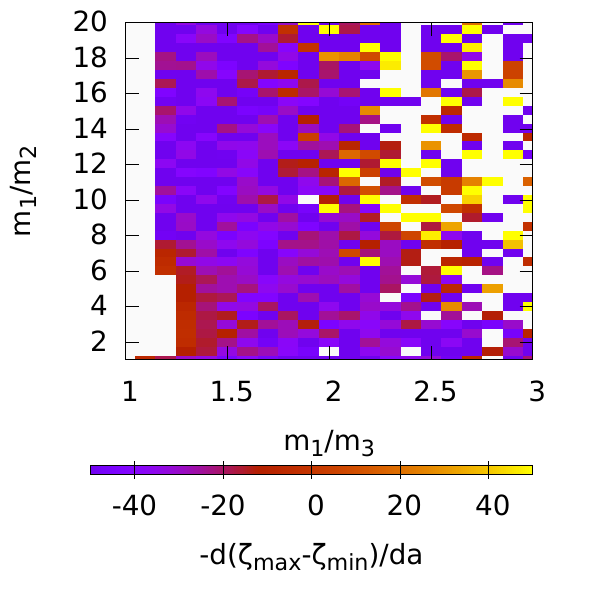}\includegraphics[width=0.49\linewidth]{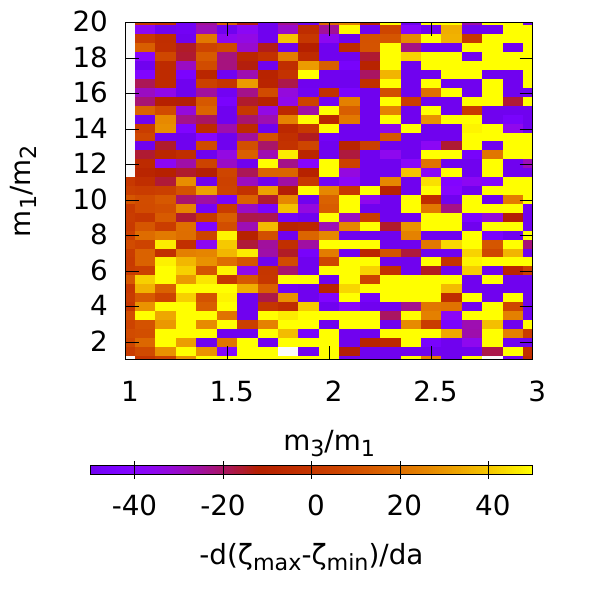}
  \setlength{\unitlength}{1cm}
\begin{picture}(.001,0.001)
\put(-0.3,2.9){\rotatebox{90}{$3:2$}}
\put(-0.3,7.4){\rotatebox{90}{$4:3$}}
\put(-2.9,9.2){$m_3$ inside}
\put(1.5,9.2){$m_3$ outside}
\end{picture}
\caption{\label{fig:rchain_zete} Evolution of the amplitude of the libration of the resonant angle $\zeta_{max}-\zeta_{min}$ (with respect to the evolution of the semi-major axis) once the co-orbitals are captured in a mean-motion resonance with another planet. The systems are the same as in Fig. \ref{fig:rchain_zetf}.  Purple colour indicates that the co-orbitals keep converging toward the exact resonance ($L_4$ or $L_5$), while yellow indicates that the amplitude is increasing post-capture. White pixels represent the systems for which the co-orbitals were not in the intended resonance with $m_3$ at the end of the simulation. See the text for more details.  }
\end{center}
\end{figure}
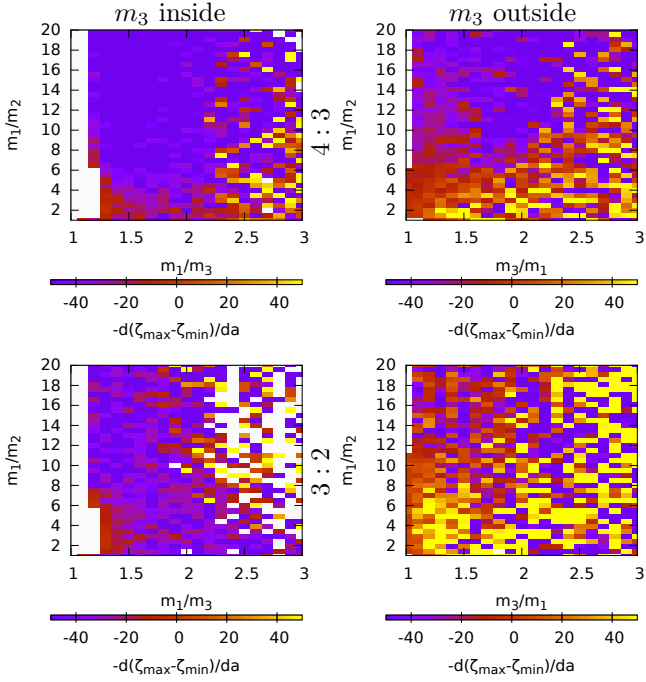

The existence of trojans with small amplitudes of libration in the synthetic planetary systems described in Sect. \ref{sec:pop_outcome} prompted us to study the effect of resonant chains on the co-orbital configuration. A complete study of the stability of co-orbitals in resonant chains is beyond the scope of this paper, so we restrain this analysis to the effect of a single planet inside or outside of a co-orbital configuration, trapped in a 3:2 or 4:3 mean motion resonance (MMR).

To begin with, we look at three examples where two co-orbitals ($m_1/m_2=6$) are captured in a 4:3 MMR with an outer planet $m_3$ such that $m_3/m_1=[1,1.6,2.5]$. Figure \ref{fig:rchain_ex} shows the evolution of the amplitude of libration as well as the eccentricity of the co-orbitals after their capture. In these three cases, the amplitude of libration, which was initially at $\zeta_{max}-\zeta_{min}=320^\circ$, is quickly reduced down to a few tens of degree by the capture into the 4:3 MMR with the 3rd planet. However, past the first $2 \times 10^5$ years, the chosen examples exhibit three different behaviours: the blue case sees its amplitude of libration monotonically decrease over time, while the purple one keeps increasing after the first phase of the capture. In both of these cases, the eccentricities of the co-orbitals reach an equilibrium value, typical for two planets migrating in a first order MMR. The red case displays a more complex behaviour for both its amplitude of libration and eccentricities.\\

The effect of the relative masses between the co-orbitals $m_1$ and $m_2$, and the 3rd planet $m_3$ is studied by integrating a grid of initial conditions, taking for the co-orbitals $a_1=a_2=1$ au, $\zeta=20^\circ$ (which yield an amplidue of libration of $320^\circ$, hence a horseshoe configuration), and masses $m_1/m_2$ in the $[1:20]$ range. Both the 3:2 and 4:3 MMR are studied. The mass of the 3rd planet is $m_3=3\times 10^{-5}m_0$ when it is at a larger semi-major axis than the co-orbitals, and $m_3=1\times 10^{-5}m_0$ when it is at a smaller semi-major axis. 
%$m_1$ and $m_2$ can be deduced from the x and y axis of the Figures \ref{fig:rchain_zetf} and \ref{fig:rchain_zete}. 
All eccentricities and inclinations are initially set to $0$. $\tau_{a_j}=10/m_j$, and  $\tau_{e_j}=\tau_{a_j}/150$, which are consistent with the parameters of the disc described in Sect. \ref{sec:gavintype1} and \cite{TaWa2004}. 
We set $K=0$ ($\tau_{a_j}$ independent from $a_j$). Each initial condition (each set of masses) is integrated for $1.6\times 10^{5}$ years, which corresponds to a migration down to $\approx .4$ au for the co-orbitals, depending on the chosen masses. 
%Taking $K=0$ implies that non-equal mass co-orbital would diverge from the exact resonance if they were evolving on their own (see Fig. \ref{fig:tauacritTana}).

Results are displayed in Fig. \ref{fig:rchain_zetf} and \ref{fig:rchain_zete}. In both figures white pixels represent the systems for which the co-orbitals were not in the intended resonance with $m_3$ at the end of the simulation. For the systems close to $m_3=m_1$, the planet did not converge or did not converge fast enough to reach the desired resonance during the simulation. For the other white pixels, the desired resonance was crossed but the capture did not happen, or did not hold. 

In Fig. \ref{fig:rchain_zetf}, each pixel represents the final amplitude of libration of the co-orbital configuration for that set of masses. In almost all the studied cases, the capture into a mean-motion resonance with another planet led the co-orbital configuration to greatly reduce its own amplitude of libration, going from horseshoe to trojan configuration.

Figure \ref{fig:rchain_zete} shows the evolution of the amplitude of libration of the co-orbitals once they are captured in the MMR with $m_3$. The set of integrations is the same as in Fig. \ref{fig:rchain_zetf}. The quantity $-d(\zeta_{max}-\zeta_{min})/da$ is obtained by comparing the amplitude of libration of the co-orbitals between $t=[t_{max}/3:2t_{max}/3]$ and $[2t_{max}/3:t_{max}]$. Areas of blue or purple pixels follow a similar behaviour to the blue example of Fig. \ref{fig:rchain_ex}, where the amplitude of libration keeps decreasing post-capture. Orange or yellow pixels show mass ratios for which the amplitudes of libration are, on average, increasing post-capture. For comparison, in absence of $m_3$ these co-orbital configurations would be diverging from the equilibrium for all values of $m_1/m_2=(1:20]$, since in these examples, $K=0$ (see Fig. \ref{fig:tauacritTana}). The dependency of the stability with respect to the mass ratios is obviously complex and will be the object of a future study. We can nonetheless see from Fig. \ref{fig:rchain_zetf} and \ref{fig:rchain_zete} that resonant chains can have a stabilising effect on the co-orbital configuration.

\section{Forces partial derivatives}
\label{ap:Forcespd}

 \begin{figure}
\begin{center}
\includegraphics[width=0.99\linewidth]{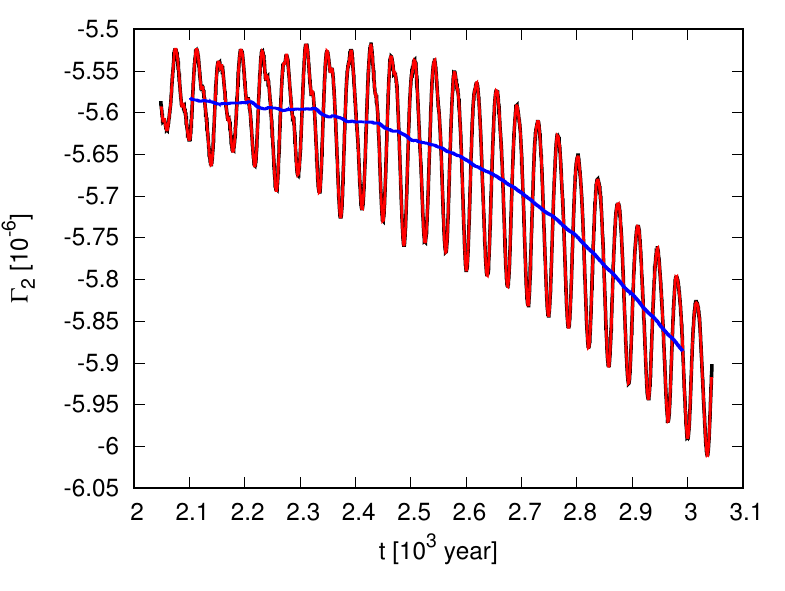}
\caption{\label{fig:Ftevol_12Me} Torque applied on the $12M_\oplus$ planet in the $6M_\oplus$ $12M_\oplus$ case. }
\end{center}
\end{figure}

 \begin{figure}
\begin{center}
\includegraphics[width=0.99\linewidth]{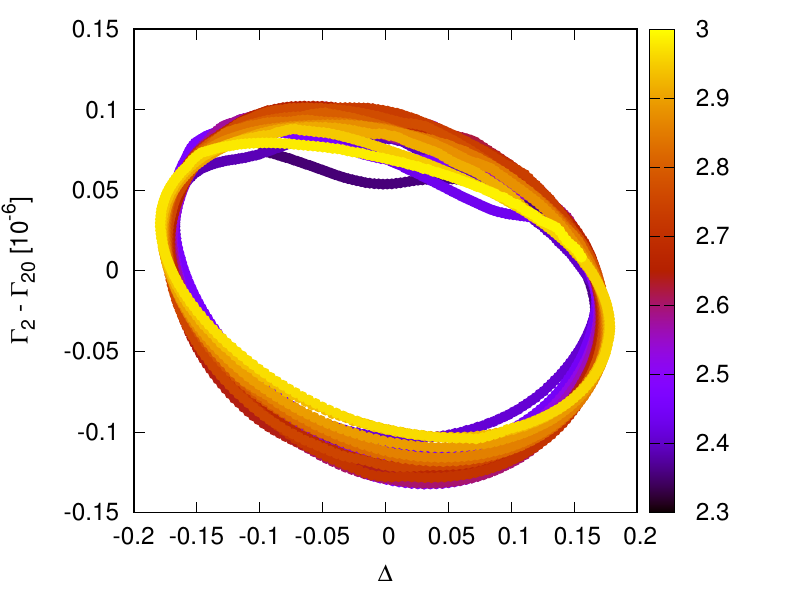}\\
\caption{\label{fig:IFt_12Me} $\Gamma_2$ vs $\Delta$ in the $6M_\oplus$ $12M_\oplus$ case. The colour code is the time in [$10^3$  year]. }
\end{center}
\end{figure}

 \begin{figure}
\begin{center}
\includegraphics[width=0.99\linewidth]{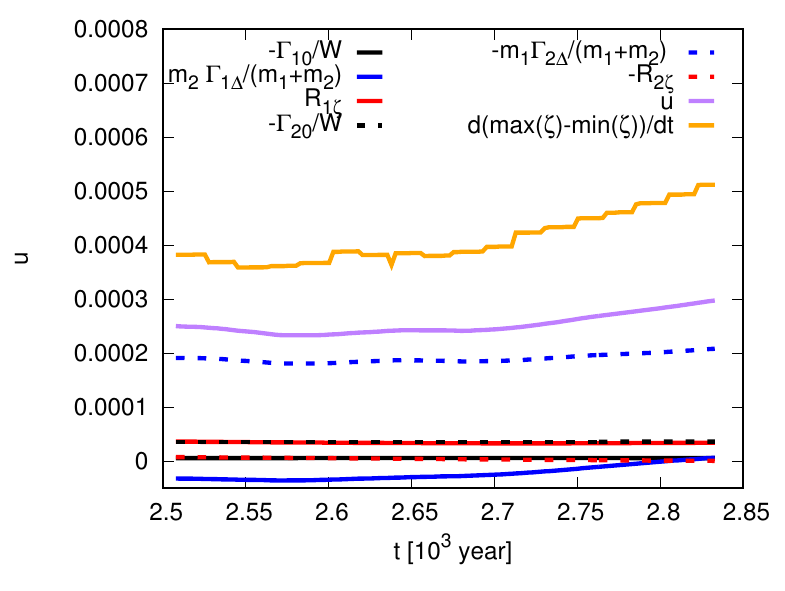}\\
\caption{\label{fig:u_12Me} Temporal evolution of the terms of $u$ in the  $6M_\oplus$ $12M_\oplus$ case. }
\end{center}
\end{figure}

The forces applied by the disc on each planet during the hydrodynamical simulations were saved with a time-step of 0.05 year. To compute the partial derivatives $\Gamma_{j\Delta}=\frac{\partial \Gamma_j}{\partial \Delta}$ and $R_{j\zeta}=\frac{\partial R_j}{\partial \zeta}$ we first performed a sliding averaging of the quantities $ \Gamma_j$ and $R_j$ over the local orbital period, in order to get rid of short terms effect, notably the oscillations due to the eccentricities. The raw torque and the result of this averaging are displayed in black and red in Fig. \ref{fig:Ftevol_12Me}, respectively, in the case $m_1=6M_\oplus$ (leading),  $m_2=12M_\oplus$ (trailing), discussed in Sect. \ref{sec:type1} . 

Then we computed $\Gamma_{j0}$ and $R_{j0}$ by performing an additional sliding average over the local libration period. This period was computed by frequency analysis of the resonant angle $\zeta$. For our example, the instantaneous value of $\Gamma_{20}$ is displayed in blue in Fig. \ref{fig:Ftevol_12Me}. The quantity $\Gamma_{j}-\Gamma_{j0}$ is shown in Fig. \ref{fig:IFt_12Me}, with respect to the variable $\Delta$ and time (colour code). As we made a linear approximation in the neighbourhood of $\Delta=z=0$ in the analytical part of the study, we fitted $\Gamma_{jI}\Delta+\Gamma_{j\zeta}  z$ to the quantity $\Gamma_{j}-\Gamma_{j0}$. Here again, this fit is done over a sliding window of width 1-libration period. 
 
\section{Evolution of eccentricity under mass change}
\label{ap:emdot}

In the conservative case, described in Sect. \ref{sec:conse}, the diagonalisation is obtained by a change of variable $ X= P_X {\cal X}$, where the columns of the matrix $P_X$ are proportional to the eigenvectors of the matrix $M_X$ (this change is thus not unique). In the case of evolving masses, the matrix $P_X$ is not constant. We obtain the following relation:
 \begin{equation}
 \begin{aligned}
\dot{{\cal X}}= M_{{\cal X}} {\cal X} = \left( P_X^{-1} M_X P_X - P_X^{-1} \dot P_X \right) \cal X
    \end{aligned}
   \label{eq:dcalX}
   \end{equation}
We hence look for a change of variables ${\cal X} = P_{\cal X} X$ that diagonalises $M_{{\cal X}}$ at second order in $\eps$. To do so, we look for a change of basis $\eps$ close to $P_X$:  $P_{\cal X}= P_X+ \eps P'_X$. Noting $P_{\cal X}[j,k]$ the $k^{th}$ element of the $j^{th}$ line, the chosen change of basis is:
\begin{equation}
\begin{aligned}
P_{{\cal X}}[1,1] &=m_2 \operatorname{e}^{i \frac{\pi}{3}} \\
P_{{\cal X}}[1,2] &= \operatorname{e}^{i \frac{\pi}{3}} +  \frac{\eps\operatorname{e}^{-i \frac{\pi}{6}}}{g_{L_4}} \left( \frac{\dot X_{1,d}}{X_1}- \frac{\dot X_{2,d}}{X_2} \right)\\
 P_{{\cal X}}[2,1] &=-m_1 + \frac{ \eps i }{g_{L_4}} \left[ \dot m_1 - \frac{m_1}{m_2} \dot m_2 + m_1 \left( \frac{\dot X_{1,d}}{X_1}- \frac{\dot X_{2,d}}{X_2} \right) \right]\\
P_{{\cal X}}[2,2] &= 1\\
\end{aligned}
     \label{eq:PcX}
   \end{equation}
while $ M_{{\cal X}} $ reads:
\begin{equation}
\dot M_{{\cal X}} =
 \begin{pmatrix}
 g_1 + r_{x1}  & 0 \\
0  & g_2 + r_{x2} \\
   \end{pmatrix}
      \begin{pmatrix}
 z_1 \\
 z_2
     \end{pmatrix}
     \label{eq:diagX}
   \end{equation}
where
\begin{equation}
\begin{aligned}
g_1 & =\eps i \frac{27 (m_1+m_2)\eta_L}{8m_0}\, , \\
r_{x1} &= \eps^2 \frac{m_2\frac{ \dot X_{1,d}}{X_1}+m_1 \frac{\dot X_{2,d}}{X_2}-\dot m_1 - \dot m_2}{m_1+m_2}\, , \\
g_2 & =0\, , \\
r_{x2} &= \eps^2 \frac{m_1\frac{ \dot X_{1,d}}{X_1}+m_2 \frac{\dot X_{2,d}}{X_2}}{m_1+m_2}\, , \\
\end{aligned}
     \label{eq:diagec}
   \end{equation}
The temporal evolution of the variable ${\cal X}_j$ is hence simply given by $ {\cal X}_j(t)={\cal X}_j(0)\operatorname{e}^{(g_j+r_{xj})t}$. Orbits along the eccentric Lagrangian equilibria are defined by ${\cal X}_1=0$, while those along the anti-Lagrangian equilibria are defined by ${\cal X}_2=0$. For both of these configurations, the evolution of the eccentricities can be estimated from the quantity $X_j\bar X_j=e_j^2$:
\be
2e_j\dot e_j=\dot X_j \bar X_j+\dot{\bar{X_j}}  X_j
\label{eq:edot}
\ee
where $ \bm{X}= P^{-1}_{\cal X} {\cal X}$, and  $ \dot{\bm{X}}= \dot P^{-1}_{\cal X} {\cal \bm{X}}+ P^{-1}_{\cal X} \dot{\cal \bm{X}}$.\\

\end{document}